%% file: HParxivsub.tex
\documentclass[10pt]{article}
\usepackage{jheppub}

\usepackage{indentfirst}
\usepackage{graphicx,wrapfig,float,slashed}
\usepackage{amsmath,amssymb,dsfont,epsfig,graphicx,xcolor}
\usepackage{epstopdf}
\usepackage{ragged2e}
\usepackage{comment}
\usepackage{enumerate}
\usepackage{lipsum}
\usepackage{xcolor}
\usepackage{tabu}
\usepackage[toc]{appendix}
\usepackage[normalem]{ulem}
\usepackage[font=small,skip=1pt]{caption}
\allowdisplaybreaks
\usepackage{pifont}
\usepackage{subcaption}
\usepackage{cancel}
\def\VEV#1{\left\langle #1 \right\rangle}

\newcommand{\bea}{\begin{eqnarray}}
\newcommand{\eea}{\end{eqnarray}}
\newcommand{\beq}{\begin{equation}}
\newcommand{\eeq}{\end{equation}}
\newcommand{\non}{\nonumber}
\newcommand{\MPsi}[1]{M_{\Psi_{#1}}}
\newcommand{\ptcut}{p_T^{\rm cut}}
\input{FeynTikZ.tex}
\title{Composite Higgs at high transverse momentum}
\author{Andrea Banfi$^1$, Barry M. Dillon$^2$, Wissarut Ketaiam$^1$, and Sandra Kvedarait\.e$^1$}

\affiliation{\textit{\small{$^1$ Department of Physics and Astronomy, University of Sussex, BN1 9QH Brighton, UK}}\newline
\textit{\small{$^2$ Department of Theoretical Physics, Jozef Stefan Institute, 1000 Ljubljana, Slovenia}}}
\emailAdd{a.banfi@sussex.ac.uk}
\emailAdd{barry.dillon@ijs.si}
\emailAdd{w.ketaiam@sussex.ac.uk}
\emailAdd{S.Kvedaraite@sussex.ac.uk}

\abstract{In this paper we explore composite Higgs scenarios through
  the effects of light top-partners in Higgs+Jet production at the
  LHC.  The pseudo-Goldstone boson nature of the Higgs field means
  that single-Higgs production via gluon fusion is insensitive to the
  mass spectrum of the top-partners. However in associated production this is
  not the case, and new physics scales may be probed.  
   In the course of the work we consider scenarios with both one and two light top-partner multiplets in the spectrum of composite states.
   We study corrections to the Higgs couplings and the effects that the light top-partner
  multiplets have on the transverse momentum spectrum of the Higgs.
  Interestingly, we find that the
  corrections to the Standard Model expectation depend strongly on the
  representation of the top-partners in the global symmetry.  
 }

\begin{document}

\maketitle
\flushbottom

\section{Introduction}

\noindent The Standard Model (SM) description of particle physics does not
contain a dynamical description of electroweak symmetry breaking
(EWSB), or a natural explanation of the vast hierarchies in the Yukawa
couplings of the theory.  There are many proposed extensions to the SM
which provide these explanations and in turn predict deviations in the
Higgs properties with respect to what one would expect from the SM
description.  Thus far the production and decay rates of the scalar
resonance discovered in 2012 by the ATLAS and CMS
collaborations~\cite{Aad:2012tfa,Chatrchyan:2012xdj} fit the
description of those of the SM Higgs boson, providing constraints on
these beyond-the-SM (BSM) models.  However, as more data is collected,
it is possible that new physics effects could present themselves soon.

Composite Higgs
models~\cite{Kaplan:1983fs,Kaplan:1983sm,Georgi:1984af,Dugan:1984hq,Contino:2003ve}
are one particular class of models which provide an elegant
description of a dynamical origin for EWSB and of hierarchical Yukawa
couplings.  These new physics scenarios posit that the Higgs boson is
a composite state of some strongly coupled gauge theory with a
confinement scale of the order of one TeV.  Along with the Higgs boson
we expect a plethora of new composite states laying near or above the
confinement scale, and we refer to these collectively as the strong
sector.  The hierarchy between the Higgs mass and the confinement
scale is neatly accommodated in the theory when the Higgs field arises
as a set of Goldstone bosons generated by the spontaneous breaking of
some global symmetry of the UV theory.  The different composite Higgs
models are labeled by the coset representing the spontaneous symmetry
breaking which gives rise to the Higgs doublet, with the minimal
models being $SU(3)\times U(1)/SU(2)\times U(1)$ and
$SO(5)\times U(1)/SO(4)\times
U(1)$~\cite{Agashe:2004rs,Contino:2006qr}. In this paper we will focus
on the latter model because the unbroken global symmetry contains the
custodial group, and thus bounds from electroweak precision and the
$Z\bar{b}b$ measurements are less
stringent~\cite{Agashe:2006at}.  Given that the Higgs field is
composed of Goldstone bosons, it has no potential at tree-level and
thus the electroweak symmetry must be broken radiatively, making the
Higgs boson a pseudo-Nambu-Goldstone-Boson (pNGB).

The mechanism at the heart of both EWSB and the Yukawa coupling
hierarchies is partial compositeness, through which
the SM fermions couple linearly to composite fermions from the strong
sector.  
These couplings necessarily break the global
symmetry of the strong sector, because the SM fermions cannot transform as complete multiplets of this global symmetry, and a Higgs potential is generated
radiatively.  
Given that the top quark has the strongest coupling with
the Higgs one would rightly assume that this field, and the composite
states to which it couples, provide important contributions to the Higgs potential.  
The composite states that couple to the top-quark through partial-compositeness are known as `top-partners' and in explicit composite Higgs
constructions, i.e. those in which the Higgs potential is finite, it
is invariably found that the top-partner masses are directly related
to the fine-tuning required in the Higgs potential to achieve the
observed Higgs mass and vacuum expectation value, with lighter masses
corresponding to less fine-tuning.  A variety of explicit
constructions have been developed over the years, most notably the
holographic models~\cite{Hosotani:1983xw,Hosotani:1983vn,Maldacena:1997re,Witten:1998qj,Randall:1999ee,Rattazzi:2000hs,Huber:2000ie,Contino:2003ve,Contino:2004vy}.
There has also been much success with the discrete site
models~\cite{Panico:2011pw} and models employing Weinberg sum
rules~\cite{Marzocca:2012zn}.  
It is well known by now that in order to reduce fine-tuning in the Higgs potential in composite Higgs models light to-partners are generically required.
This was first observed in \cite{Contino:2006qr} and subsequently studied in more detail in \cite{Matsedonskyi:2012ym} where the relation between the Higgs mass, the top-partner mass, and compositeness scale was understood analytically.
In \cite{Panico:2012uw} the authors confirmed this behaviour for a variety of embeddings of the top-partners in the global symmetry, noting that the case in which the right-handed top quark is composite belongs to a class of minimally tuned composite Higgs scenarios.
It has been noted in \cite{Croon:2015wba} that the holographic models, as studied in this paper, do in fact allow for an alleviation of the tension with light top-partners observed in \cite{Contino:2006qr} by reducing the UV scale of the 5D model.
In the 4D picture this is analogous to a modification in the number of colours in the strongly coupled gauge theory underlaying the compositeness sector.
Phenomenological bounds on the masses of new
vector-like fermionic states are readily available, the absence of
any direct detection of top-partner states at the LHC puts a lower
bound on their mass at around $1400$ GeV~\cite{ATLAS:2016cuv,CMS:2017jfv,CMS:2017wwc,CMS:2018pma,Sirunyan:2019tib,CMS:2018vou,CMS:2018haz,Aaboud:2017qpr,Aaboud:2017zfn,Sirunyan:2017ynj,Sirunyan:2017pks,Sirunyan:2018fjh,Sirunyan:2018omb,Aaboud:2018uek,ATLAS:2018qxs}.
Through the measurement of Higgs couplings to gauge bosons a direct lower
bound on the decay constant of the Higgs field is found to be
$600\!-\!800$ GeV, depending on the specific model~\cite{Sanz:2017tco}.

In this paper we will investigate the effects that top-partners in
composite Higgs models have on the Higgs+Jet production process at the
LHC.  When studying top-partner phenomenology in composite Higgs
models it is necessary to use simplified models to capture the
relevant features of the top-partner states without over-complicating
the parameter space.  In this spirit we follow closely the simplified
phenomenological models described
in~\cite{DeSimone:2012fs,Marzocca:2012zn}.  The effects of top-partner
states on single-Higgs production via gluon fusion have been studied
in detail, however in this case the pNGB nature of the Higgs boson
leads to a cancellation of new physics effects dependent on the
top-partner masses in the production
cross-section~\cite{Kniehl:1995tn,Azatov:2011qy,Low:2010mr}.  To probe
the top-partners in gluon initiated Higgs production the produced
Higgs must be allowed to recoil off a gluon, and for this the study of
Higgs production in association with a jet is useful.  This process
has been studied in some detail
already~\cite{Banfi:2013yoa,Grojean:2013nya,Azatov:2016xik}.  In this paper we take
the study of Higgs+Jet production further in two ways; first of all we
show how different top-partner models result in different signatures
in the $p_T$ spectra of the Higgs, and secondly we go beyond the
simplified models in~\cite{DeSimone:2012fs} and study the effects of a
second light top-partner.  In the course of this work we also study
the effects of CP violating couplings of the top quark and
top-partners in the Higgs+Jet process.

In the next section we will introduce the simplified models we use in
our analysis and discuss partial compositeness in the top sector of
the Minimal Composite Higgs Model (MCHM) based on the $SO(5)/SO(4)$
coset.  After describing the different top-partner states that we have
in these models we give a brief overview of the relevant collider
bounds and discuss how they constrain the parameter space we study.
In section~\ref{sec:massyuk} we study the mass spectrum and Yukawa
couplings of the top-partners when we have more than one light
multiplet.  In section~\ref{sec:HiggsProd} we will discuss the main
features of single-Higgs production via gluon fusion and in
association with a jet, highlighting the differences between the two
processes.  Finally in section~\ref{sec:results} we will present the
results of our analysis on each of the simplified models in scenarios
with both one and two light top-partner states.

\section{Top-partners in the Minimal Composite Higgs Model} \label{sec:background}

\noindent The starting point in writing down the MCHM is the
assumption that the strongly coupled gauge theory underlying the
composite dynamics has a global $SO(5)\times U(1)_X$ symmetry, which
at the confinement scale $f$ is spontaneously broken to
$SO(4)\times U(1)_X$.  The four Goldstone bosons arising from this
symmetry breaking pattern form an $SO(4)$ fourplet in the
$SO(5)/SO(4)$ coset which we identify with the Higgs field.  The fact
that this breaking preserves the custodial symmetry has important
consequences for the phenomenological bounds on the
model~\cite{Agashe:2006at}, as mentioned in the Introduction.
Following~\cite{DeSimone:2012fs,Marzocca:2012zn} we assume that all
the SM fields enter as elementary fields with the exception of the
right-handed top quark, $t_R$, which we assume to be a chiral bound
state of the strong sector.  The gauge fields are coupled to the
strong sector through the gauging of the $SU(2)_L\times U(1)_Y$ subset
of $SO(4)\times U(1)_X$ symmetry, with the hypercharge generator being
associated with the diagonal generator of $SU(2)_R$ plus the $X$
generator, i.e. $Y=T^{3}_R+X$.  The SM fermions are coupled to the
strong sector through the partial compositeness mechanism, where
operators containing SM quarks are coupled to operators of the strong
sector.  The SM quark doublet cannot fill a complete $SO(4)$ multiplet
without the introduction of additional external states while the
states in the strong sector can, thus some of the components of this
multiplet will be spurious and lead to explicit breaking of the
$SO(5)$ symmetry.

We will use the standard CCWZ~\cite{Coleman:1969sm,Callan:1969sn} toolkit to determine the
structure of our top-partner effective field theory (EFT) given the
$SO(5)/SO(4)$ coset.  These techniques were first used for top-partner
studies in~\cite{Marzocca:2012zn}, and for a detailed account of the application
of this formalism to the models we employ here we refer the reader to
\cite{DeSimone:2012fs}.  The main objects we require in our analysis are the
Goldstone boson matrix $U$ and the $d_\mu$ vector used to construct
the kinetic term of the Goldstone boson Lagrangian.  Under $SO(5)$
rotations the Goldstone matrix transforms non-linearly as
$U\rightarrow gUh^\dagger$, with $g\in SO(5)$ and $h\in SO(4)$,
whereas $d_\mu$ transforms linearly as a fourplet of $SO(4)$.  In
unitary gauge the Goldstone boson matrix can be expressed
as \begin{equation}
  U=\begin{pmatrix}&&&0&0\\&\mathbb{I}_{3\times3}&&0&0\\&&&0&0\\0&0&0&c_h&-s_h\\0&0&0&s_h&c_h \end{pmatrix},
\end{equation} where $s_h=\sin\tfrac{h}{f}$ and $c_h=\cos\tfrac{h}{f}$.  When
the Higgs is expanded around a vacuum expectation value $\VEV h$ we
take $h(x)=\langle h\rangle+\rho(x)$ and fix
$f\sin\tfrac{\langle h\rangle}{f}=v$, with
$v\simeq 246\,\mathrm{GeV}$, such that the electroweak gauge boson
masses are the same as in the SM.  We also define $\epsilon\equiv\tfrac{\langle h\rangle}{f}$, and use the short-hand notation $\sin\epsilon\equiv s_\epsilon$ and
$\cos\epsilon\equiv c_\epsilon$.

If we now construct top-partner states $\Psi$ in representations of
$SO(4)$ we can promote these to $SO(5)$ representations using the
Goldstone boson matrix $U$.  As in~\cite{DeSimone:2012fs}, we will
study top-partners in either the $\mathbf{1_{2/3}}$ or
$\mathbf{4_{2/3}}$ representations of $SO(4)\times U(1)_X$, while the
SM doublet quarks will be embedded in either a $\mathbf{5_{2/3}}$ or
$\mathbf{14_{2/3}}$ of $SO(5)\times U(1)_X$.  The right-handed top
quark will always be defined as a $\mathbf{1_{2/3}}$ of
$SO(4)\times U(1)_X$, since it is being treated as a composite chiral
state.  The SM left-handed quark doublets in the $\mathbf{5}$ and the
$\mathbf{14}$ are written in the form
\begin{equation}
Q_L^{\mathbf5}=\frac{1}{\sqrt{2}}\begin{pmatrix} ib_L \\ b_L \\ it_L\\ -t_L\\ 0 \end{pmatrix},~~Q_L^\mathbf{14}= \frac{1}{\sqrt{2}} \begin{pmatrix} 0 & 0 & 0 & 0 & ib_L \\ 0 & 0 & 0 & 0 & b_L \\ 0 & 0 & 0 & 0 & it_L \\ 0 & 0 & 0 & 0 & -t_L \\ ib_L & b_L & it_L & -t_L & 0 \end{pmatrix}.
\end{equation}
The right-handed top quark can be embedded in an $SO(5)$ fiveplet as
\begin{equation}
t_R^{\mathbf1}=\begin{pmatrix} 0 & 0 & 0 & 0 & t_R \end{pmatrix}^T.
\end{equation}
In writing down the effective Lagrangian for the Higgs field, the top quark, and the top-partners, it is useful to also write the vector-like top-partners as embeddings in $SO(5)$ multiplets.  The top-partners in $\mathbf{4}$ and $\mathbf{1}$ representations of $SO(4)$ can be embedded in $SO(5)$ fiveplets as
\begin{equation}
\Psi^{\mathbf4}=\frac{1}{\sqrt{2}}\begin{pmatrix} i B-i X_{5/3}\\ B+ X_{5/3}\\i T+i X_{2/3}\\- T+ X_{2/3} \\0  \end{pmatrix},~~\Psi^{\mathbf1}=\begin{pmatrix} 0 \\ 0 \\ 0 \\ 0 \\ T \end{pmatrix},
\end{equation}
respectively.  Embeddings in a $\mathbf{14}$ of $SO(5)$ follow
similarly.  The embedding of the SM quarks ensures that the theory
includes the SM $q_L=(b_L,t_L)$ doublet with $Y=1/6$ hypercharge and a
right-handed top quark with $Y=X=2/3$.  In fact the hypercharge of the
right-handed top quark fixes the $U(1)_X$ charge assignments of all
the fermionic fields described above, and the singlet top-partner has
the same SM charges as the SM right-handed top quark.  However the
quarks from the fourplet form two $SU(2)_L$ doublets, $Q=( T, B)$, has
the same SM charge assignment as the SM quark doublet and
$( X_{5/3}, X_{2/3})$ is an exotic doublet where the subscript denotes
the electromagnetic charge.

\subsection{The models}  \label{SecModels}

\noindent In this work we employ simplified
models, as outlined in~\cite{DeSimone:2012fs}, which serve to capture
the features of light top-partner states relevant for phenomenological
purposes.  These models are not complete concrete realisations and
there is not enough structure to compute a finite Higgs potential or
determine the level of fine-tuning present in the Higgs potential.  Due to this
we will assume that the Higgs mass takes its observed value and that the fine-tuning in the Higgs potential is smaller for smaller top-partner masses.  
We will however be able to calculate the top quark mass from the mixing
between the SM top quark and top-partners and this will serve as a
constraint on the parameters of the Lagrangian.

Composite Higgs models predict many new composite resonances of differing spin with masses near the compositeness scale, which we define as $m_*$.
If $m_*$ is sufficiently large one can write down an effective field theory where states above that mass scale have been integrated out.
However, in order to obtain a natural EWSB scenario we know that light top-partners are required, therefore it would be natural to suspect that the lightest top-partners have masses which lay below the scale $m_*$ and cannot be integrated out.
The approach taken in~\cite{DeSimone:2012fs} and in other simplified models, including those used by the ATLAS and CMS collaborations, assumes that only one top-partner lays below the scale $m_*$.
Allowing more than one light top-partner could drastically change the collider phenomenology as the possibility of additional cascade decays opens up and the relationship between the top-partner masses, couplings, and $f$ changes.
We will discuss the current collider bounds in section~\ref{sec:collider} however this will not be the focus of the current paper.

The effective field theory for the models we use here are constructed using the same power counting rules as in~\cite{DeSimone:2012fs} which in turn follows the `SILH' approach~\cite{Giudice:2007fh}.
Given the choices of top-partner states and SM quark embeddings described in the previous section we see that there are four top-partner models to study: $\mathbf{M4_5},~\mathbf{M4_{14}},~\mathbf{M1_5}$, and $\mathbf{M1_{14}}$.

\subsubsection*{~~~$\mathbf{M4_5}$}

\noindent With a light top-partner transforming as a $\mathbf{4_{2/3}}$ of $SO(4)$ and the SM left-handed top-bottom doublet embedded in a $\mathbf{5}$ of $SO(5)$ the relevant effective action for the SM plus the top-partner, after the states heavier than $m_*$ have been integrated out, is
\begin{align}
  \label{eq:M45L}
\mathcal{L}_{\mathbf{M4_5}}=~&i\bar{q}_L\cancel{D}q_L+i\bar{t}_R\cancel{D}t_R+i\bar{\Psi}^{\mathbf{4}}\cancel{D}\Psi^{\mathbf{4}}-M_{\Psi}\bar{\Psi}^{\mathbf{4}}\Psi^{\mathbf{4}}+ic_1\bar{\Psi}^{\mathbf{4}}_Rd_{\mu}\gamma^\mu t_R \non\\
&+yf\bar{Q}_L^{\mathbf{5}}U\Psi^{\mathbf{4}}_R+yf c_2\bar{Q}_L^{\mathbf{5}}Ut_R^{\mathbf{1}}~+~\text{h.c.}
\end{align}
where the $SO(5)$ embedding of the top-partner states is assumed.  The
$y$ in Eq.~\eqref{eq:M45L} is the coupling that mixes the elementary
and strong sectors, and $c_{1,2}$ are expected to be $\mathcal{O}(1)$
coefficients arising from integrating out the heavier states.  Notice
that the coupling proportional to $c_1$ does not carry a $y$
dependence since $t_R$ is treated as a composite state.
Fixing to unitary gauge and expanding the Higgs field around its vacuum expectation value the following mass matrix is found for the top and top-partners
\begin{equation}
\label{eq:Mass-M45}
\begin{pmatrix}\bar{t}_L\\ \bar{ T}_L\\ \bar{X}_{2/3,L}\end{pmatrix}^T\begin{pmatrix}
-\tfrac{yf c_2}{\sqrt{2}}s_{\epsilon}&\frac{y}{2}f (1+c_{\epsilon})&\frac{y}{2}f (1-c_{\epsilon}) \\ 0&-M_{\Psi}&0\\0&0&-M_{\Psi}
\end{pmatrix}\begin{pmatrix}t_R\\  T_R\\  X_{2/3,R}\end{pmatrix}.
\end{equation}
An orthogonal rotation of the $T$ and $X_{2/3}$ states reduces the above mass matrix to a mixing between just one linear combination of the top-partners, and leaves the kinetic and vector-like mass terms invariant.
If we did not do this rotation now then it would simply be part of the mass matrix diagonalization done later.
This transformation can be written as
\begin{equation}
\begin{pmatrix}t\\ T\\ X_{2/3}\end{pmatrix}\rightarrow\frac{1}{N}\begin{pmatrix}N&0&0\\0&1+c_{\epsilon}&1-c_{\epsilon}\\0&-1+c_{\epsilon}&1+c_{\epsilon}\end{pmatrix}\begin{pmatrix}t\\ T\\ X_{2/3}\end{pmatrix},~~N=\sqrt{2+2c_{\epsilon}^2},
\end{equation}
and the resultant mass matrix is
\begin{equation}
\begin{pmatrix}\bar{t}_L\\ \bar{T}_L\end{pmatrix}^T\begin{pmatrix}
-\tfrac{yf c_2}{\sqrt{2}}s_{\epsilon}&\frac{y}{2}f \sqrt{3+c_{2\epsilon}} \\ 0&-M_{\Psi}
\end{pmatrix}\begin{pmatrix}t_R\\ T_R\end{pmatrix}
\end{equation}
with the $X_{2/3}$ state is now decoupled from the top quark and the Higgs.
Upon diagonalising this mass matrix the mass  of the $T$ top-partner gets shifted away from the vector-like mass, however the masses of both the $X_{2/3}$ and $X_{5/3}$ state remain degenerate at $M_\Psi$.

\subsubsection*{~~~$\mathbf{M4_{14}}$}

\noindent For a light top-partner transforming as a $\mathbf{4_{2/3}}$ of $SO(4)$ and the SM left-handed top-bottom doublet embedded in a $\mathbf{14}$ of $SO(5)$ the relevant effective action is
\begin{align}
  \label{eq:M414L}
\mathcal{L}_{\mathbf{M4_{14}}}=~&i\bar{q}_L\cancel{D}q_L+i\bar{t}_R\cancel{D}t_r+i\bar{\Psi}^{\mathbf{4}}\cancel{D}\Psi^{\mathbf{4}}-M_\Psi \bar{\Psi}^{\mathbf{4}}\Psi^{\mathbf{4}}+ic_1\bar{\Psi}^{\mathbf{4}}_Rd_\mu \gamma^\mu t_R \non\\
&+yf\text{Tr}\left(\bar{Q}^{\mathbf{14}}_LU\Psi^{\mathbf{4}'}_RU^T\right)+yf\text{Tr}\left(\bar{Q}^{\mathbf{14}}_LUt_R^{\mathbf{1}'}U^T\right)
\end{align}
where $\Psi^{\mathbf{4}'}$ is defined as the direct product of the $SO(5)$ breaking VEV, $\Sigma_0=(0,0,0,0,1)$, and $\Psi^{\mathbf{4}}$.
In this way the invariant in the Lagrangian can be written as $(\bar{Q}_L^{\mathbf{14}})^{IJ}U_{IM}U_{JN}\Sigma^N_0(\Psi^{\mathbf{4}})^M$, in accordance with \cite{DeSimone:2012fs}.
We also use an analogous definition of $t_R^{\mathbf{1}'}$.
Because the
top-partners transform in a $\mathbf{4}$ of $SO(4)$ the particle
content here is the same as in the $\mathbf{M4_5}$ model, however the
mass matrix differs slightly due to the embedding of the SM doublet,
\begin{equation}
\label{eq:Mass-M414}
\begin{pmatrix}\bar{t}_L\\ \bar{T}_L\\ \bar{X}_{2/3L}\end{pmatrix}^T\begin{pmatrix}
-\tfrac{yf c_2}{2\sqrt{2}}s_{2\epsilon}&\tfrac{yf}{2} (c_{\epsilon}+c_{2\epsilon})&\tfrac{yf}{2} (c_{\epsilon}-c_{2\epsilon}) \\ 0&-M_{\Psi}&0\\0&0&-M_{\Psi}
\end{pmatrix}\begin{pmatrix}t_R\\ T_R\\ X_{2/3R}\end{pmatrix}.
\end{equation}
Analogously to the previous model we can also rotate the top-partner states such that only one of the top-partners couples to the SM doublet and the Higgs, with the transformation being
\begin{equation}
\begin{pmatrix}t\\ T\\ X_{2/3}\end{pmatrix}\rightarrow\frac{1}{N}\begin{pmatrix}N&0&0\\0&c_{\epsilon}+c_{2\epsilon}&c_{\epsilon}-c_{2\epsilon}\\0&-c_{\epsilon}+c_{2\epsilon}&c_{\epsilon}+c_{2\epsilon}\end{pmatrix}\begin{pmatrix}t\\ T\\ X_{2/3}\end{pmatrix},~~N=\sqrt{2+c_{2\epsilon}+c_{4\epsilon}},
\end{equation}
leaving the resultant mass matrix as
\begin{equation}
\begin{pmatrix}\bar{t}_L\\ \bar{T}_L\end{pmatrix}^T\begin{pmatrix}
-\tfrac{yf c_2}{2\sqrt{2}}s_{2\epsilon}&\tfrac{yf}{2} \sqrt{2+c_{2\epsilon}+c_{4\epsilon}} \\ 0&-M_{\Psi}
\end{pmatrix}\begin{pmatrix}t_R\\ T_R\end{pmatrix}.
\end{equation}
The $X_{2/3}$ state has decoupled in the same way as in the $\mathbf{M4_5}$ model and has a mass degenerate with the exotic $X_{5/3}$ top-partner.

\subsubsection*{~~~$\mathbf{M1_5}$}

\noindent For a light top-partner transforming as a $\mathbf{1_{2/3}}$ of $SO(4)$ and the SM left-handed top-bottom doublet embedded in a $\mathbf{5}$ of $SO(5)$ the relevant effective action is
\begin{align}
  \label{eq:M15L}
\mathcal{L}_{\mathbf{M1_5}}=~&i\bar{q}_L\cancel{D}q_L+i\bar{t}_R\cancel{D}t_R+i\bar{\Psi}^{\mathbf{1}}\cancel{D}\Psi^{\mathbf{1}}-M_{\Psi}\bar{\Psi}^{\mathbf{1}}\Psi^{\mathbf{1}} \non\\
&+yf\bar{Q}_L^{\mathbf{5}}U\Psi^{\mathbf{1}}_R+yf c_2\bar{Q}_L^{\mathbf{5}}Ut_R^{\mathbf{1}}~+~\text{h.c.}
\end{align}
where the term proportional to $c_1$ is now absent.
With singlet top-partners we only have one top-partner state with charges equal to that of the right-handed top quark.
The mass matrix in this case is simpler than with fourplet top-partners, and is written as
\begin{equation}
\label{eq:Mass-M15}
\begin{pmatrix}\bar{t}_L\\ \bar{{T}}_L\end{pmatrix}^T\begin{pmatrix}
-\tfrac{yf c_2}{\sqrt{2}}s_{\epsilon}&\tfrac{yf}{\sqrt{2}} s_{\epsilon} \\ 0&-M_{\Psi}
\end{pmatrix}\begin{pmatrix}t_R\\ {T}_R\end{pmatrix}.
\end{equation}

\subsubsection*{~~~$\mathbf{M1_{14}}$}

\noindent For a light top-partner transforming as a $\mathbf{1_{2/3}}$ of $SO(4)$ and the SM left-handed top-bottom doublet embedded in a $\mathbf{14}$ of $SO(5)$ the relevant effective action is
\begin{align}\label{M114L}
\mathcal{L}_{\mathbf{M1_{14}}}=~&i\bar{q}_L\cancel{D}q_L+i\bar{t}_R\cancel{D}t_R+i\bar{\Psi}^{\mathbf{1}}\cancel{D}\Psi^{\mathbf{1}}-M_{\Psi}\bar{\Psi}^{\mathbf{1}}\Psi^{\mathbf{1}} \non\\
&+yf\text{Tr}\left(\bar{Q}_L^{\mathbf{14}}U\Psi^{\mathbf{1}'}_RU^T\right)+yf c_2\text{Tr}\left(\bar{Q}_L^{\mathbf{14}}Ut_R^{\mathbf{1}'}U^T\right)~+~\text{h.c.}
\end{align}
where the singlet composite states are embedded in $\mathbf{14}$ representations of $SO(5)$ when coupled to the SM doublet.
The mass matrix is similar to the $\mathbf{M1_5}$ case,
\begin{equation}
\label{eq:Mass-M114}
\begin{pmatrix}\bar{t}_L\\ \bar{{T}}_L\end{pmatrix}^T\begin{pmatrix}
-\tfrac{yf c_2}{2\sqrt{2}}s_{2\epsilon}&\tfrac{yf}{2\sqrt{2}} s_{2\epsilon} \\ 0&-M_{\Psi}
\end{pmatrix}\begin{pmatrix}t_R\\ {T}_R\end{pmatrix}.
\end{equation}

\subsection{Additional light top-partner multiplets}

\noindent Introducing additional light top-partner multiplets can be
done in a straightforward way.  To keep the models simple we will
assume that all top-partner states couple to the SM with the same
strength, with their masses determining their influence on the top
mass and Yukawa coupling.  We label our top-partner multiplets as
$\Psi_i^{\mathbf{4}}$ and $\Psi_i^{\mathbf{1}}$, and their masses as
$M_{\Psi_i}$, whereas the components of these multiplets we label as
$T^i$, $B^i$, $X_{2/3}^i$, $X_{5/3}^i$.

Introducing additional multiplets in the $\mathbf{M1_5}$ and
$\mathbf{M1_{14}}$ is straightforward since we are dealing with
singlet top-partners.  For example the mass matrices for these models
with one additional singlet each can be written as
\begin{align}
&\begin{pmatrix}\bar{t}_L\\ \bar{{T}}_L^1\\ \bar{{T}}_L^2\\\end{pmatrix}^T\begin{pmatrix}
-\tfrac{yf c_2}{\sqrt{2}}s_{\epsilon}&\tfrac{yf}{\sqrt{2}} s_{\epsilon} &\tfrac{yf}{\sqrt{2}} s_{\epsilon} \\ 0&-M_{\Psi_1}&0\\0&0&-M_{\Psi_2}
\end{pmatrix}\begin{pmatrix}t_R\\ {T}_R^1 \\ {T}_R^2\end{pmatrix} \text{~~~for $\mathbf{M1_{5}}$ and}
\non\\
&\begin{pmatrix}\bar{t}_L\\ \bar{{T}}_L^1\\ \bar{{T}}_L^2\\\end{pmatrix}^T\begin{pmatrix}
-\tfrac{yf c_2}{2\sqrt{2}}s_{2\epsilon}&\tfrac{yf}{2\sqrt{2}} s_{2\epsilon} &\tfrac{yf}{2\sqrt{2}} s_{2\epsilon} \\ 0&-M_{\Psi_1}&0\\0&0&-M_{\Psi_2}
\end{pmatrix}\begin{pmatrix}t_R\\ {T}_R^1 \\ {T}_R^2\end{pmatrix} \text{~~~for $\mathbf{M1_{14}}$}.
\end{align}
When the top partners are in fourplets all we need to do is to rotate each $(T^i,X_{2/3}^i)$ pair separately such that only one linear combination of quarks from each multiplet couples to the top quark and the Higgs.
For one additional top-partner in the fourplet models this can be done using the orthogonal transformations
\begin{align}
\begin{pmatrix}t\\ T^1\\ X_{2/3}^1\\T^2\\X_{2/3}^2\end{pmatrix}\rightarrow
\frac{1}{N}
\begin{pmatrix}
N&0&0&0&0\\
0&1+c_{\epsilon}&1-c_{\epsilon}&0&0\\
0&-1+c_{\epsilon}&1+c_{\epsilon}&0&0\\
0&0&0&1+c_{\epsilon}&1-c_{\epsilon}\\
0&0&0&-1+c_{\epsilon}&1+c_{\epsilon}
\end{pmatrix}
\begin{pmatrix}t\\ T^1\\ X_{2/3}^1\\T^2\\X_{2/3}^2\end{pmatrix}
\end{align}
for $\mathbf{M4_{5}}$ with $N=\sqrt{2+2c_{\epsilon}^2}$, and
\begin{align}
\begin{pmatrix}t\\ T^1\\ X_{2/3}^1\\T^2\\X_{2/3}^2\end{pmatrix}\rightarrow
\frac{1}{N}
\begin{pmatrix}
N&0&0&0&0\\
0&c_{\epsilon}+c_{2\epsilon}&c_{\epsilon}-c_{2\epsilon}&0&0\\
0&-c_{\epsilon}+c_{2\epsilon}&c_{\epsilon}+c_{2\epsilon}&0&0\\
0&0&0&c_{\epsilon}+c_{2\epsilon}&c_{\epsilon}-c_{2\epsilon}\\
0&0&0&-c_{\epsilon}+c_{2\epsilon}&c_{\epsilon}+c_{2\epsilon}
\end{pmatrix}
\begin{pmatrix}t\\ T^1\\ X_{2/3}^1\\T^2\\X_{2/3}^2\end{pmatrix}
\end{align}
for $\mathbf{M4_{14}}$ with $N=\sqrt{2+c_{2\epsilon}+c_{4\epsilon}}$.
Adding more top-partners requires analogous rotations of the form above.
The important point is that we can completely decouple the $X_{2/3}$ states from the top quark and the Higgs irrespective of how many top-partners we have.
The mass matrices for these models with one additional light top-partner can then be written as
\begin{align}
&\begin{pmatrix}\bar{t}_L\\ \bar{{T}}_L^1\\ \bar{{T}}_L^2\\\end{pmatrix}^T
\begin{pmatrix}
-\tfrac{yf c_2}{\sqrt{2}}s_{\epsilon}&\tfrac{yf}{2} \sqrt{3+c_{2\epsilon}} &\tfrac{yf}{2} \sqrt{3+c_{2\epsilon}} \\ 
0&-M_{\Psi_1}&0
\\0&0&-M_{\Psi_2}
\end{pmatrix}
\begin{pmatrix}t_R\\ {T}_R^1 \\ {T}_R^2\end{pmatrix} \text{~~~for $\mathbf{M4_{5}}$ and}
\non\\
&\begin{pmatrix}\bar{t}_L\\ \bar{{T}}_L^1\\ \bar{{T}}_L^2\\\end{pmatrix}^T
\begin{pmatrix}
-\tfrac{yf c_2}{2\sqrt{2}}s_{2\epsilon}&\tfrac{yf}{2} \sqrt{2+c_{2\epsilon}+c_{4\epsilon}} &\tfrac{yf}{2} \sqrt{2+c_{2\epsilon}+c_{4\epsilon}} \\ 
0&-M_{\Psi_1}&0\\
0&0&-M_{\Psi_2}
\end{pmatrix}
\begin{pmatrix}t_R\\ {T}_R^1 \\ {T}_R^2\end{pmatrix} \text{~~~for $\mathbf{M4_{14}}$}.
\end{align}
One can see from this construction that adding an arbitrary number of
light top-partners can be implemented in a straightforward way.  There
is also no need for the light top-partners to be in the same $SO(4)$
representation as each other, one could just as well have a light
singlet and fourplet in the spectrum and there would be no extra
complication.

\section{Mass spectrum and Yukawa couplings}
\label{sec:massyuk}

\noindent
The purpose of this section is to study how the
masses and Yukawa couplings vary with the input parameters for
scenarios with both one and two light top-partner multiplets in each
of the scenarios discussed in the previous section.

The first thing to discuss is the effect of the operators in
Eq.~\eqref{eq:M45L} and Eq.~\eqref{eq:M414L} which are preceded by the
$c_1$ coefficients.  After writing the $d_\mu$ term in unitary gauge
we have
\begin{equation}
d_\mu^i=\delta^{i4}\sqrt{2}\,\frac{\partial_\mu \rho}{f}+\dots\,,   
\end{equation}
and thus the top-partners have a derivative coupling with the Higgs
boson.  Via a field re-definition we can recast this derivative
coupling to a CP-odd Yukawa term, which scales as $\text{Im}(c_1)$,
plus operators that involve higher powers of the Higgs boson field, or
different fermionic fields, and hence are not relevant for
single-Higgs production.

The general
EFT Lagrangian that contains the interactions between the top quark
$t_{L,R}$ and the charge $2/3$ top-partners $T_{L,R}$ which mix with
the top quark is
\begin{align}
  \label{eq:EFT}
  \mathcal{L}_{\text{EFT}}\supset &
                             -m_t\bar{t}t-m_b\bar{b}b-m^j_T\bar{T}_jT_j-\kappa_t\frac{m_t}{v}\bar{t}th-\kappa_b\frac{m_b}{v}\bar{b}bh\non\\
&+\kappa^j_T\frac{m^j_T}{v}\bar{T}_jT_jh+i\tilde{\kappa}_t\frac{m_t}{v}\bar{t}\gamma_5th +i\tilde{\kappa}^j_T\frac{m^j_T}{v}\bar{T_j}\gamma_5T_jh \,,
\end{align}
where the sums over $j$ indicate sums over top-partner multiplets, and
in this work we will consider at most two multiplets.  
The mixing of the bottom quark with the composite sector is assumed to be small therefore we do not include the bottom partners in the EFT.
The $\kappa_i$'s are defined such that in the SM we have $\kappa_{b,t}=1$, and
$\kappa_T=\tilde{\kappa}_{t,T}=0$.  The CP-odd couplings in the second
line of Eq.~\eqref{eq:EFT} will only exist for the ${\bf M4_{5}}$
and ${\bf M4_{14}}$ models, and will be functions of the mixing angles
and $\text{Im}(c_1)$.  

\subsection{One light top-partner multiplet}

\noindent
In the case where we have only one light top-partner multiplet the
Yukawa couplings in the mass eigenbasis can be written down
analytically.
In general, the mass-mixing matrix can be written in the form
\begin{equation}
\label{eq:Mass-general}
-\begin{pmatrix}\bar{t}_L\\ \bar{{T}}_L\end{pmatrix}^T\begin{pmatrix}
m&\Delta \\ 0&M_{\Psi}
\end{pmatrix}\begin{pmatrix}t_R\\ {T}_R\end{pmatrix}.
\end{equation}
diagonalization of the matrix is achieved via a double rotation with
left-handed and right-handed mixing angles $\theta_L$ and\ $\theta_R$
respectively.  This gives us the mass eigenstates with top mass $m_t$
and top-partner mass $M_T$, and consequently a relation between
$m,\Delta,M_{\Psi}$ and the parameters $m_t,M_T,\theta_L, \theta_R$:
\begin{equation}
  \label{eq:parameter-relation}
  \begin{split}
    m & = \frac{\cos\theta_R}{\cos\theta_L} m_t = \frac{\sin\theta_R}{\sin\theta_L} M_T\,,\qquad
    M_{\Psi}  = \frac{\sin\theta_L}{\sin\theta_R} m_t= \frac{\cos\theta_L}{\cos\theta_R} M_T\,,\\
    \Delta & = \frac{\sin^2 \theta_L-\sin^2\theta_R}{\sin\theta_L \sin \theta_R} \tan\theta_L m_t = \frac{\sin^2 \theta_L-\sin^2\theta_R}{\sin\theta_L \sin \theta_R} \tan\theta_R M_T\,, 
  \end{split}
\end{equation}
where the two mixing angles are related through
\begin{equation}
  \label{eq:thL-thR}
  \tan \theta_L= \frac{M_{T}}{m_{t}}\tan \theta_R\,.
\end{equation}
It is also useful to have expressions for the mixing angles in terms of the inputs $(m_t,M_{\Psi},y,f)$,
\beq\label{anglesInputs}
\sin^2\theta_R=\frac{\Delta^2}{m_t^2+\Delta^2\left(\frac{M_{\Psi}}{m_t}\right)^2+M_{\Psi}^2\left(\left(\frac{M_{\Psi}}{m_t}\right)^2-2\right)},~~~\sin^2\theta_L=\left(\frac{M_{\Psi}}{m_t}\right)^2\sin^2\theta_R.
\eeq
Although $m_t$ is obtained as a result of diagonalising the mass matrix, we fit the top mass using $c_2$, and so it makes more sense to trade this to take $m_t$ as the input.
From this we can deduce that
\beq
\Delta=m_t\frac{\sin\theta_R}{\cos\theta_L}\left(\left(\frac{M_{\Psi}}{m_t}\right)^2-1\right).
\eeq
In each of the models we have computed the Yukawa couplings of the top
and top-partner to be
\begin{subequations}
  \label{eq:1TPyuks}
  \begin{align}
  \label{eq:1TPyuks-M15}
    \mathbf{M1_5:}~~~&\kappa_t=  c_{\epsilon}  \cos^{2} \theta_L \non \\
	&\kappa_{T}= c_{\epsilon} \sin^{2} \theta_L \\
  \label{eq:1TPyuks-M114}
    \mathbf{M1_{14}:}~~~&\kappa_t=\frac{c_{2\epsilon}}{c_{\epsilon}} \cos^{2} \theta_L  \non \\
	&\kappa_{T}=\frac{c_{2\epsilon}}{c_{\epsilon}} \sin^{2} \theta_L \\
  \label{eq:1TPyuks-M45}
    \mathbf{M4_{5}:}~~~&\kappa_t=  c_{\epsilon} \left(\cos^{2} \theta_R - \frac{s^2_{\epsilon}}{1+c^2_{\epsilon}} \left(\cos^{2} \theta_L - \cos^{2} \theta_R\right) \right) \non \\
                    &\kappa_T= c_{\epsilon} \left(\sin^{2} \theta_R - \frac{s^2_{\epsilon}}{1+c^2_{\epsilon}} \left(\sin^{2} \theta_L - \sin^{2} \theta_R\right) \right) \\
  \label{eq:1TPyuks-M414}
    \mathbf{M4_{14}:}~~~&\kappa_t=\left(\frac{c_{2\epsilon}}{ c_{\epsilon}} 
    \cos^{2} \theta_R-
    \frac{s_{\epsilon} \left(s_{2\epsilon}+2 s_{4\epsilon}\right)}{2\left(c^2_{\epsilon}+c^2_{2\epsilon}\right)}
    \left(\cos^{2} \theta_L-\cos^2\theta_R\right)\right)  \non\\
	&\kappa_T=\left(\frac{c_{2\epsilon}}{c_{\epsilon}} \sin^{2} \theta_R-\frac{s_{\epsilon} \left(s_{2\epsilon}+2 s_{4\epsilon}\right)}{2\left(c^2_{\epsilon}+c^2_{2\epsilon}\right)}\left(\sin^{2} \theta_L-\sin^{2} \theta_R\right)\right).
\end{align}
\end{subequations}

The CP-odd $\tilde{\kappa}$ coefficients can be calculated from the interaction
\begin{align}
\mathcal{L}\supset&\> ic_{1}\left[\bar{X}_{2/3}-\bar{T}\right] d_\mu\gamma^\mu t_R+\text{~h.c.}= ic_1\frac{\partial_\mu \rho}{f}\left[\bar{X}_{2/3}-\bar{T}\right]\gamma^\mu t_R+\text{~h.c.} + \dots
\end{align}
Since this term depends only on the fluctuation of the Higgs field
$\rho$, we can diagonalise the mass matrix independent of it. After
doing that, we find
\begin{align}
\mathcal{L}_{\mathbf{M4_{5}}}&\supset \frac{4c_\epsilon}{\sqrt{2+2c_\epsilon^2}}\text{Im}(c_1)\sin\theta_R\cos\theta_R\frac{\partial_\mu \rho}{f}\left(\bar{t}_R\gamma^\mu t_R-\bar{T}_R\gamma^\mu T_R\right) \,, \non\\
\mathcal{L}_{\mathbf{M4_{14}}}&\supset \frac{4(1-2s_\epsilon^2)}{\sqrt{2+c_{2\epsilon}+c_{4\epsilon}}}\text{Im}(c_1)\sin\theta_R\cos\theta_R\frac{\partial_\mu \rho}{f}\left(\bar{t}_R\gamma^\mu t_R-\bar{T}_R\gamma^\mu T_R\right)\,.
\end{align}
where we have neglected terms which mix the states in the mass eigenbasis because these terms do not contribute to Higgs production.
Now we perform the field redefinitions
\begin{equation}
  \label{eq:tT-redef}
  t_R \to \left(1+\tilde c\frac{\rho}{f}\right)t_R\,\qquad
  T_R \to \left(1-\tilde c\frac{\rho}{f}\right)T_R\,,
\end{equation}
with $\tilde c$ to be determined in such a way that the Lagrangian
does not contain any terms linear in $\partial_\mu \rho$. In this way,
the coupling of $\bar{t}t$ and $\bar{T}T$ to the Higgs derivative are
recast into couplings to higher powers of the Higgs boson and CP-odd
Yukawa couplings described by
\begin{subequations}
\label{eq:1TPCPoddyuks}
  \begin{align}
\label{eq:1TPCPoddyuks-M45}
    \mathbf{M4_5}:&~~~~~\tilde{\kappa_t}=-\tilde{\kappa}_T=\frac{4c_\epsilon s_\epsilon}{\sqrt{2+2c_\epsilon^2}}\text{Im}(c_1)\sin\theta_R\cos\theta_R \,.\\
\label{eq:1TPCPoddyuks-M414}
    \mathbf{M4_{14}}:&~~~~~\tilde{\kappa_t}=-\tilde{\kappa}_T=\frac{4s_\epsilon(1-2s_\epsilon^2)}{\sqrt{2+c_{2\epsilon}+c_{4\epsilon}}}\text{Im}(c_1)\sin\theta_R\cos\theta_R\,.
\end{align}
\end{subequations}
The addition of more light top-partners in the model will prevent us
from obtaining simple analytical solutions such as the ones above, and
the $\tilde{\kappa}_t=-\tilde{\kappa}_T$ is certainly spoiled.

The bottom quark mass is also generated via partial compositeness,
although the mixing of the bottom quark with the composite sector is
much milder and the right-handed bottom is certainly not composite.
The bottom quark Yukawa couplings are shifted by the same factors of
$s_\epsilon$ as the top quark, and we can assume that the mixing
angles with the composite sector are negligible.  Therefore we have
only two scenarios for the bottom quark, $\kappa_b^{\bf 5}=c_\epsilon$
and $\kappa_b^{\bf 14}=\frac{c_{2\epsilon}}{c_\epsilon}$.  Given that
the CP-odd terms are also proportional to the mixing with the
composite sector, these can also be taken to be absent for the bottom
quark.

\subsubsection{Perturbativity and relevant limits}
\label{sec:perturbativity}

\noindent Before discussing some relevant limits of the couplings in
eqs.~(\ref{eq:1TPyuks})-(\ref{eq:1TPCPoddyuks}), we investigate which
values of $m_t,M_T$, and $f$ correspond to values of the mixing
paramter $y$ that are in the perturbative regime, which we take to be
$y<3$~\cite{DeSimone:2012fs}. To do this, it is useful to re-write the
off-diagonal term $\Delta$ in terms of a single mixing-angle. Using
the relations
\begin{align}
\cos^2\theta_R=&\frac{M_T^2 \cos^2\theta_L}{M_T^2 \cos^2\theta_L+m_t^2\sin^2\theta_L}, 
\qquad\cos^2\theta_L=\frac{m_t^2\cos^2\theta_R}{m_t^2\cos^2\theta_R+M_T^2\sin^2\theta_R}	
\end{align}
we obtain
\begin{equation}
  \label{eq:Delta-angles}
  \Delta=\frac{M_T^2-m_t^2}{2\sqrt{m_t^2 \cos^2\theta_R+M_T^2\sin^2\theta_R}}\sin(2\theta_R)=
  \frac{M_T^2-m_t^2}{2\sqrt{M_T^2\cos^2\theta_L+m_t^2\sin^2\theta_L}}\sin(2\theta_L)\,.
\end{equation}
We now want to assess what values of the top-partner masses and mixing
angles are consistent with the perturbativity of the interaction terms
mixing the top with the vector-like quarks.  The perturbativity bound
restricts parameters in a different way for singlet and fourplet
models, due to the different scaling of $\Delta$ with respect to $v$
and $f$. For singlet models we have that $\Delta\sim y v$, thus, for
moderate mixing angle $\theta_L\lesssim \pi/4$ and $M_T\gg m_t$,
Eq.~\eqref{eq:Delta-angles} implies that $y\sim
M_T/v\tan\theta_L$. The only possibility to have very large
top-partner masses is then to make the angle $\theta_L$, and hence
$\theta_R$, increasingly small. This is a physically sensible limit,
because it means that a very heavy top partner essentially
decouples. This feature occurs irrespectively of the value of $f$.  In
the case of fourplet models we have that $\Delta\sim y
f$, thus large values of top-partner
masses and moderate mixing angles are still consistent with
perturbativity, provided $f$ is sufficiently large. These basic
considerations show that simplified models determined just in terms of
a mixing angle and the mass of a top-partner may not result from a
perturbative composite Higgs model. Therefore, in the following
sections, whenever we fix mixing angle and top-partner mass, we will
always compute the corresponding value of $y$ and we will exclude all
values of parameters that are not consistent with perturbativity
bounds.

The perturbativity bounds we have described do not allow us to take
the mass of one top partner to infinity by fixing all other
parameters. We can instead take the $f\rightarrow\infty$ limit within
perturbativity bounds, in which case we have:
\begin{subequations}
\label{eq:kappa-largef}
  \begin{align}
\label{eq:kappa-largef-M1}
    \mathbf{M1_5, M1_{14}}:&~~~~~\kappa_t=\cos^2\theta_L\,,\qquad \kappa_T=\sin^2\theta_L \,.\\
\label{eq:kappa-largef-M4}
    \mathbf{M4_5, M4_{14}}:&~~~~~\kappa_t=\cos^2\theta_R\,,\qquad \kappa_T=\sin^2\theta_R\,,
\end{align}
\end{subequations}
and all CP-odd couplings vanish. For fourplet models, this corresponds
to the limit considered in~\cite{Banfi:2013yoa}. Note that, for
singlet models, perturbativity restricts the possible values of $M_T$
even for $f\to \infty$, whereas for fourplet models, this limit
corresponds to $y\to 0$, with $yf$ fixed. In this limit, further
constraints on the mixing angle are posed by experimental bounds on
$\kappa_t$~\cite{Aaboud:2018urx}, which set the $2\sigma$ lower bound
$\kappa_t\gtrsim 0.8$. This excludes $\sin^2\theta_L > 0.2$ for
singlet models, and $\sin^2\theta_R > 0.2$ for fourplet models. For
finite $f$, this bound is more stringent for singlet models, because
$\kappa_t \le \cos^2\theta_L$. Fourplet models are less constrained due to cancellations occurring between two different contributions (see eqs.~(\ref{eq:1TPyuks-M45}),~(\ref{eq:1TPyuks-M414})).

\subsection{Two light top-partner multiplets}
\label{sec:2tps}

\noindent In the case of two top partners $T^1$ and $T^2$, we take a
different approach with respect to the single top-partner case, in
that we study the relationship between the fundamental parameters of
each model (i.e.\ the vector-like masses, the couplings and the decay
constant $f$) and the physical top-partner masses and Yukawa
couplings. In particular, we take as free parameters
$y,f,M_{\Psi_1},M_{\Psi_2}$, as well as the CP-odd couplings, with
$c_2$ being used to fix the top quark mass to $\sim 173\,$GeV.

\begin{figure}[htbp!]
\centering
\includegraphics[width=.9\textwidth]{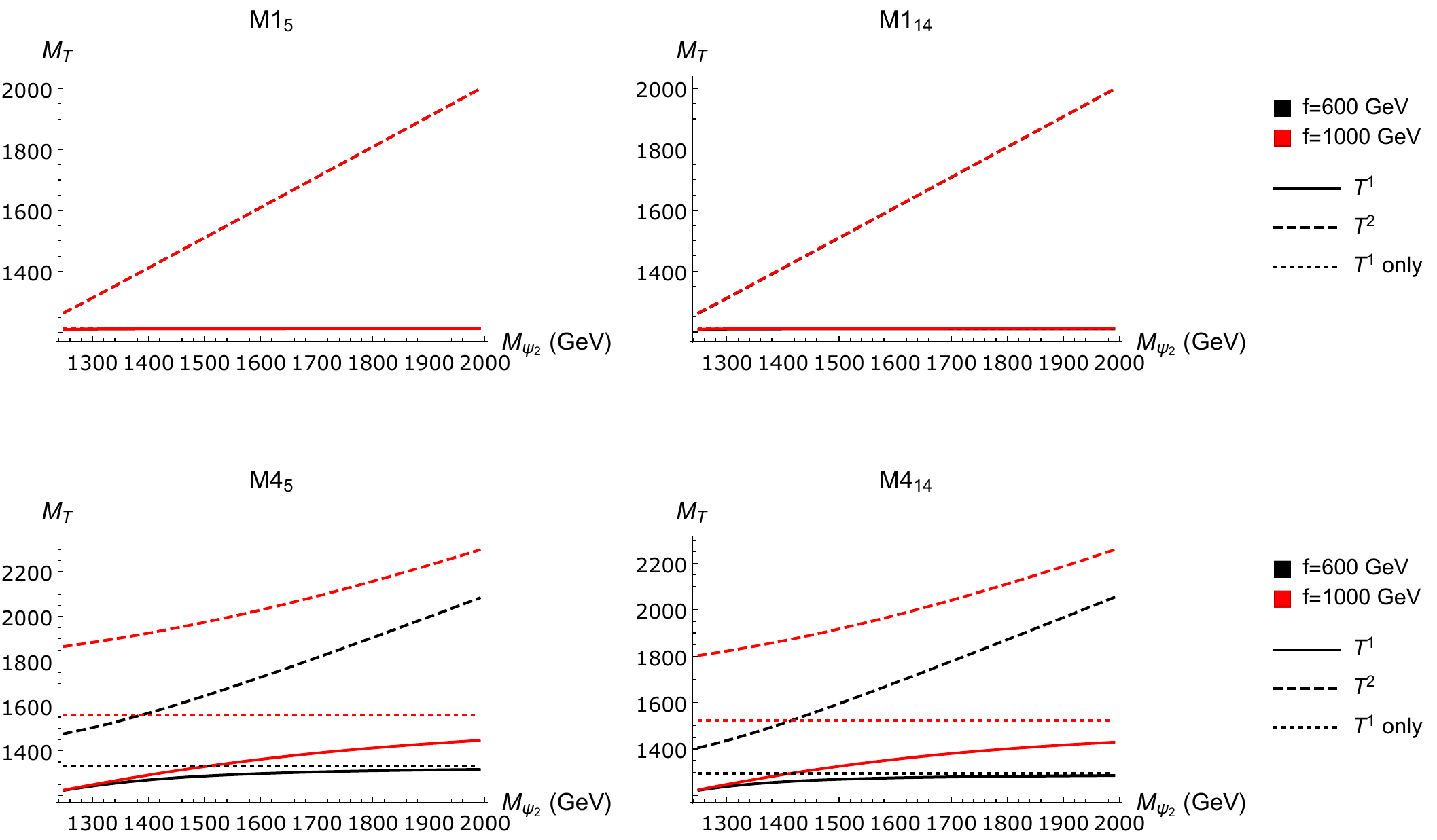}
\vspace{5mm}
\caption{The masses of the two light top-partners  $T^1$ or $T^2$ as functions of the heavier vector-like mass, for $M_{\Psi_1}=1200\,$GeV, $y=1$, and $f=600/1000\,$GeV.}
  \label{fig:masTops}
\end{figure}

\begin{figure}[htbp!]
  \centering \includegraphics[width=.9\textwidth]{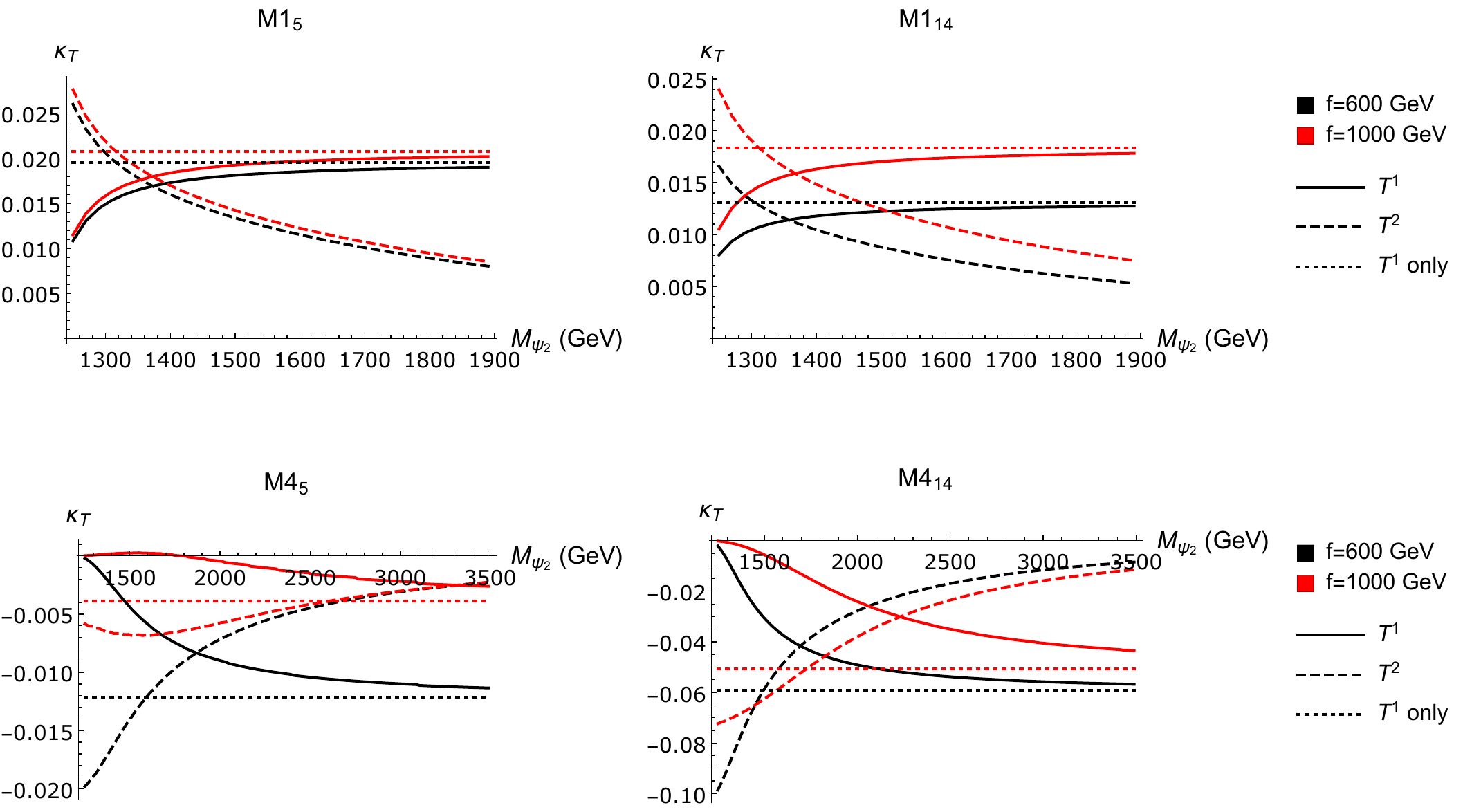}
  \vspace{5mm}
  \caption{The Yukawa couplings of the two light $T^1$ or $T^2$
    top-partners as functions of the heavier vector-like mass, for $M_{\Psi_1}=1200$ GeV, $y=1$, and $f=600/1000$ GeV.  The `$T^1$ only' results can easily be verified from Eq.~\eqref{anglesInputs} and Eq.~\eqref{eq:1TPyuks}.}
    \label{fig:yukTops}
\end{figure}

In Figures~\ref{fig:masTops} and~\ref{fig:yukTops} we plot the masses
and Yukawa couplings of $T^1$ or $T^2$ as a function of the heavier
vector-like mass for $M_{\Psi_1}=1200$ GeV, $y=1$, and $f=600/1000$
GeV. In each plot we also show the masses and couplings of a single
top partner (labelled $T^{1}$ only), corresponding to the same values
of $y$ and $f$, and $M_{\Psi}=M_{\Psi_1}$.
We stress that, everywhere, $T^{1}$ is the lighter top partner. The
first thing we notice when looking at Figure~\ref{fig:masTops} is
that, in the singlet top-partner models, there is almost a degeneracy
between the vector-like mass $\MPsi{2}$ and the mass of the $T^2$
state.  There is also no difference between the $f=600$ GeV and
$f=1000$ GeV scenarios for the singlet models, this is because in
these models the mass matrix is largely insensitive to $f$, a feature
not shared by the fourplet models.  In fourplet models instead, this
occurs only as one of the vector-like masses is made much larger than
the other.  Also, when considering fourplet models, we should keep in
mind that $M_{\Psi_{1,2}}$ are in fact the masses of the
$X_{2/3}^{1,2}$ and $X_{5/3}^{1,2}$ states. Therefore, for
$\MPsi{2}\gg \MPsi{1}$, $T^2$ has the same mass as $X_{2/3}^{2}$ and
$X_{5/3}^{2}$.
The behaviour of the Yukawa couplings shown of
Figure~\ref{fig:yukTops} presents some interesting features, as we see
that, in some circumstances, the heavier top-partner can have a larger
coupling to the Higgs than the lighter one.  This effect is only
present when when $\MPsi{2}$ is close to $\MPsi{1}$, and as the gap
between the two masses widens we can see the heavier top-partner
beginning to decouple from the Higgs and the Yukawa coupling of the
lighter top-partner approaching the value expected when only one light
top-partner is present.  For the fourplet models, in particular
$\mathbf{M4_5}$, there is a large region in which the heavier
top-partner holds the dominant coupling to the Higgs boson.  What is
striking here is that the coupling of the lighter top-partner to the
Higgs boson can be suppressed if there is a heavier top-partner of the
same charge with a slightly heavier mass.

\begin{figure}[htbp!]
\centering
\begin{minipage}[l]{.45\textwidth}
\begin{flushleft}
\includegraphics[width=\textwidth]{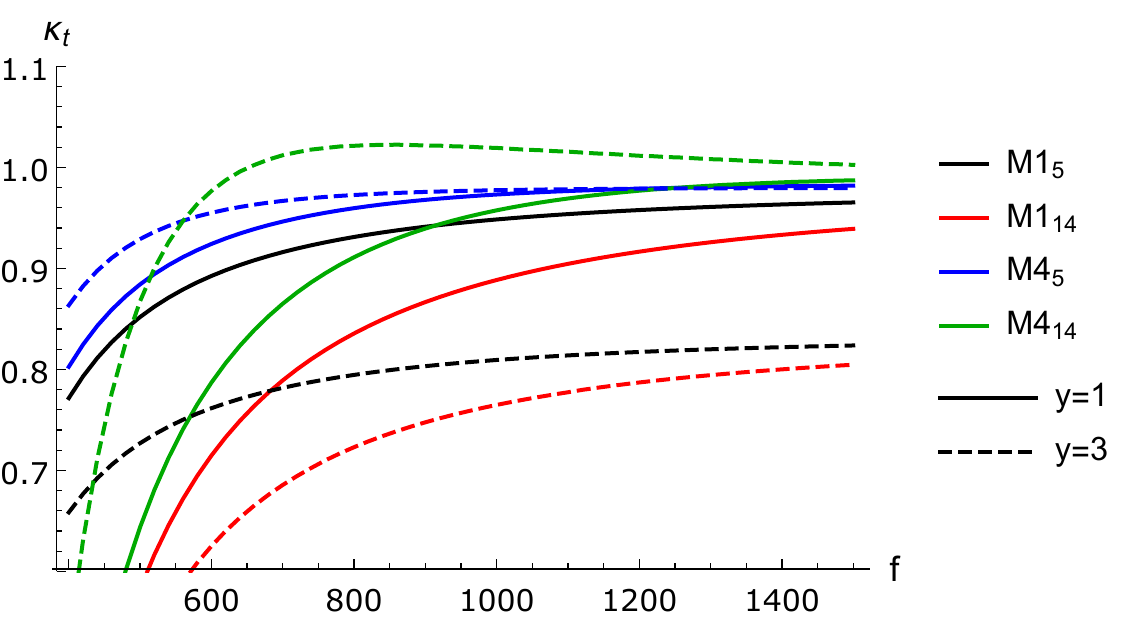}
\end{flushleft}
\end{minipage}
\begin{minipage}[r]{.45\textwidth}
\begin{flushright}
\includegraphics[width=\textwidth]{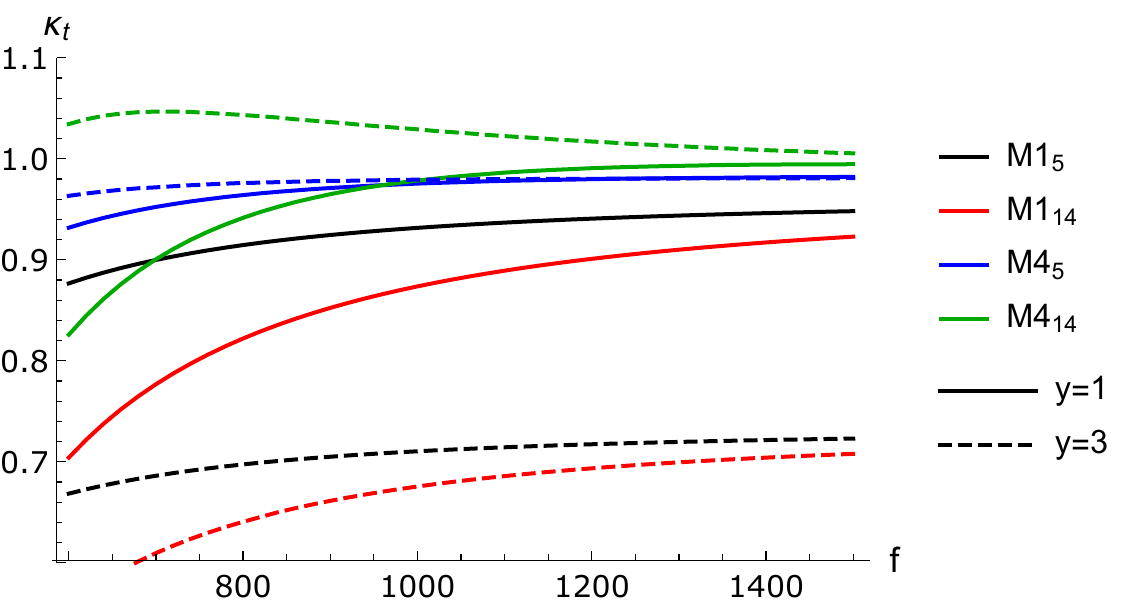}
\end{flushright}
\end{minipage}
\vspace{5mm}
\caption{The top-quark Yukawa coupling in each of the models as a function of $f$, for $M_{\Psi_1}=1200$ GeV, $M_{\Psi_2}=1300$ GeV, and $y=1,3$, for one top partner (left) and two top partners (right).}
  \label{fig:fYukDep}
\end{figure}

We then direct our attention to the top quark Yukawa couplings in
Figure~\ref{fig:fYukDep}, and we plot their values as a function of
the compositeness scale $f$ in the interesting case of
quasi-degenerate vector-like masses $\MPsi{1}=1200\,$GeV and
$\MPsi{2}=1300\,$GeV, and for two different values of $y$. In the case
of a single top-partner, we set $M_{\Psi}=\MPsi{1}$. As can be seen
from Fig.~\ref{fig:yukTops}, this represents very well the case of two
top partners when $M_{\Psi_1}\ll M_{\Psi_2}$. We observe that, except
for small values of $f$, the top-quark anomalous Yukawa coupling
depends very mildly on $f$ since at large $f$ the corrections are
dominated by the mixing angles between the top and the top
partners. The dependence on $y$ is similar both for one top partner
(the left panel of Figure~\ref{fig:fYukDep}) and for two top partners
(the right panel of Figure~\ref{fig:fYukDep}). In particular, the
$y$-dependence is much stronger for singlet models than for fourplet
models, with a 30\% suppression with respect to the SM for larger
values of $y$. As expected, for low values of $f$ we see drastic
deviations from the SM value, particularly in the singlet models.  In
fact, the recent observation of Higgs production in association with a
top-antitop pair by the ATLAS experiment~\cite{Aaboud:2018urx} sets
the $2\sigma$ lower bound $\kappa_t\gtrsim0.8$, which excludes the
singlet models with $y=3$ in both the one and two top-partner cases
for the whole range of values of $f$ shown. Therefore, in the
following, when comparing singlet and fourplet models, we will
restrict ourselves to the case $y=1$.

\begin{figure}[htbp!]
\centering
\includegraphics[width=.9\textwidth]{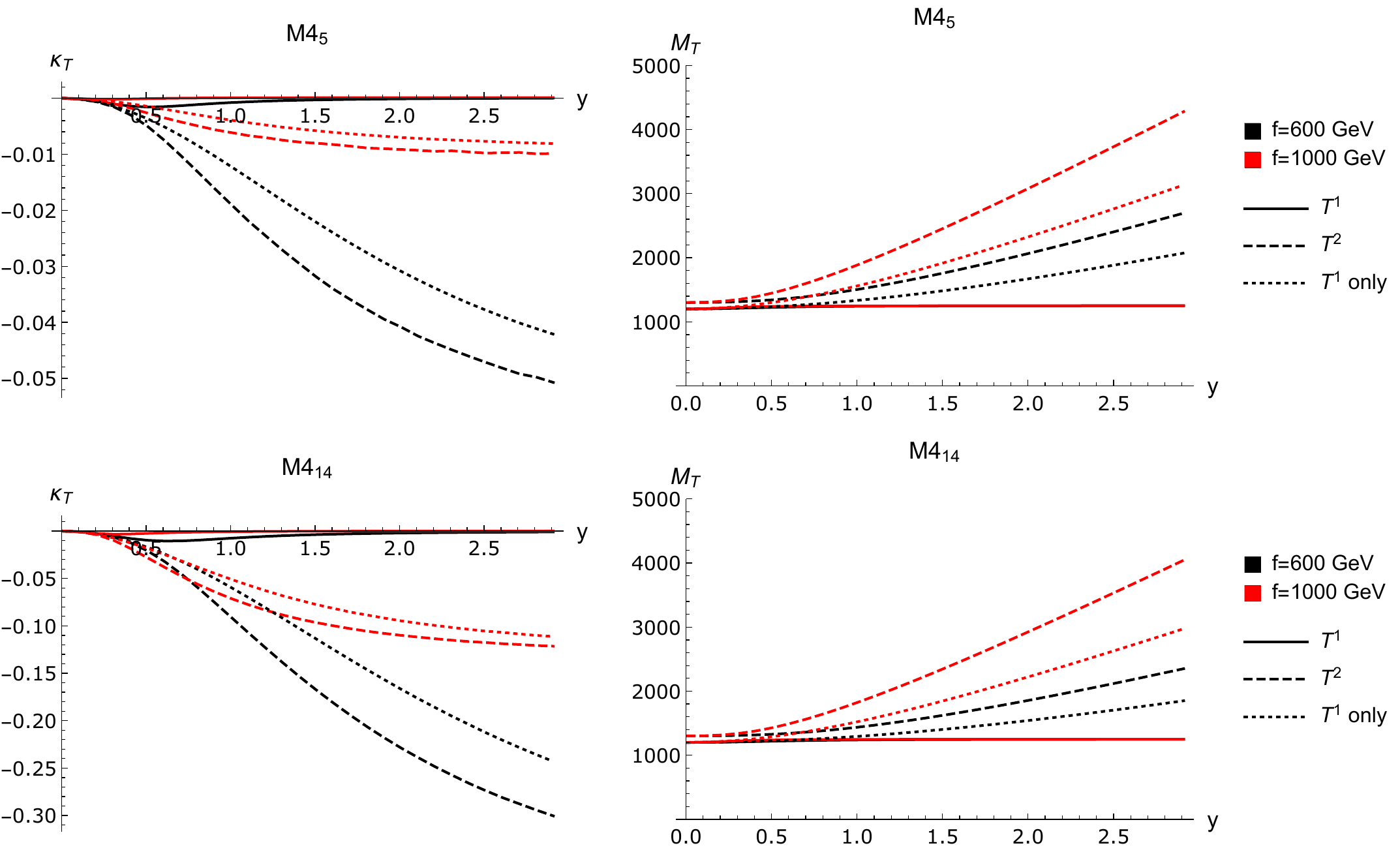}
\vspace{5mm}
\caption{The masses and Yukawa couplings of the two light $T$ top-partners from the $\mathbf{M4_5}$ and $\mathbf{M4_{14}}$ models as functions of the coupling $y$, for $M_{\Psi_1}=1200$ GeV, $M_{\Psi_2}=1300$ GeV, and $f=600/1000$ GeV.}\label{fig:yDep}
\end{figure}

In order to show how the composite-elementary mixing parameter $y$
affects the physical masses and Yukawa couplings we have plotted them
in Figure~\ref{fig:yDep} for the $\mathbf{M4_5}$ and
$\mathbf{M4_{14}}$ models with $M_{\Psi_1}=1200\,$GeV and
$M_{\Psi_2}=1300\,$GeV. The first feature to note is the behaviour at
small $y$: in this region the Yukawa couplings of the top-partners
become very small and the masses approach $M_{\Psi_{1,2}}$.  The
reason for this is because we are decoupling them from the top quark
and the Higgs, while the top quark mass is being kept at its observed
value by the $c_2$ coupling. As $y$ increases, the differences between
$T^1$ and $T^2$ grow, the mass of the lighter state stays close to the
vector-like mass $M_{\Psi{1}}$ and the mass of the heavier state is
enhanced.  The behaviour of the Yukawa couplings as $y$ varies is less
trivial and depends strongly on the value of $f$.

\begin{figure}[htbp]
\centering
\includegraphics[width=.9\textwidth]{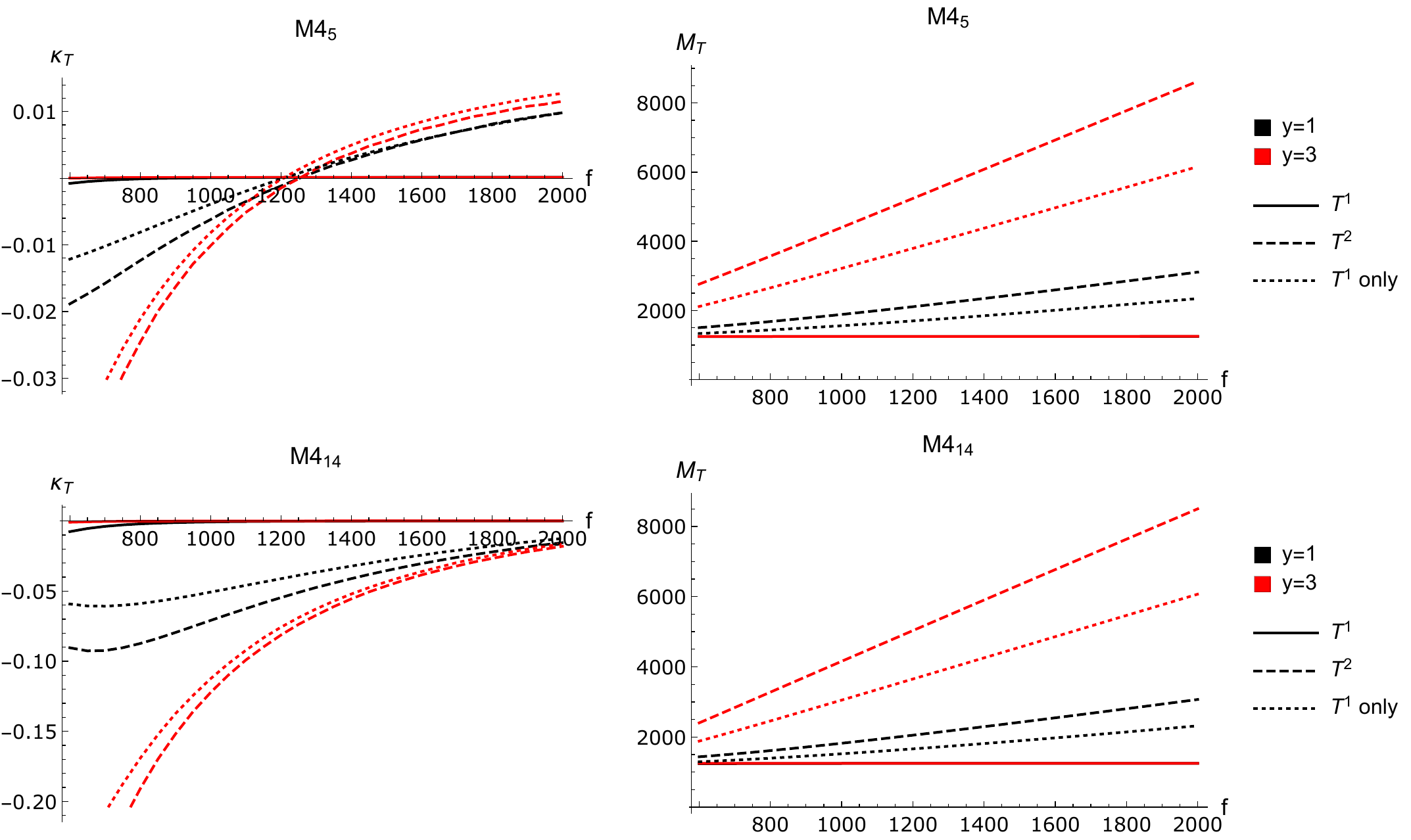}
\vspace{5mm}
\caption{The masses and Yukawa couplings of the two light top-partners
  from the $\mathbf{M4_5}$ and $\mathbf{M4_{14}}$ models, as functions
  of the coupling $f$, for $M_{\Psi_1}=1200$ GeV, $M_{\Psi_2}=1300$
  GeV, and $y=1,3$.}
  \label{fig:fDep}
\end{figure}

In Figure~\ref{fig:fDep} we show how these masses and couplings depend
on the decay constant $f$.  We have again used the $\mathbf{M4_5}$ and
$\mathbf{M4_{14}}$ models as an example, with $M_{\Psi_1}=1200$ GeV
and $M_{\Psi_2}=1300$ GeV, $y=1$ or $3$, and plotted the anomalous
Yukawa couplings and masses as a function of $f$.  Interestingly, we
see that one top-partner effectively remains light as $f$ is increased
and decouples from the Higgs.  The other increases in mass as $f$
increases, and its Yukawa coupling approaches that of the top-partner
in the single top-partner case.  We should note that, as $f$ is
increased, the ratio $v/f$ becomes small and indicates that the Higgs
potential will require more fine-tuning to reproduce the observed
Higgs mass and vacuum expectation value.

\begin{figure}[htbp!]
\centering
\includegraphics[width=.9\textwidth]{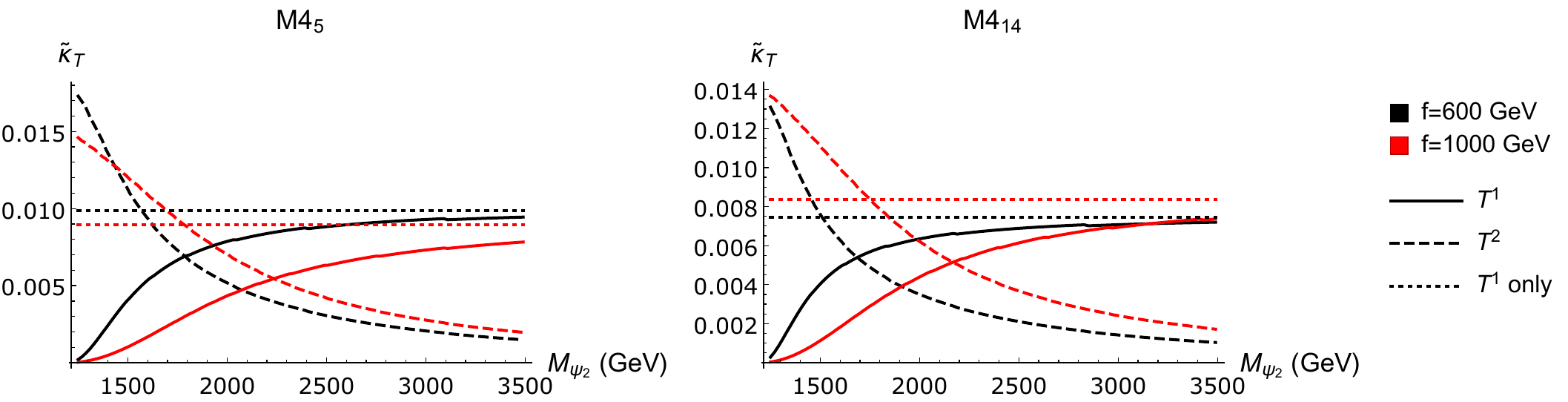}
\vspace{5mm}
\caption{The CP-odd Yukawa couplings of the two top-partners $T^1$ and $T^2$ as functions of the heavier vector-like mass $M_{\Psi_2}$, for $M_{\Psi_1}=1200$ GeV, $y=1$, $c=0.2$ and $f=600/1000$ GeV.}
  \label{fig:CPo2}
\end{figure}

Lastly, we need to study the couplings of the top-partners to the
derivatives of the Higgs in the ${\bf M4_5}$ and ${\bf M4_{14}}$
cases, which with two top-partners these can be written as
\begin{align}
\mathcal{L}\supset& -ic_{1,1}\bar{T}^1d_\mu\gamma^\mu t_R- ic_{1,2}\bar{T}^22d_\mu\gamma^\mu t_R +\text{~h.c.}
= -i\frac{\partial_\mu h}{f}\bar{\psi}C\psi+\text{~h.c.}
\end{align}
with
\begin{equation}
\psi=\begin{pmatrix}t_R\\T^1_R\\T^2_R\end{pmatrix}~~~~\text{and}~~~~C=\begin{pmatrix}0&0&0\\c_{1,1}&0&0\\c_{1,2}&0&0\end{pmatrix}.
\end{equation}
At this point, we can calculate the interactions between the quarks
and the Higgs derivative in the mass eigenbasis by rotating the matrix
$C$ with the rotations required to diagonalise the mass matrix.  In
the mass eigenbasis we denote $\tilde{C}=O_R(C-C^\dagger)O_R^T$, where
$O_R$ is the orthogonal matrix that diagonalises the right-handed side
of the top-partner mass matrix, and the couplings relevant for Higgs
production via gluon fusion are the diagonal elements.  Just as in the
case with one top-partner we can perform field redefinitions which
recast the diagonal couplings of the quarks to the Higgs derivative
into couplings to higher powers of the Higgs boson plus a set of
CP-odd Yukawa couplings given by
\begin{align}
\mathbf{M4_5}:&~~~~~\tilde{\kappa}_t=\frac{2c_\epsilon s_\epsilon}{\sqrt{2+2c_\epsilon^2}}\tilde{C}_{11},~~\tilde{\kappa}_{T,1}=\frac{2c_\epsilon s_\epsilon}{\sqrt{2+2c_\epsilon^2}}\tilde{C}_{22},~~\tilde{\kappa}_{T,2}=\frac{2c_\epsilon s_\epsilon}{\sqrt{2+2c_\epsilon^2}}\tilde{C}_{33} \non\\
\mathbf{M4_{14}}:&~~~~~\tilde{\kappa}_t=\frac{2s_\epsilon(1-2s_\epsilon^2)}{\sqrt{2+c_{2\epsilon}+c_{4\epsilon}}}\tilde{C}_{11},~~\tilde{\kappa}_{T,1}=\frac{2s_\epsilon(1-2s_\epsilon^2)}{\sqrt{2+c_{2\epsilon}+c_{4\epsilon}}}\tilde{C}_{22},~~\tilde{\kappa}_{T,2}=\frac{2s_\epsilon(1-2s_\epsilon^2)}{\sqrt{2+c_{2\epsilon}+c_{4\epsilon}}}\tilde{C}_{33}
\end{align}
It is also useful to see how the CP-odd couplings scale as a function
of the input parameters.
\begin{figure}[htbp]
\centering
\includegraphics[width=.9\textwidth]{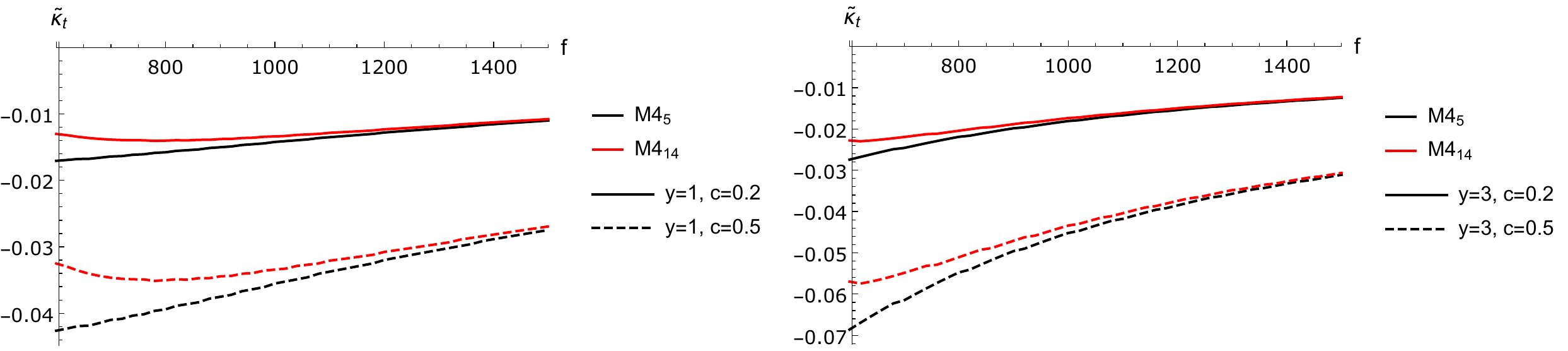}
\vspace{5mm}
\caption{The CP-odd top-quark Yukawa coupling varies in the fourplet
  models as a function of $f$, for $M_{\Psi_1}=1200$ GeV,
  $M_{\Psi_2}=1300$ GeV, and for different values of the parameters
  $y$ and $c$.}
	\label{fig:CPo1}
\end{figure}
In Figures~\ref{fig:CPo1} and~\ref{fig:CPo2} we plot the CP-odd top
and top-partner Yukawa couplings, respectively, in a scenario in which
the parameters determining the CP-odd couplings are universal, i.e.\
$\text{Im}(c_{1,1})=\text{Re}(c_{1,1})=\text{Im}(c_{1,2})=\text{Re}(c_{1,2})=c$.
For the top, we examine the dependence of the CP-odd coupling on $f$
for different values of $y$ and $c$, and for the top-partners we
examine the dependence on the vector-like masses $\MPsi{1,2}$, with
$y=1$ and two different values of $f$.  Note that when we take one
top-partner mass to be very heavy, that top-partners CP-odd coupling
diminishes, and we find that the CP-odd couplings of the top quark and
the light top-partner are equal and opposite, as in the case with one
top-partner. This same behaviour occurs in the CP-even Yukawa
couplings as the vector-like masses are varied.

\subsection{Brief summary of experimental bounds}
\label{sec:collider}
 
\noindent The first experimental constraint to mention is that on the
decay constant $f$, which through the analysis in~\cite{Sanz:2015sua}
is constrained to be larger than $\sim 600\,$GeV.  These bounds are
derived from Higgs decays to vector bosons and Higgs production.
Recent bounds on top-partner masses have been obtained through
analyses at $\sqrt{s}=13$ TeV by the ATLAS
collaboration~\cite{ATLAS:2016cuv,CMS:2017jfv,CMS:2017wwc,CMS:2018pma,CMS:2018vou,CMS:2018haz,Aaboud:2017qpr,Aaboud:2017zfn,Sirunyan:2017ynj,Sirunyan:2017pks,Sirunyan:2018fjh,Sirunyan:2018omb,Aaboud:2018uek,ATLAS:2018qxs}.
The first point to note is that these analyses only consider the
presence of one light top-partner state, and thus these bounds are
relevant to our lightest state.  Including heavier states opens up
possibilities for much more intricate signatures involving cascade
decays.  The lower mass bounds on the $T$ and $X_{2/3}$ partners from
the fourplet models are quoted at $1350\,$GeV, and the lower mass
bound on $T$ for singlet models is quoted at $1170\,$GeV. However, the
latter bound assumes that Br$(T\rightarrow Wb)=100\%$.
These bounds are
weakened if one considers sizeable branching ratios into multiple channels.
More interesting and intricate signatures arise in twin Higgs models~\cite{Chacko:2005pe,Chacko:2005vw,Chacko:2005un} which have QCD-like dark sectors with Higgs portal couplings to the SM.
Much work has been done in developing these models 
\cite{Craig:2015pha,Craig:2016kue,Barbieri:2016zxn,Geller:2014kta,Serra:2017poj,Csaki:2017jby,Dillon:2018wye} and studying their phenomenology 
\cite{Burdman:2014zta,Curtin:2015fna,Chacko:2015fbc,Ahmed:2017psb,Chacko:2017xpd}.
Translating these collider constraints into bounds on the top-partner
models presented in the previous sections is beyond the scope of this
paper, and in our analyses we will use a lower limit of $1200$ GeV for
the lightest vector-like mass.

Constraints on the $c_{1,1}$ and $c_{1,2}$ parameters have been
derived from electron and neutron Electric Dipole Moment (EDM)
experiments~\cite{Panico:2017vlk}.  These results indicate that with the
top-partner masses at the TeV scale, the imaginary values of these
parameters are constrained to lay $\lesssim 0.2$.  It is not the goal
of the present paper to study the effects of these parameters on the
EDMs, therefore we will simply constrain
$\text{Re}(c_{1,1}),~\text{Re}(c_{1,1}),~\text{Im}(c_{1,2}),~\text{Im}(c_{1,2})<
0.2$ in our work.  Future electron EDM experiments will introduce much
more stringent constraints on these parameters.  The remaining
parameter space that we wish to study in this paper is summarised by
$1.2~\text{TeV}<M_T<2.2~\text{TeV}$,
$600~\text{GeV}<f<1.2~\text{TeV}$, and $y<3$.

 \section{Higgs production}
\label{sec:HiggsProd}

\subsection{Total Higgs cross section}

\noindent It is well known that the production cross-section of the
Higgs boson via gluon fusion is insensitive to the mass spectrum of
top-partners in composite Higgs models.  This low energy theorem
arises due to the pseudo-Goldstone boson origin of the Higgs field in
composite Higgs models.  This insensitivity has been explored in many
studies. In~\cite{Low:2010mr}, the effects from new coloured fermions
in composite Higgs models to gluon fusion Higgs production, along with
other less transparent effects from new physics, were studied by means
of an effective Lagrangian. The new physics effects were inspected by
analysing the following higher dimensional operators constructed from
SM fields:
\begin{equation}
\label{eq:Three_operators}
\mathcal{O}_{H}=\partial^{\mu} (H^{\dagger} H) \partial_{\mu} (H^{\dagger} H), \quad \mathcal{O}_{y} = H^{\dagger} H \bar{\psi}_{L} H \psi_{R}, \quad \mathcal{O}_{g}= H^{\dagger} H G^a_{\mu \nu} G^{\mu \nu}_a.
\end{equation}
Through an explicit calculation, the authors of
ref.~\cite{Low:2010mr} showed that the gluon fusion production rate of
the composite Higgs depended only on the decay constant $f$ of
the model, not on the top-partners mass spectrum.

In ref.~\cite{Azatov:2011qy}, a different approach was used to show
the insensitivity of the cross section in composite Higgs model to
mass of the top partners. This work considered a generic Lagrangian
for a composite Higgs model with a top-partner multiplet in the
$\mathbf{5}$ representation, but their argument holds for all the
models discussed in section~\ref{sec:background}.  Consider the
contribution to the partonic cross section
$\hat \sigma_{gg \rightarrow H}$ gluon fusion Higgs production arising
from a loop of a number of fermions
\begin{equation}
  \label{eq:Higgs-xsct}
\hat \sigma_{gg \rightarrow H} = \frac{\alpha^{2}_{s} m^{2}_{H}}{576 \pi} \left|\sum_{j} \frac{Y_{jj}}{M_{j}} A_{1/2} (\tau_{j})\right|^{2} \delta (\hat{s}-m^{2}_{H})\,.
\end{equation}
In the above equation, $Y_{jj}$ is the Yukawa coupling of fermion $j$
of mass $M_j$ to the Higgs boson, $\hat s$ is the partonic
centre-of-mass energy squared, and $A_{1/2}(\tau_j)$ is the following
function of $\tau_j=m_H^2/(4 M_j^2)$:
\begin{equation}
  \label{eq:A12}
  A_{1/2}(\tau) = -2 \left[\tau+(\tau-1) f(\tau)\right]/\tau^2\,,\qquad
  f(\tau) = \left\{
    \begin{split}
      & \arcsin^2{\sqrt{\tau}} \qquad\qquad\quad\,\, \tau \le 1  \\
      -\frac{1}{4}
      &\left[\ln\frac{1+\sqrt{1-\tau^{-1}}}{1-\sqrt{1-\tau^{-1}}}\right]^2
       \quad \tau >1
    \end{split}
  \right.\,.
\end{equation}
The contribution to the total cross section from all fermions that are
heavier than the Higgs boson can be approximated as
\begin{align}
\label{eq:Asatov_diff_log_det}
  \sum_{j} \frac{Y_{jj}}{M_{j}} - \sum_{M_{j} < m_{H}} \frac{Y_{jj}}{M_{j}} = \text{Tr}(Y M^{-1}) - \sum_{M_{j} < m_{H}} \frac{Y_{jj}}{M_{j}}\,,
\end{align}
where $M$ is a matrix whose eigenvalues are the masses of the fermions
and $Y$ incorporates the corresponding Yukawa couplings. Furthermore,
one can show~\cite{Azatov:2011qy} that
\begin{equation}
  \label{eq:Azatov_log_det}
  \text{Tr}(Y M^{-1}) = \frac{\partial \log(\det M)}{\partial \VEV{h}}\,.
\end{equation}
If we repeat the analysis of ref.~\cite{Azatov:2011qy} for our
composite Higgs models we find that, for the models $\mathbf{M1_5}$ and
$\mathbf{M4_5}$, we have
\begin{equation}
  \frac{\partial \log(\det M)}{\partial \VEV h} = \frac{1}{f} \cot\left(\frac{\left\langle h \right\rangle}{f}\right)=\frac{c_\epsilon}{v}\,,
\end{equation}
whereas for the models $\mathbf{M1_{14}}$ and $\mathbf{M4_{14}}$ we obtain
  \begin{equation}
    \frac{\partial \log(\det M)}{\partial \VEV h} = \frac{2}{f} \cot\left(\frac{2 \left\langle h \right\rangle}{f}\right)=\frac{c_{2\epsilon}}{v\,c_\epsilon}\,,
\end{equation}
  which are independent of the masses and couplings of the top
  partners. For a single top partner, the above results can be checked
  explicitly by computing the Higgs partonic cross section as in
  Eq.~(\ref{eq:Higgs-xsct}) using the Yukawa couplings obtained from
  Eq.~(\ref{eq:1TPyuks}).

\subsection{Higgs production with an additional jet}

\noindent In contrast to the case of single-Higgs production from gluon fusion,
Higgs production with an additional jet $pp \rightarrow h+j$ has been
shown to have some dependence on the mass of a top partner in
composite Higgs models. In ref.~\cite{Banfi:2013yoa} it was shown how
the low energy theorem rendering the cross section insensitive to the
masses of fermions in the loop no longer holds when the transverse
momentum of one of the final states is large. For Higgs plus one extra
parton (quark or gluons), this happens at high $p_T$, the transverse
momentum of either the Higgs or the jet.  Let us consider one of the partonic subprocesses contributing to $pp \rightarrow h+j$, namely
$gg \rightarrow h+g$. The $gg \rightarrow h+g$ matrix element
$\mathcal{M}_{\lambda_{1} \lambda_{2} \lambda_{3}}$, where
${\lambda_{i}}=\pm$ denotes the helicities of the 3 gluons, for one
fermion species in the loop with mass $m_f$ and Yukawa coupling
$\frac{m_f}{v} \kappa_{f}$ will have a different behaviour
according to the size of $p_T$. For instance, for the amplitude $\mathcal{M}_{+++}$, in the limit $p_T \gg m_f, m_H$ we have~\cite{Banfi:2013yoa}
\begin{equation}
\label{eq:high_pt_element}
\mathcal{M}_{+++} \propto \frac{m_f^2 \kappa_{f}}{p_{T}} \left(A_{0}+A_{1} \ln\left(\frac{p_{T}^{2}}{m_f^{2}}\right) + A_{2} \ln^{2} \left(\frac{p_{T}^{2}}{m_f^{2}}\right)\right)\,,
\end{equation}
where $A_{0}, A_{1}, A_{2}$ are combinations of constants and
logarithms that are independent of $m_f$. On the other hand, for low
$p_{T}$ we have~\cite{Banfi:2013yoa} 
\begin{equation}
  \label{eq:M+++-lowpt}
\mathcal{M}_{+++} \propto \kappa_{f} p_{T}\,.
\end{equation}
where there is no dependence on the fermion mass, and the result is
proportional to what you would obtain for $gg \rightarrow h$.  If we
now consider a top quark, with mass $m_{t}$ and Yukawa coupling
$\frac{m_{t}}{v} \kappa_{t}$, and a top partner with mass $M_{T}$ and
Yukawa coupling $\frac{M_T}{v} \kappa_{T}$, for final states with low
$p_{T}$ the low energy theorem still applies. However, if the
transverse momentum is increased to the range
$m_{t}\ll p_{T}\ll M_{T}$, one can approximate the top and top partner
contributions to be in high-$p_{T}$ and low-$p_{T}$ limits
respectively, and obtain, in this kinematic region,
\begin{equation}
  \label{eq:M+++-higpt}
  \mathcal{M}_{+++} \propto \frac{m_{t}^{2} \kappa_{t}}{p_{T}} \left(A_{0}+A_{1} \ln\left(\frac{p_{T}^{2}}{m_{t}^{2}}\right) + A_{2} \ln^{2} \left(\frac{p_{T}^{2}}{m_{t}^{2}}\right)\right) + \kappa_{T} p_{T}\,,
\end{equation}
The expression above is only sensitive to the top mass and Yukawa
coupling, and to the top partner Yukawa coupling. The dependencies on
the top partner mass will be present if we increase $p_{T}$ further to the
region $p_{T} \gg m_{t}, m_{H}, m_{T}$, where both the top quark and top
partner contributions will approximately be in the high-$p_{T}$ limit
form given in Eq.~\eqref{eq:high_pt_element}. This behaviour of the
matrix element was also confirmed numerically~\cite{Banfi:2013yoa}.

\section{Higgs plus one jet production at the LHC}
\label{sec:results}

\noindent The difference between the differential cross section
$d\sigma/dp_T$ of a SM Higgs and that of a composite Higgs is
certainly a very useful probe of the compositeness of the Higgs.
  This was the observable considered in
  ref.~\cite{Banfi:2013yoa}. However, two $p_T$-spectra might be
  different just by an overall factor because of different total cross
  sections. Then, as explained in section~\ref{sec:HiggsProd}, such
  difference gives no information at all about the presence of top
  partners, but only on the compositeness scale, and can be already
  appreciated by looking at deviations in the total rate for Higgs
  production. In order to decorrelate the two effects, we prefer to
  work with ratios of cross sections. Therefore, in this work we
propose to employ a net Higgs plus jet efficiency, i.e. the fraction
of events for which the Higgs (or at least one jet) has a transverse
momentum larger than $p_{T}^{\rm cut}$
\begin{equation}
  \label{eq:efficiency}
  \epsilon(p_{T} > p_{T}^{\text{cut}})=\frac{1}{\sigma}\int_{p_{T}^{\text{cut}}} d p_{T} \frac{d \sigma}{d p_{T}}\,.
\end{equation}
In this case, an overall normalisation of the cross section cancels
between numerator and denominator in eq.~\eqref{eq:efficiency}, so
that this quantity is most sensitive to the mass of top-partner and
the corresponding Yukawa couplings. We now assess the deviation of the
one-jet efficiency from its SM value using the variable
\begin{equation}
\label{eq:delta_definition}
\delta(p_T^{\rm cut}) \equiv \frac{\epsilon_{\rm BSM} (p_{T} > p_{T}^{\text{cut}})}{\epsilon_{\rm SM} (p_{T} > p_{T}^{\text{cut}})} -1\,.
\end{equation}
In the above definition, $\epsilon_{\rm SM}$ denotes the SM
efficiency, while $\epsilon_{\rm BSM}$ is the efficiency of any of the
composite Higgs models studied in this work.  As a last remark, in this work
we compute $\delta(p_T^{\rm cut})$ using tree-level cross sections
only. However, it has recently been found that the K-factor (NLO/LO)
for the total Higgs cross section and the Higgs $p_T$ distribution
with full top-mass dependence are very similar, roughly a factor of 2
over a wide range of values of $p_T$~\cite{Jones:2018hbb}. Therefore,
we believe that our estimate of $\delta(p_T^{\rm cut})$ will be
basically unchanged after the inclusion of higher-order corrections.

\subsection{One light top-partner multiplet}\label{sec:onetpres}

\noindent We first consider the case of a single top partner in all of
the four models considered in section~\ref{sec:background}. The Yukawa
couplings of the top quark and top-partners to the Higgs boson are
given, in their analytical form, in Eq.~\eqref{eq:1TPyuks}. We compute
$p_T$ distributions at the LHC with a centre-of-mass energy
$\sqrt s = 14\,$TeV, the same as in the high-luminosity phase of the
LHC. As remarked in ref.~\cite{Banfi:2018pki}, with $3\,$ab$^{-1}$ of
integrated luminosity, one can reasonably probe transverse momenta up
to $1\,$TeV.  The calculation is performed using a modified version of
the program used in ref.~\cite{Banfi:2013yoa}, consisting of an
interface of the matrix elements of ref.~\cite{Baur:1989cm}
implemented in the program HERWIG 6.5~\cite{Corcella:2002jc}, with the
parton evolution toolkit hoppet~\cite{Salam:2008qg}. All numerical
results we present have been obtained by fixing $m_t=173.5\,$GeV,
bottom mass $m_b=4.65\,$GeV, and using MSTW2008 NLO parton
distribution functions~\cite{Martin:2009iq}, corresponding to
$\alpha_s(M_Z)=0.12$.  We present contour plots for
$\delta(p_{T}^{\text{cut}})$ (expressed as a percentage), as a
function of the mass of the top-partner $M_T$ and the compositeness
scale $f$.  The case $f \gg v$ has been already considered in
ref.~\cite{Banfi:2013yoa}. Here we focus on a range of values of $M_T$
and $f$ that are not excluded by current measurements, and where there
is no specific hierarchy between the two scales.

In the following contour plots, for the singlet models we fix the
value of $\sin^{2} \theta_{L}$, whereas for the fourplet models we fix
the value of $\sin^{2} \theta_{R}$.  The reason behind this is that,
in singlet models, the modification of Yukawa couplings depend only on
$\sin^2 \theta_L$, which, according to Eq.~\eqref{eq:thL-thR}, becomes
increasingly large with the top-partner mass. For large values of $f$,
the contribution of the top becomes smaller and smaller, and the
spectrum is dominated by the contribution of the top-partner.  The
situation of fourplet models is completely different. There, for large
$f$, the Yukawa couplings depend largely on $\sin^2
\theta_R$. However, with increasing top-partner masses and finite $f$
the Yukawa couplings contain a negative contribution proportional to
$\cos^2 \theta_L$ for the top and $\sin^2\theta_L$ for the top
partner.  It is then more reasonable to fix different mixing angles
for different representation of the top partner involved in the
models. In addition, in the following contour plots, we also indicate
the region where the value of $\kappa_t$ falls below $0.8$, which is
in tension with data, as explained in
section~\ref{sec:perturbativity}. Note that, in the limit $f \gg v$, our
predictions correspond to those presented in
ref.~\cite{Banfi:2013yoa}. In the following, we try then to choose the
same parameters as those in ref.~\cite{Banfi:2013yoa}, so as to be
able to assess the impact of a finite value of $f$ with respect to the
limit $f\gg v$ considered there.

\begin{figure}[h]
  \centering
  \includegraphics[width=.8\textwidth]{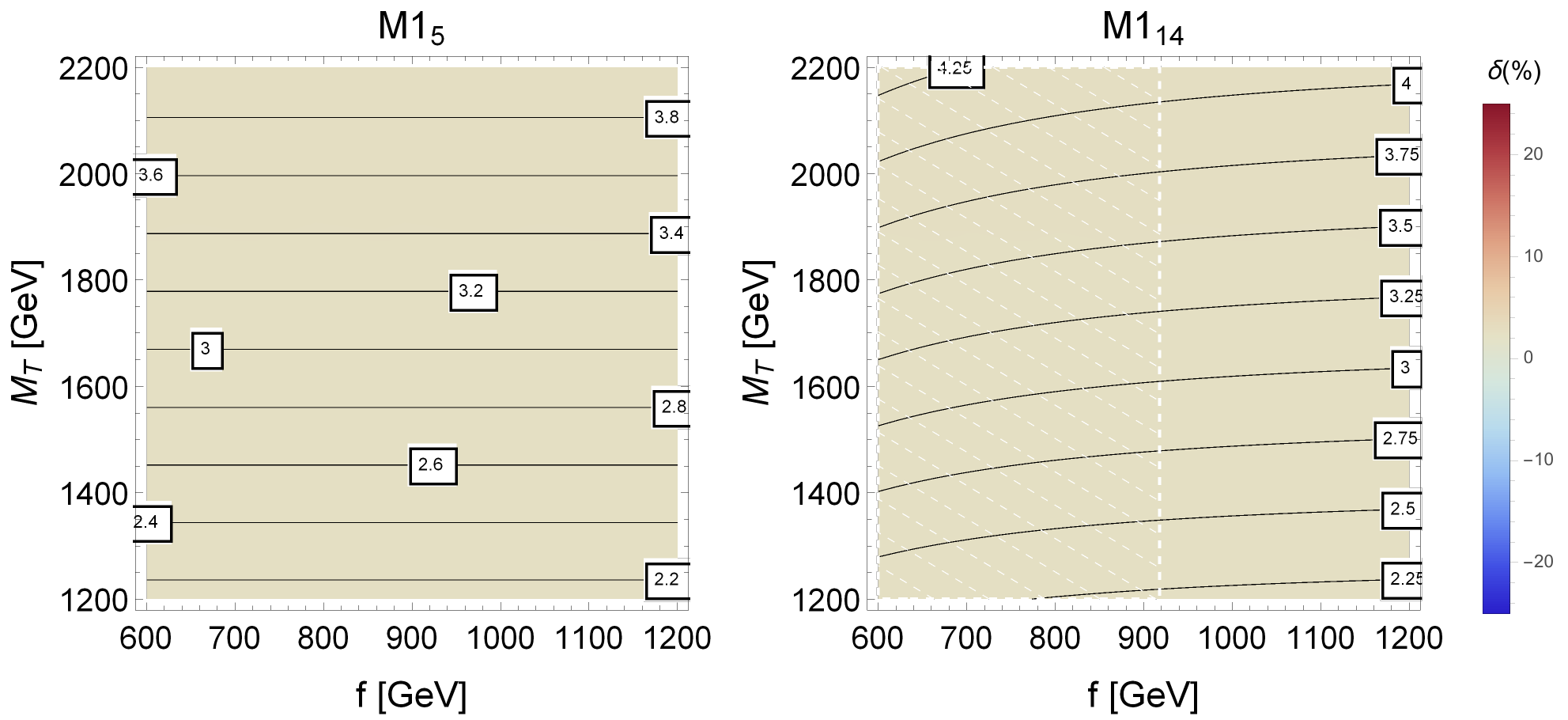}
  \vspace{5mm}
  \caption{The contour plots of $\delta(p_{T}^{\text{cut}})$ with
    $\sin^{2} \theta_{L} = 0.1$ and $p_{T}^{\text{cut}} = 200$ GeV for
    each of the singlet models with one top partner multiplet. The
    solid lines correspond to constant values of the coupling $y$.
    The region marked by dashed white lines indicates when $\kappa_t\leq0.8$.}
  \label{fig:delta-sth2_0.1-pt_200-LHC14-singlet}
\end{figure}

\begin{figure}[h]
  \centering
  \includegraphics[width=.8\textwidth]{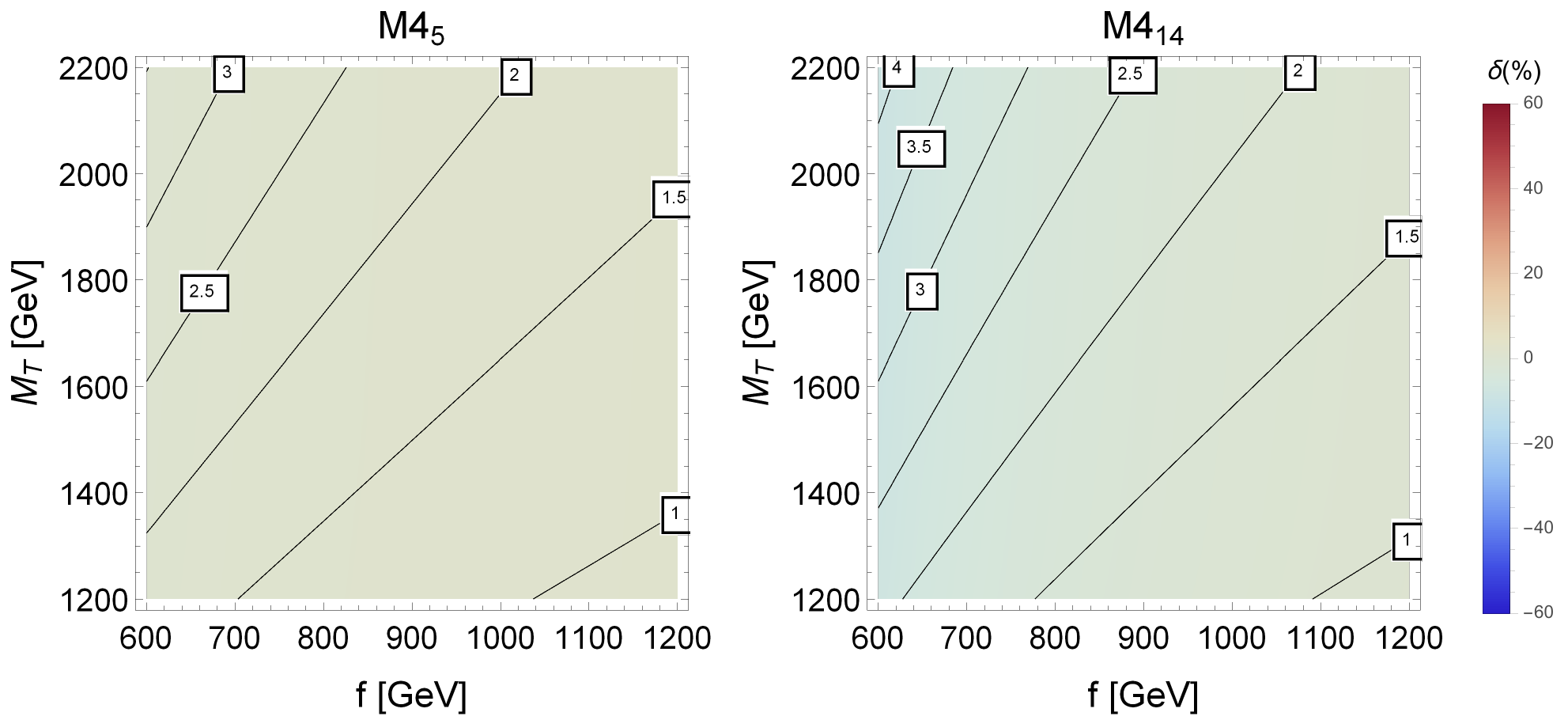}
  \vspace{5mm}
  \caption{The contour plots of $\delta(p_{T}^{\text{cut}})$ with
    $\sin^{2} \theta_{R} = 0.1$ and $p_{T}^{\text{cut}} = 200$ GeV for
    each of the fourplet models with one top partner multiplet.  The
    solid lines correspond to constant values of the coupling $y$.
    None of the parameter space on these plots result in $\kappa_t\leq0.8$.}
  \label{fig:delta-sth2_0.1-pt_200-LHC14-fourplet}
\end{figure}

In Fig.~\ref{fig:delta-sth2_0.1-pt_200-LHC14-singlet}, we show contour
plots of $\delta(p_{T}^{\text{cut}})$ for $p_{T}^{\text{cut}} = 200$
GeV and $\sin^{2} \theta_{L} = 0.1$ for singlet models. Similarly, in
Fig.~\ref{fig:delta-sth2_0.1-pt_200-LHC14-fourplet}, we show the
contour plots of $\delta(p_{T}^{\text{cut}})$ for fourplet models with
$\sin^{2} \theta_{R} = 0.1$ for the same value of
$p_{T}^{\text{cut}}$.
\begin{figure}[h]
  \centering
  \includegraphics[width=.8\textwidth]{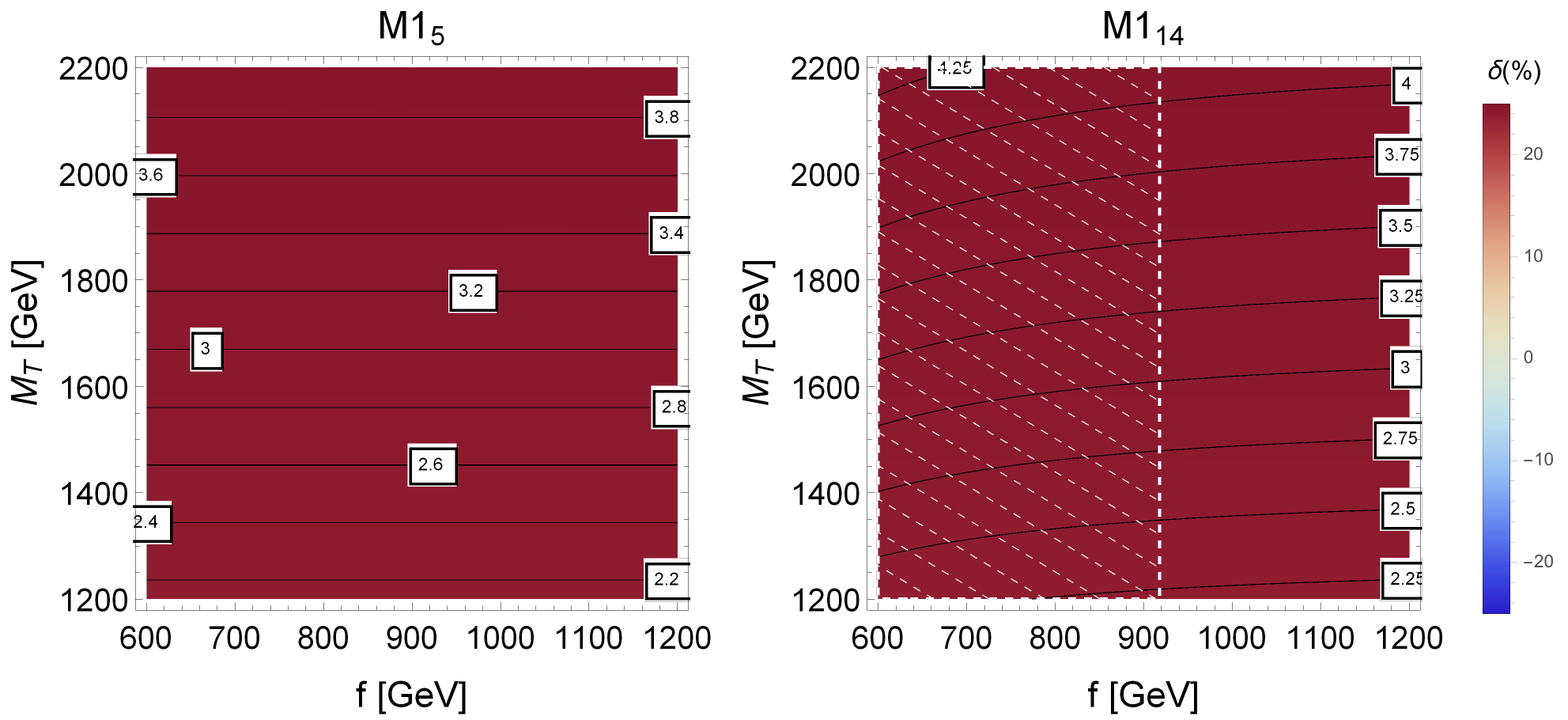}
  \vspace{5mm}
  \caption{The contour plots of $\delta(p_{T}^{\text{cut}})$ with
    $\sin^{2} \theta_{L} = 0.1$ and $p_{T}^{\text{cut}} = 600$ GeV for
    each of the singlet models with one top partner multiplet.  The
    corresponding values of $y$ are indicated by the solid lines.
     The region marked by dashed white lines indicates when $\kappa_t\leq0.8$.}
  \label{fig:delta-sth2_0.1-pt_600-LHC14-singlet}
\end{figure}
First, we observe that the deviation from the SM is not large. This is
due to the fact that the integrated transverse momentum spectrum is
dominated by the lowest values of $p_T$. There, the top still behaves
as a heavy particle in loops, therefore the cancellation between top
and top-partner contributions is still at work. Nevertheless, there is
a very different behaviour for singlet
(Fig.~\ref{fig:delta-sth2_0.1-pt_200-LHC14-singlet}) and fourplet
(Fig.~\ref{fig:delta-sth2_0.1-pt_200-LHC14-fourplet}) models. 
For singlet models, the deviation from the SM mildly increases as
$M_T$ is increased. For fourplet models the
deviations increases with increasing $f$. This behaviour arises since
negative contributions from the Yukawa coupling due to
$\sin^2 \theta_L$ and $\cos^2\theta_L$ become smaller as $f$ is
increased. Note that, for $\mathbf{M4}_{14}$, these negative
contributions dominate for small values of $f$, and one gets negative
interference between the contribution of the top and the top
partner. In these, and all remaining contour plots, we draw solid
lines that correspond to fixed values of the coupling $y$, so as to
highlight whether the corresponding choice of parameters correspond to
a perturbative composite Higgs model. We recall that perturbativity
requires $y<3$. In this respect, we observe that, for singlet models,
we cannot legitimately probe top-partner masses above $1600\,$GeV. For
fourplet modes instead, our choice of parameters leads to predictions
that are almost always within the perturbative region.

\begin{figure}[h]
  \centering
  \includegraphics[width=.8\textwidth]{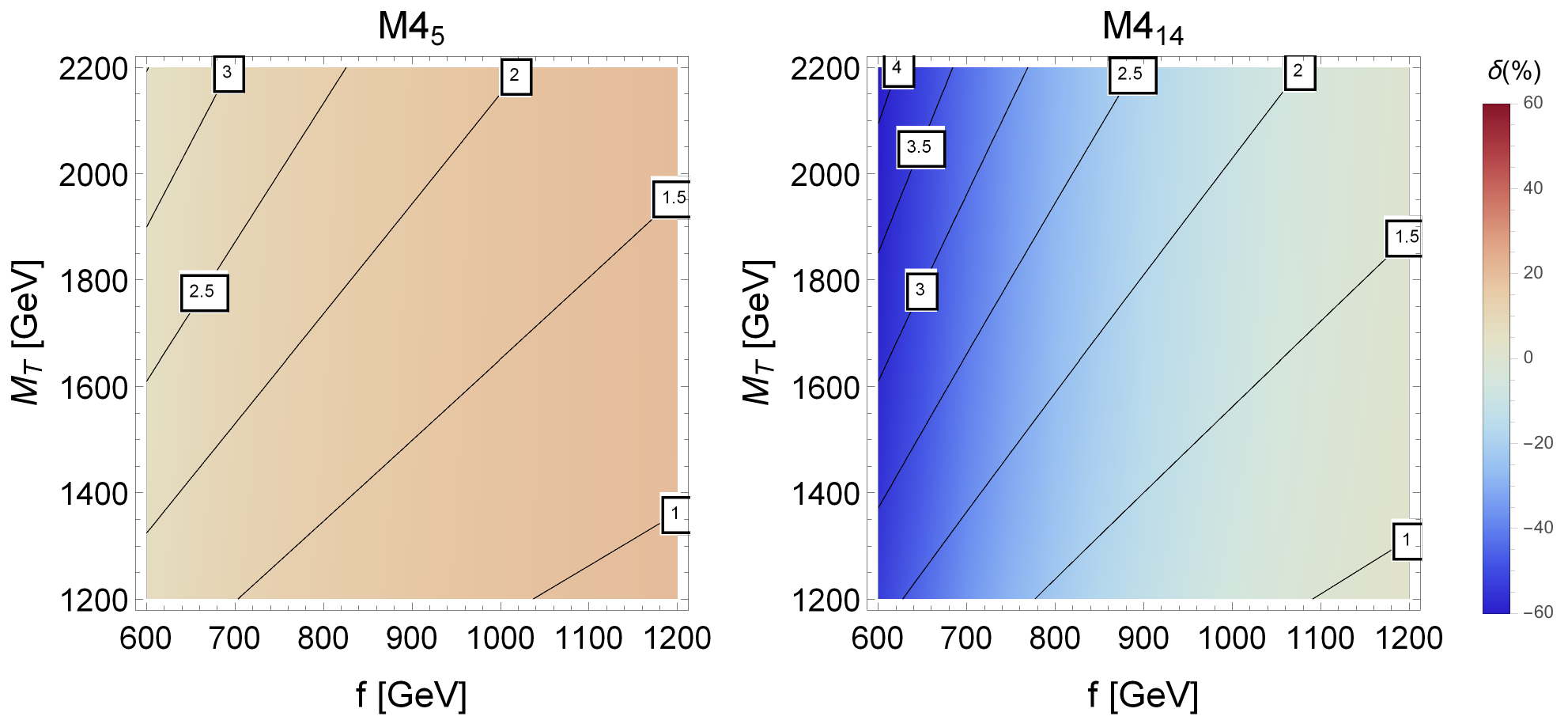}
  \vspace{5mm}
  \caption{The contour plots of $\delta(p_{T}^{\text{cut}})$ with
    $\sin^{2} \theta_{R} = 0.1$ and $p_{T}^{\text{cut}} = 600$ GeV for
    each of the fourlet models with one top partner multiplet.  The
    corresponding values of $y$ are indicated by the solid lines.
    None of the parameter space on these plots result in $\kappa_t\leq0.8$.}
  \label{fig:delta-sth2_0.1-pt_600-LHC14-fourplet}
\end{figure}

We now keep the values $\sin^2\theta_{L,R}=0.1$ and increase
$p_{T}^{\text{cut}}$ to $600\,$GeV. The corresponding contour plots
are shown in Figs.~\ref{fig:delta-sth2_0.1-pt_600-LHC14-singlet} and \ref{fig:delta-sth2_0.1-pt_600-LHC14-fourplet}, again as a function
of $M_T$ and $f$.
The $p_T$ values probed here are high enough to break the cancellation
between the contribution of a top and a top-partner in loops. This is
why, for singlet models, we observe huge deviations from the SM. For
fourplet models, we note, again, that the deviation decreases with
decreasing $f$. This is again due to the fact that for smaller $f$,
the negative contribution to the Yukawa couplings due to
$\sin^2 \theta_L$ and $\cos^2\theta_L$ becomes more important, and
vanishes for $f\to \infty$. The most striking feature occurs for
$\mathbf{M4}_{14}$ at small values of $f$, where one sees a large
negative interference between top and top-partner contributions.

\begin{figure}[h]
  \centering
  \includegraphics[width=.8\textwidth]{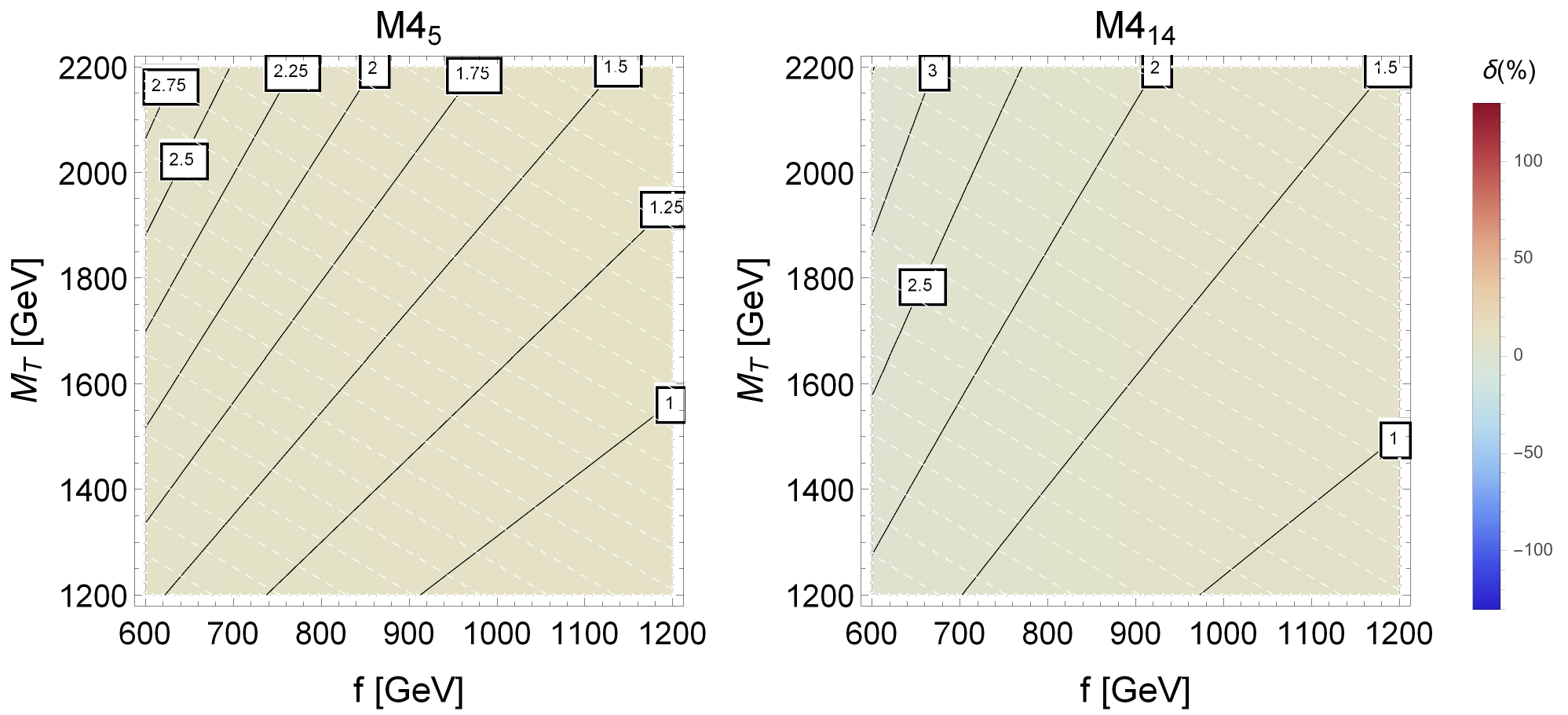}
  \vspace{5mm}
  \caption{The contour plots of $\delta$ with
    $\sin^{2} \theta_{R} = 0.4$ and $p_{T}^{\text{cut}} = 200$ GeV for
    each of the fourplet models with one top partner multiplet.  The
    corresponding values of $y$ are indicated by the solid lines.
    As indicated by the dashed white lines, all points on these plots result in $\kappa_t\leq0.8$.}
  \label{fig:delta-sth2_0.4-pt_200-LHC14-fourplet}
\end{figure}

To have a perfect parallel with ref.~\cite{Banfi:2013yoa} we should
repeat the same analysis for $\sin^{2}
\theta_{L,R}=0.4$. Unfortunately, singlet models with
$\sin^{2} \theta_{L}=0.4$ are outside the perturbative
regime. Therefore, we can only consider fourplet models with
$\sin^2 \theta_R=0.4$. Contour plots with
$p_{T}^{\text{cut}} = 200$ GeV are shown in
Fig.~\ref{fig:delta-sth2_0.4-pt_200-LHC14-fourplet}
\begin{figure}[h]
  \centering
  \includegraphics[width=.8\textwidth]{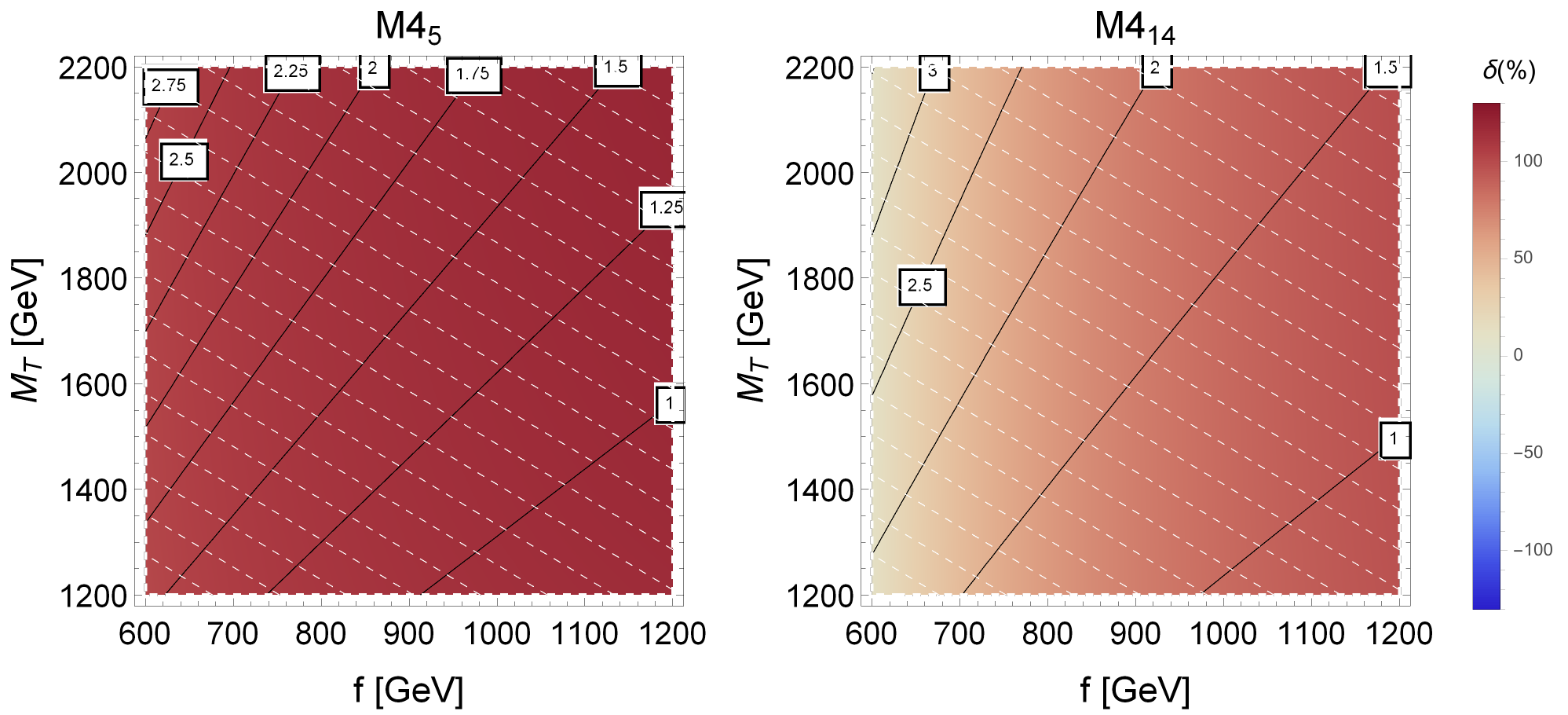}
  \vspace{5mm}
  \caption{The contour plots for $\delta$ with
    $\sin^{2} \theta_{R} = 0.4$ and $p_{T}^{\text{cut}} = 600$ GeV for
    each of the fourplet models with one top partner multiplet.  The
    corresponding values of $y$ are indicated by the solid lines.
     As indicated by the dashed white lines, all points on these plots result in $\kappa_t\leq0.8$.}
  \label{fig:delta-sth2_0.4-pt_600-LHC14-fourplet}
\end{figure}
Here the deviation from the SM is again moderate, for the same reasons
as the corresponding case with $\sin^2\theta_R=0.1$. 
Also here, $\sin^2\theta_R$ is bigger so the negative contributions to the
Yukawa couplings proportional to $\cos^2\theta_L$ and $\sin^2\theta_L$
become less important.
Finally, for fourplet models we consider the case
$p_{T}^{\text{cut}}=600\,$GeV, and $\sin^2\theta_R=0.4$, whose contour
plots are shown in Fig.~\ref{fig:delta-sth2_0.4-pt_600-LHC14-fourplet}.

The fact that $\sin^2\theta_R=0.4$ is larger actually prevents the
negative contributions from taking over. Therefore, the contribution of the
top quark in the loops becomes smaller than those with the
top-partner, giving a sizeable deviation with respect to the SM.

\begin{figure}[h]
  \centering
  \includegraphics[width=.8\textwidth]{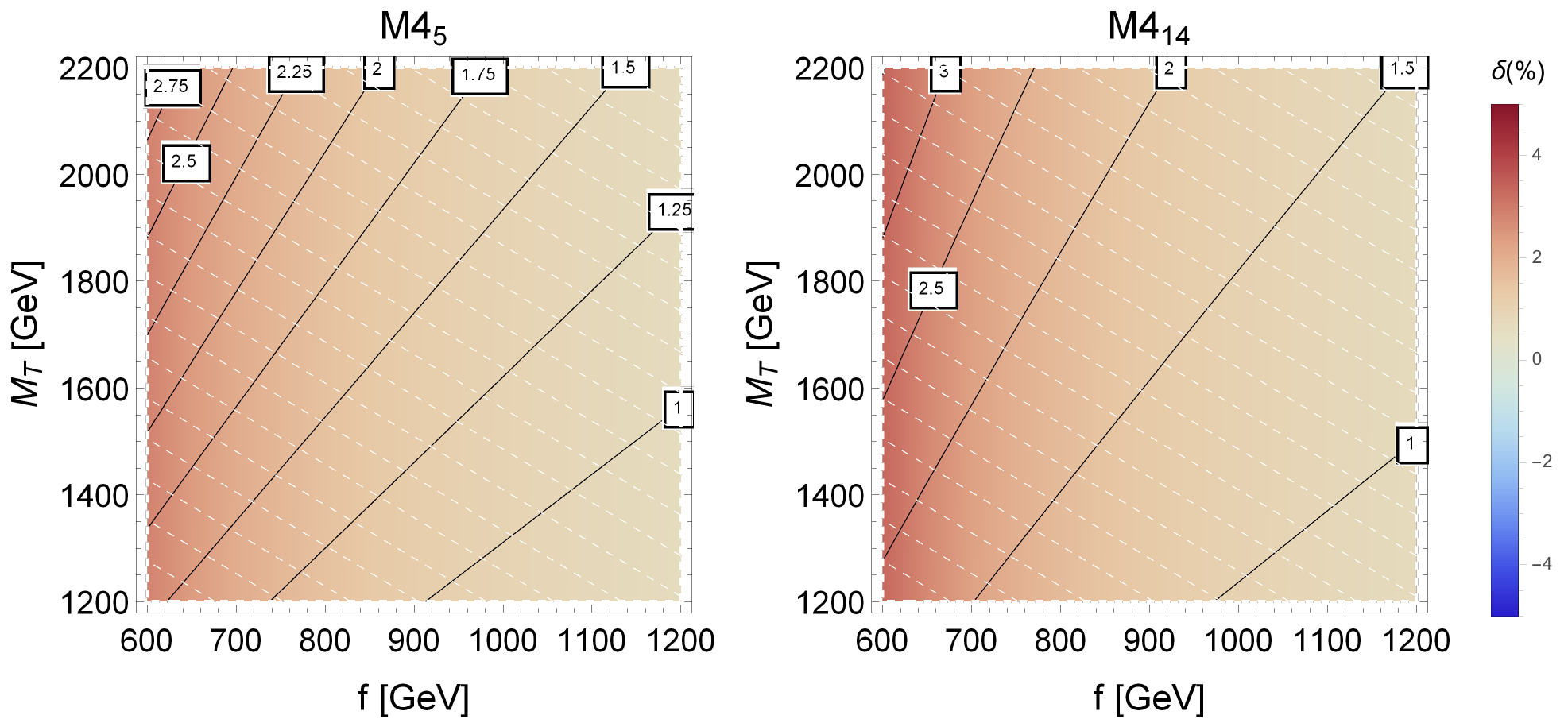}
  \vspace{5mm}
  \caption{The contour plots of the contribution to $\delta$ with
    $\sin^{2} \theta_{R} = 0.4$ and $p_{T}^{\text{cut}} =600$ GeV for
    each of the fourplet models with one top partner multiplet. In
    this figure, we take into account only the CP-odd Yukawa coupling.
    The corresponding values of $y$ are indicated by the solid lines.
    As indicated by the dashed white lines, all points on these plots result in $\kappa_t\leq0.8$.}
  \label{fig:delta-sth2_04-pt_600-LHC14-contour_cpodd}
\end{figure}

We now consider the CP-odd contributions induced by the couplings
$\tilde \kappa_t$ and $\tilde \kappa_T$ in
Eq.~(\ref{eq:1TPCPoddyuks}). They exist only for fouplet models and,
due to the fact that they cannot interfere with the SM, they are very
small. For the values of $\sin^2\theta_R$ we consider, their
contribution is at the sub-percent level, except for
$\sin^2\theta_R=0.4$ and $\ptcut=600\,$GeV, where one gets an
additional deviation of a few percent with respect to the SM. The
corresponding contour plots are shown in
Fig.~\ref{fig:delta-sth2_04-pt_600-LHC14-contour_cpodd}.

\begin{figure}[h]
  \centering
  \includegraphics[width=.8\textwidth]{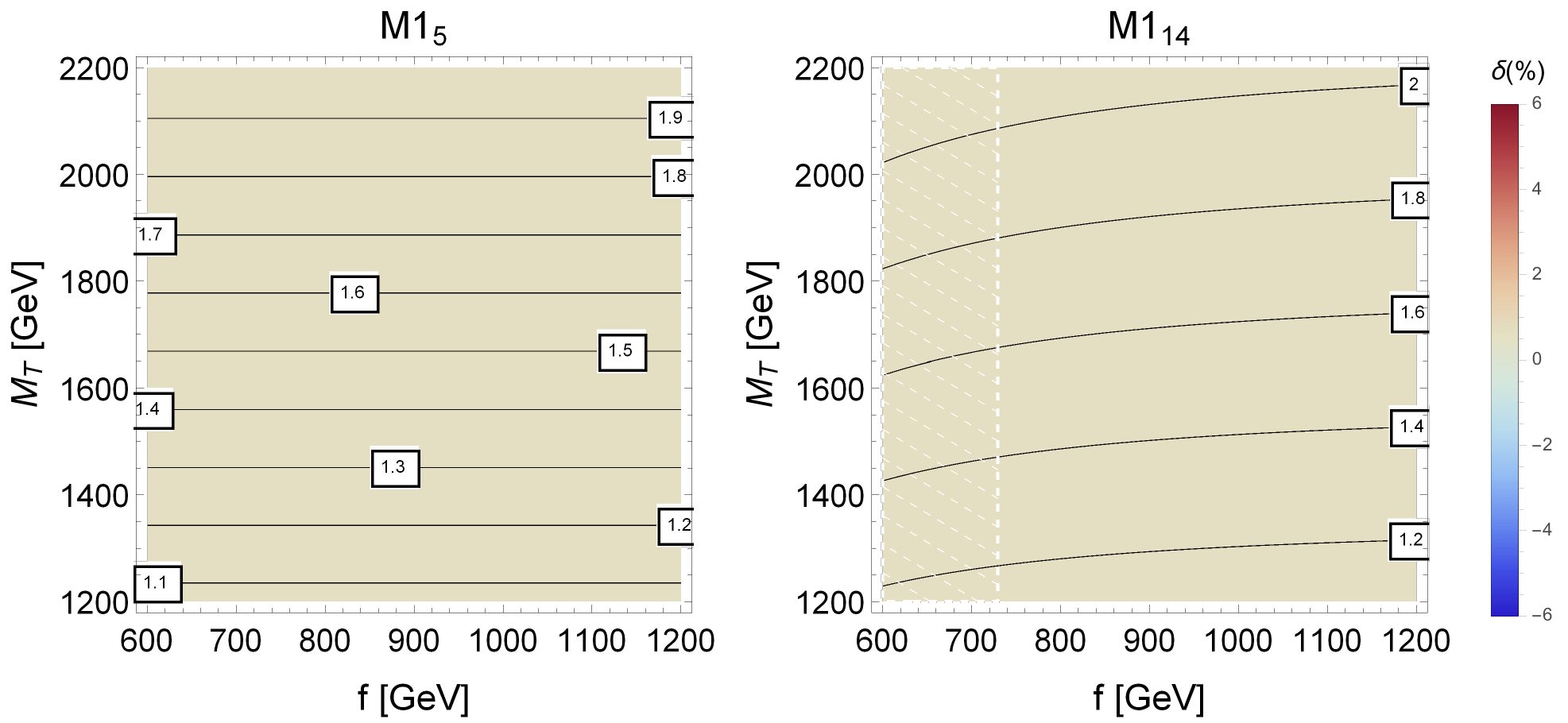}
  \vspace{5mm}
  \caption{The contour plots of $\delta(p_{T}^{\text{cut}})$ with
    $\sin^{2} \theta_{L} = 0.025$ and $p_{T}^{\text{cut}} = 200$ GeV
    for the singlet models with one top partner multiplet.  The
    corresponding values of $y$ are indicated by the solid lines.
    The region marked by dashed white lines indicates when $\kappa_t\leq0.8$.}
  \label{fig:delta-sth2_0.025-pt_200-LHC14-singlet}
\end{figure}

\begin{figure}[h]
  \centering
  \includegraphics[width=.8\textwidth]{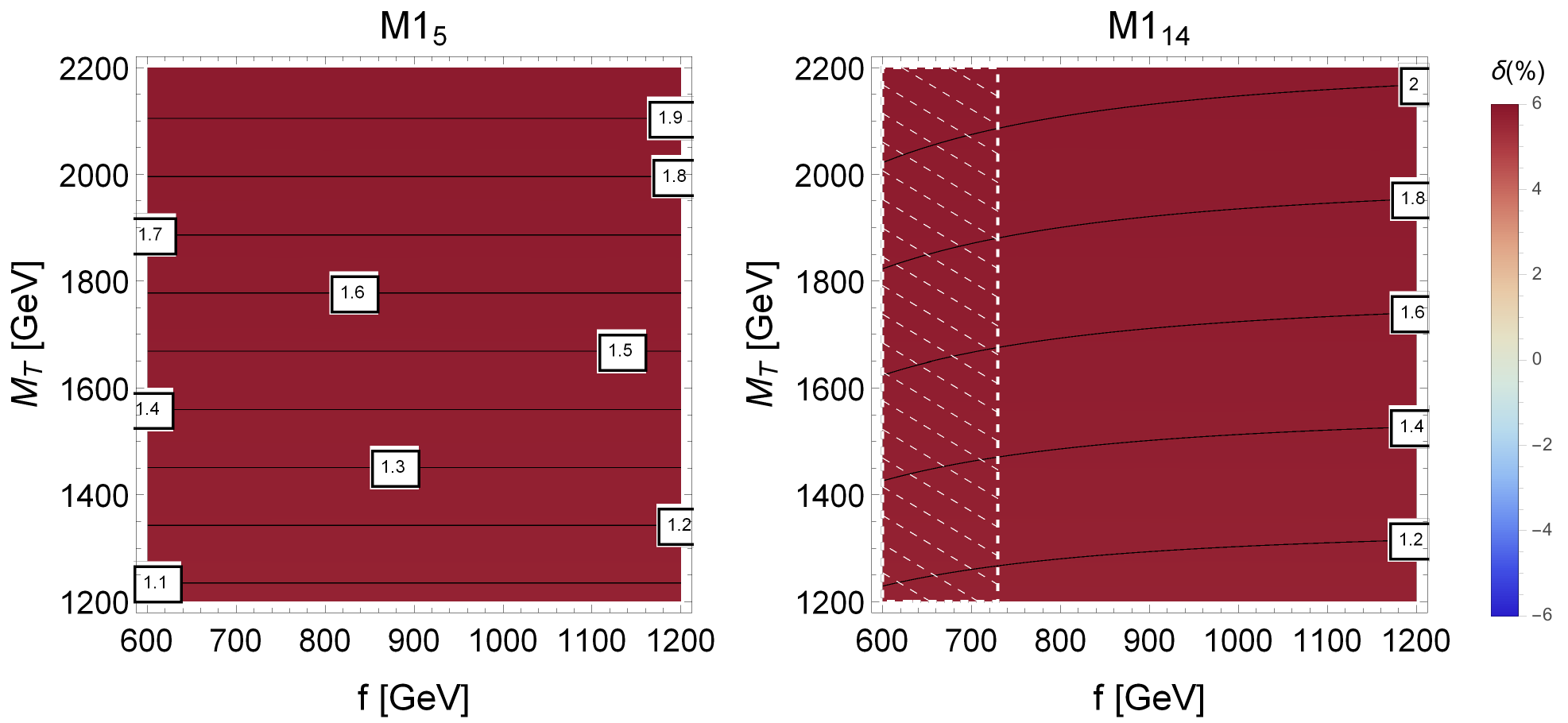}
  \vspace{5mm}
  \caption{The contour plots of $\delta(p_{T}^{\text{cut}})$ with
    $\sin^{2} \theta_{L} = 0.025$ and $p_{T}^{\text{cut}} = 600$ GeV
    for the singlet models with one top partner multiplet.  The
    corresponding values of $y$ are indicated by the solid lines.
    The region marked by dashed white lines indicates when $\kappa_t\leq0.8$.}
  \label{fig:delta-sth2_0.025-pt_600-LHC14-singlet}
\end{figure}

In order to have an extra example of deviations one might
expect for singlet models, we analyse the case where
$\sin^{2}_L=0.025$. Contour plots with $p_{T}^{\text{cut}} = 200$ GeV
are shown in Fig.~\ref{fig:delta-sth2_0.025-pt_200-LHC14-singlet}, and
those with $p_{T}^{\text{cut}} = 600$ GeV are shown in
Fig.~\ref{fig:delta-sth2_0.025-pt_600-LHC14-singlet}. Comparing these two figures we observe again that as
$p_{T}^{\text{cut}}$ is increased the cancellation between the
contribution from the top and top partner in the loop is overcome and hence the deviation from the SM is more prominent in
Fig.~\ref{fig:delta-sth2_0.025-pt_600-LHC14-singlet}. 
Also, we see in both
figures that as $f$ is increased the behaviour
approaches that of the SM.

To summarise the results of this section, with one top-partner we see
a variety of deviations from the SM, reflecting the different ways in
which the Yukawa couplings are modified according to the fundamental
parameter of each model. In particular, for singlet models, even a
mild mixing of right-handed fermions leads to huge deviations from the
SM. Therefore, the parameters of these models will be the easiest to
access through Higgs production plus one jet. For fourplet models, due
to non-trivial cancellations between different contributions to the
Yukawa couplings, the situation has to be analysed on a case-by-case
basis. The most promising situation occurs for large mixings, where,
using high values of $p_{T}^{\text{cut}}$, one expects to see sizeable
deviations from the SM.

\subsection{Two light top-partner multiplets} \label{sec:twotpres}

\noindent In this section we extend the analysis presented before to the case of
two top partners. Instead of varying the actual masses and couplings
of the top-partners, we consider a number of benchmark scenarios
obtained by fixing some of the fundamental parameters of theory as
described in section~\ref{sec:2tps}.
We first consider three benchmark scenarios with the CP-odd couplings
$c_{1,1}$ and $c_{1,2}$ set to zero, and a fourth with non-zero CP-odd couplings:
\begin{enumerate}
\item\label{Mpsivaries} $y=1$, $\MPsi{1}=1200\,$GeV,
  $1300\,\text{GeV}<\MPsi{2}<3000\,$GeV, $f=800\,$GeV (see
  Figs.~\ref{fig:masTops},~\ref{fig:yukTops}).
  \item\label{yvaries}  $0.5<y<3$, $\MPsi{1}=1200\,$GeV, $\MPsi{2}=1300\,$GeV, $f=800\,$GeV (see Fig.~\ref{fig:yDep}) (fourplet models only).
\item\label{fvaries} $y=1$, $\MPsi{1}=1200\,$GeV, $\MPsi{2}=1300\,$GeV,
  $800\,\text{GeV}<f<2000\,$GeV
  (see Figs.~\ref{fig:fYukDep},~\ref{fig:fDep}).
\item\label{fvaries-cpodd} $y=2$, $\MPsi{1}=1200\,$GeV, $\MPsi{2}=1300\,$GeV,
  $800\,\text{GeV}<f<1400\,$GeV, $c_{1,1}=c_{1,2}=0.2i$
  (see Figs.~\ref{fig:CPo1} and ~\ref{fig:CPo2}). 
\end{enumerate}
Benchmark scenario~\ref{Mpsivaries} investigates the effect of
increasing the vector-like mass $\MPsi{2}$, from the case in which
it is quasi degenerate with $\MPsi{1}$ to the case in which the
second top partner decouples, i.e.\ $\MPsi{2}\gg \MPsi{1}$. The
compositeness scale is set to $f=800\,$GeV, an intermediate value with
respect to the two shown in
Figs.~\ref{fig:masTops},~\ref{fig:yukTops}. 
\begin{figure}[htbp]
  \begin{minipage}[l]{0.5\textwidth}
  \includegraphics[width=.9\textwidth]{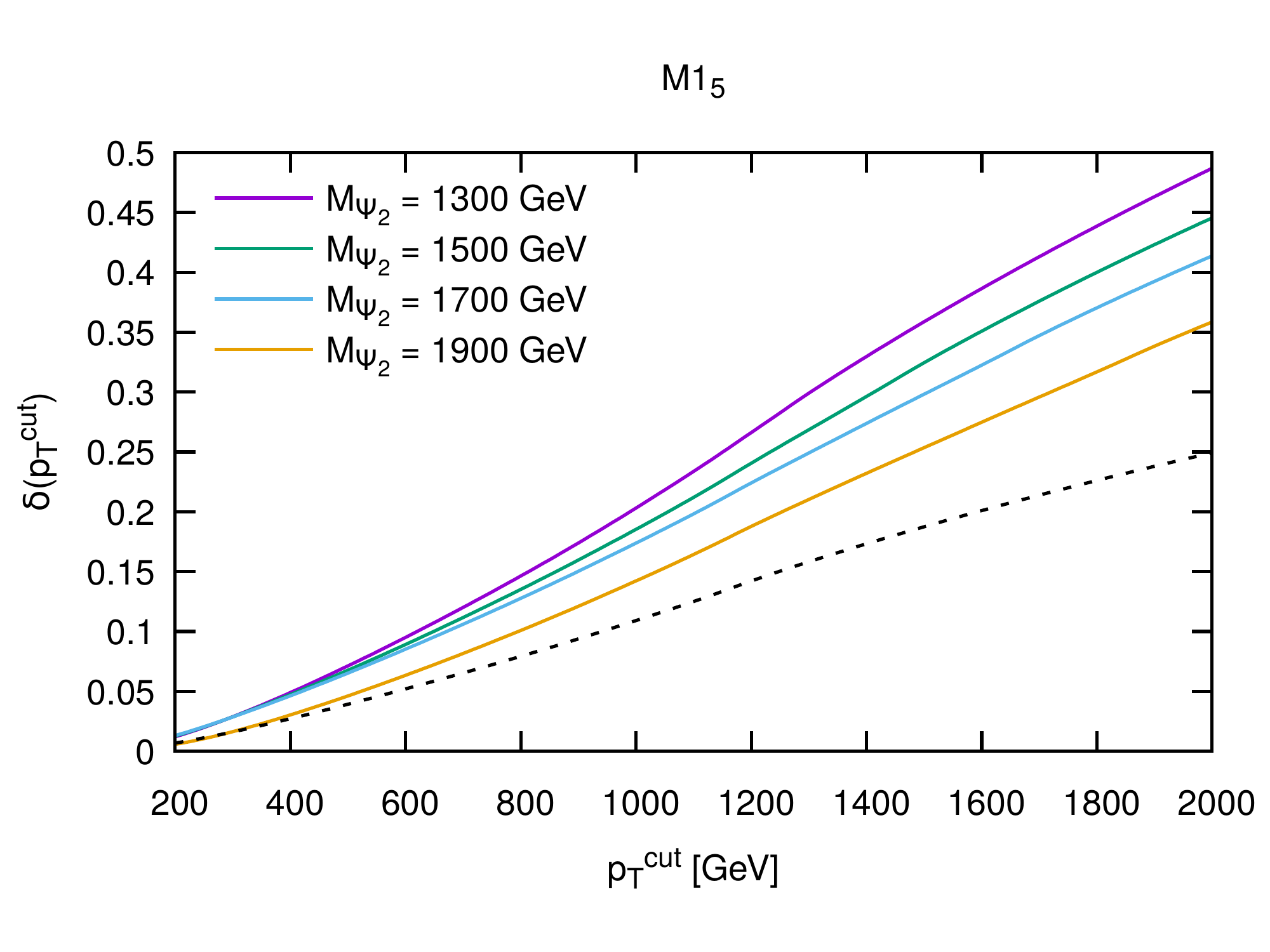}
  \end{minipage}
  \begin{minipage}[r]{0.5\textwidth}
  \includegraphics[width=.9\textwidth]{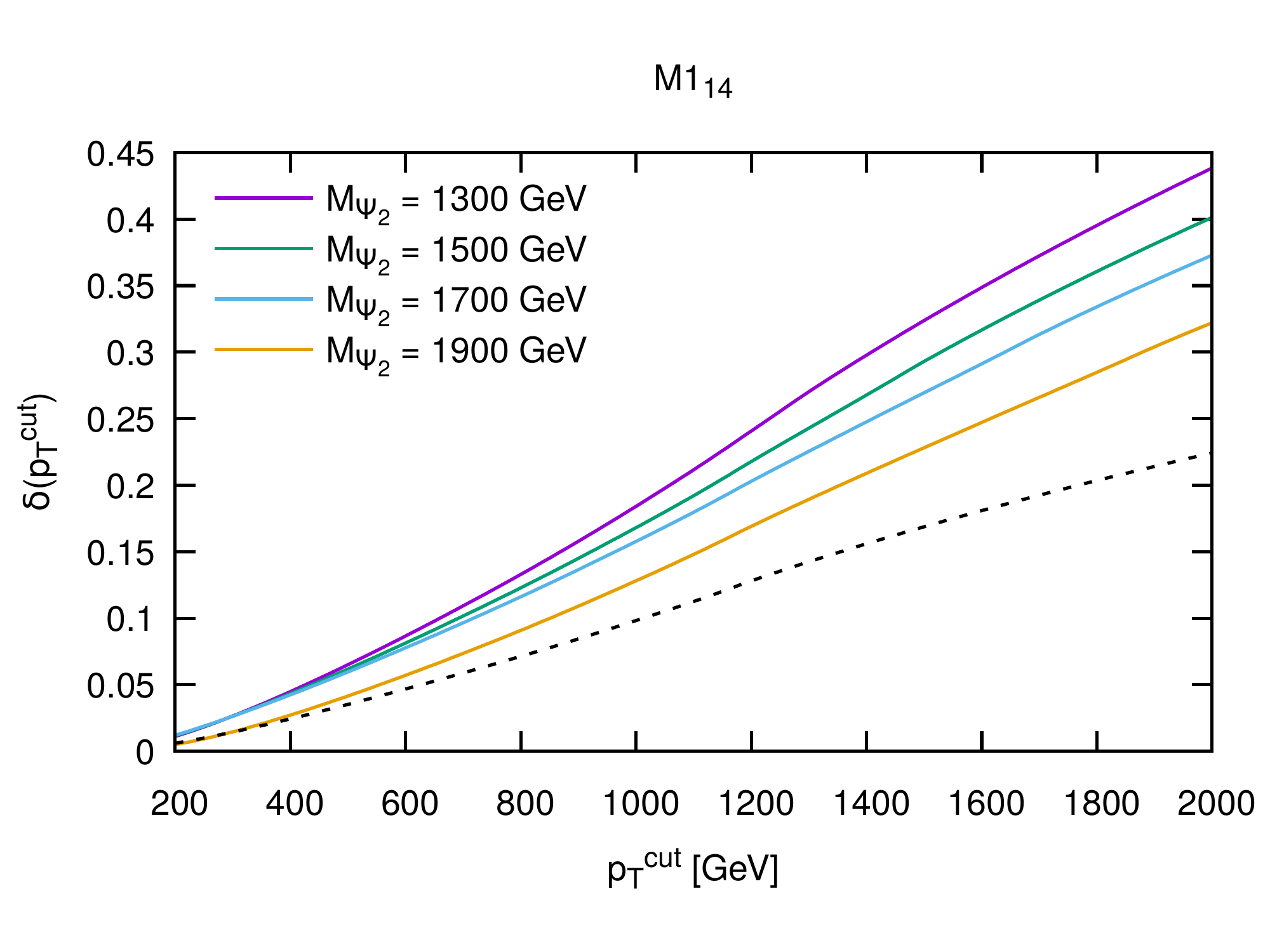}
  \end{minipage}
  \begin{minipage}[l]{0.5\linewidth}
  \includegraphics[width=.9\textwidth]{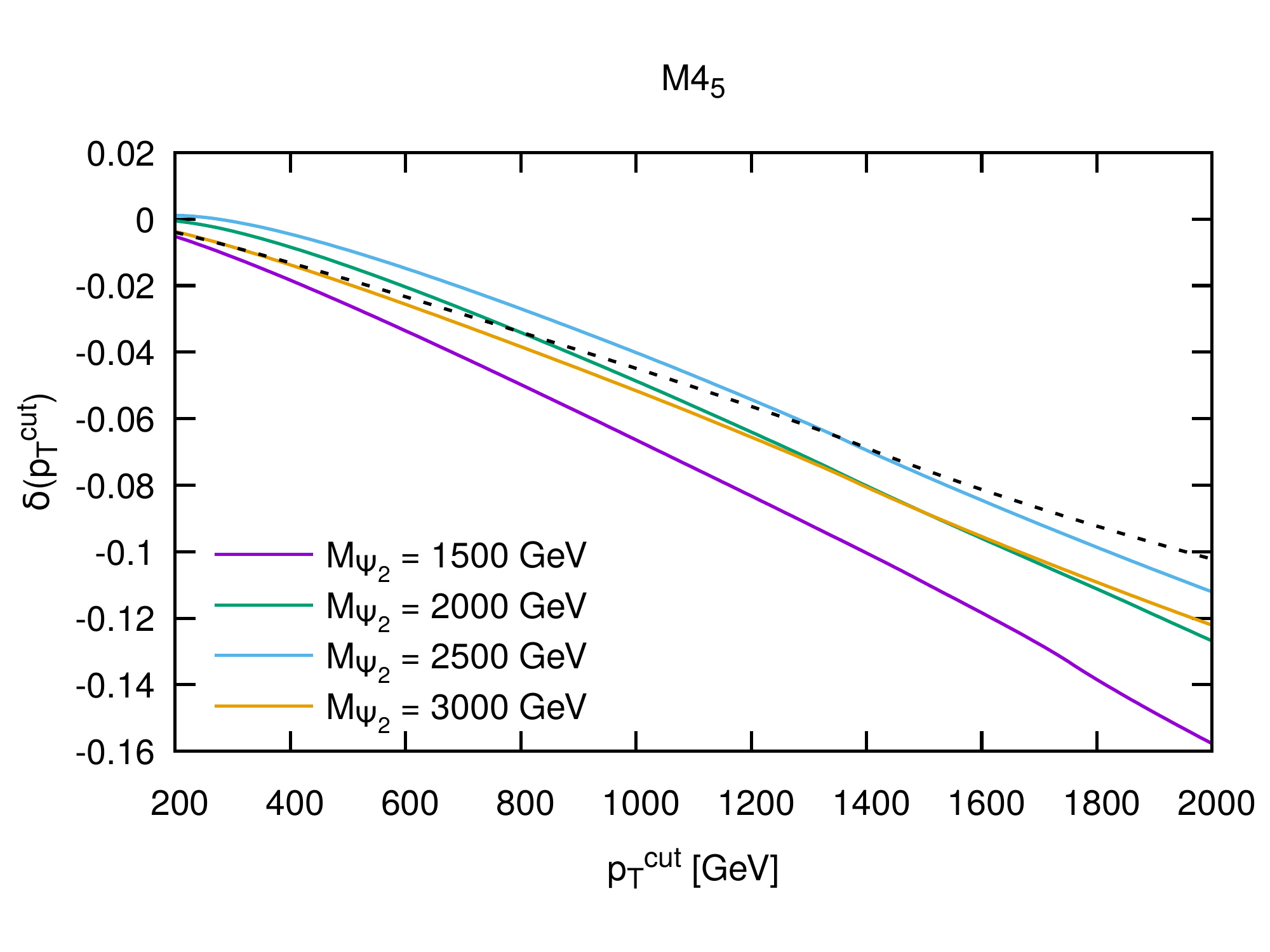}
\end{minipage}
  \begin{minipage}[r]{0.5\linewidth}
\includegraphics[width=.9\textwidth]{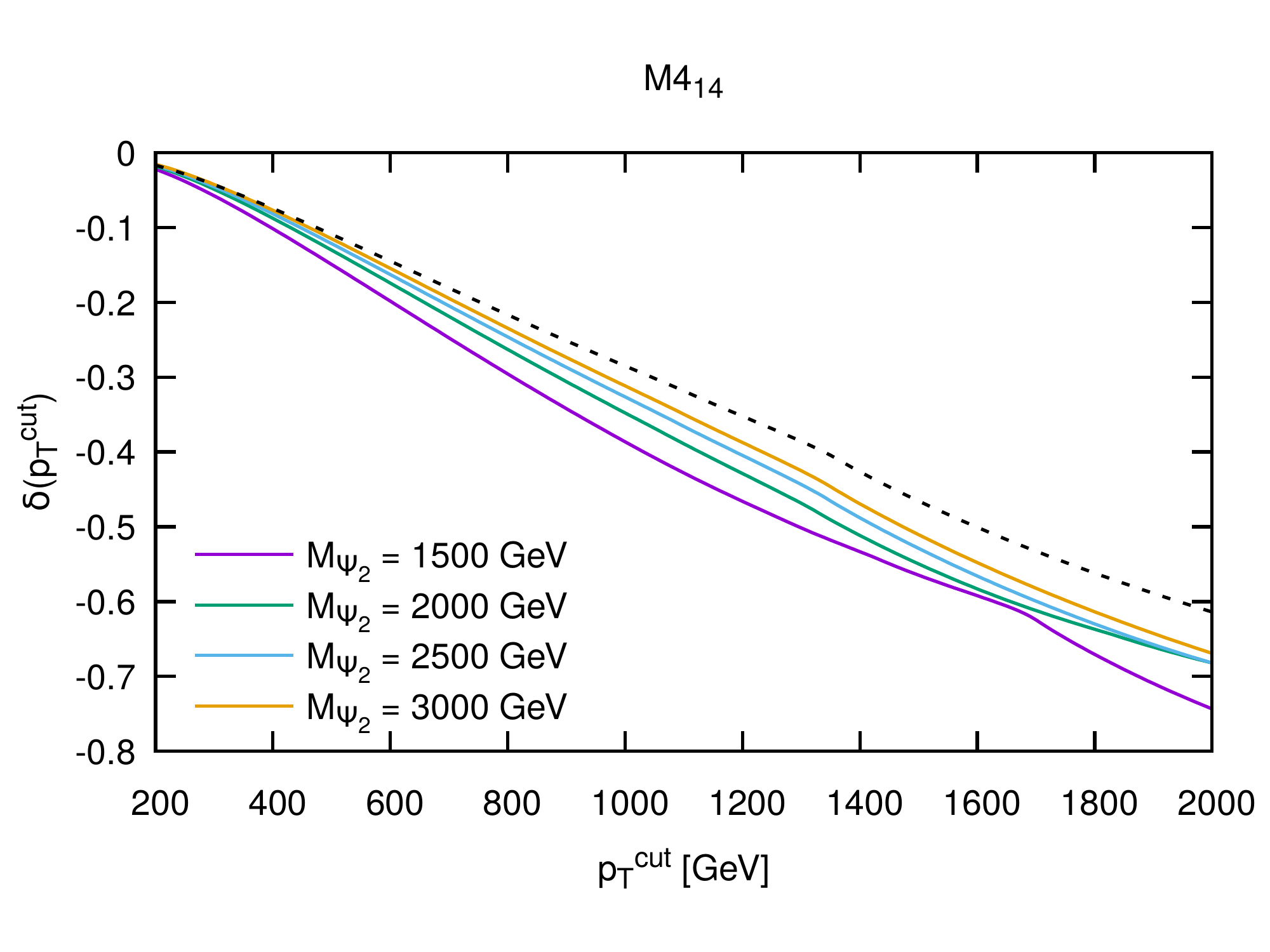}
\end{minipage}
\caption{The distribution $\delta(p_T^{\rm cut})$ for benchmark
  scenario~\ref{Mpsivaries} and the four models considered in
  section~\ref{sec:2tps}.}
  \label{fig:bench-Mpsivaries}
\end{figure}
The relative deviation from the SM $\delta(\ptcut)$ is plotted in
Fig.~\ref{fig:bench-Mpsivaries}, as a function of $\ptcut$, for
selected values of $\MPsi{2}$ (the solid curves), and for the case
with one top partner (the dashed curve), with the same value of $y$
and $M_{\Psi}=\MPsi{1}$. Here we also set $f=800\,$GeV, a value
half-way between those considered in Figs.~\ref{fig:masTops}
and~\ref{fig:yukTops}.  This benchmark scenario does not present any
unexpected features. For singlet models we have an enhancement with
respect to the SM, and for fourplet models we have a depletion due to negative
interference. We notice that there is an appreciable dependence on the
vector-like quark mass $\MPsi{2}$, in accordance with the valued of
the couplings reported in Fig.~\ref{fig:yukTops}. Also, when
$\MPsi{2}$ gets bigger, the heavier top-partner decouples, and the
deviation tends to that with a single top partner. Again, this is
expected from Figs.~\ref{fig:masTops} and~\ref{fig:yukTops}, where we
see that the masses and couplings of the lighter top-partner approach
those of the single top-partner scenario.

\begin{figure}[htbp]
  \begin{minipage}[l]{0.5\linewidth}
  \includegraphics[width=.9\textwidth]{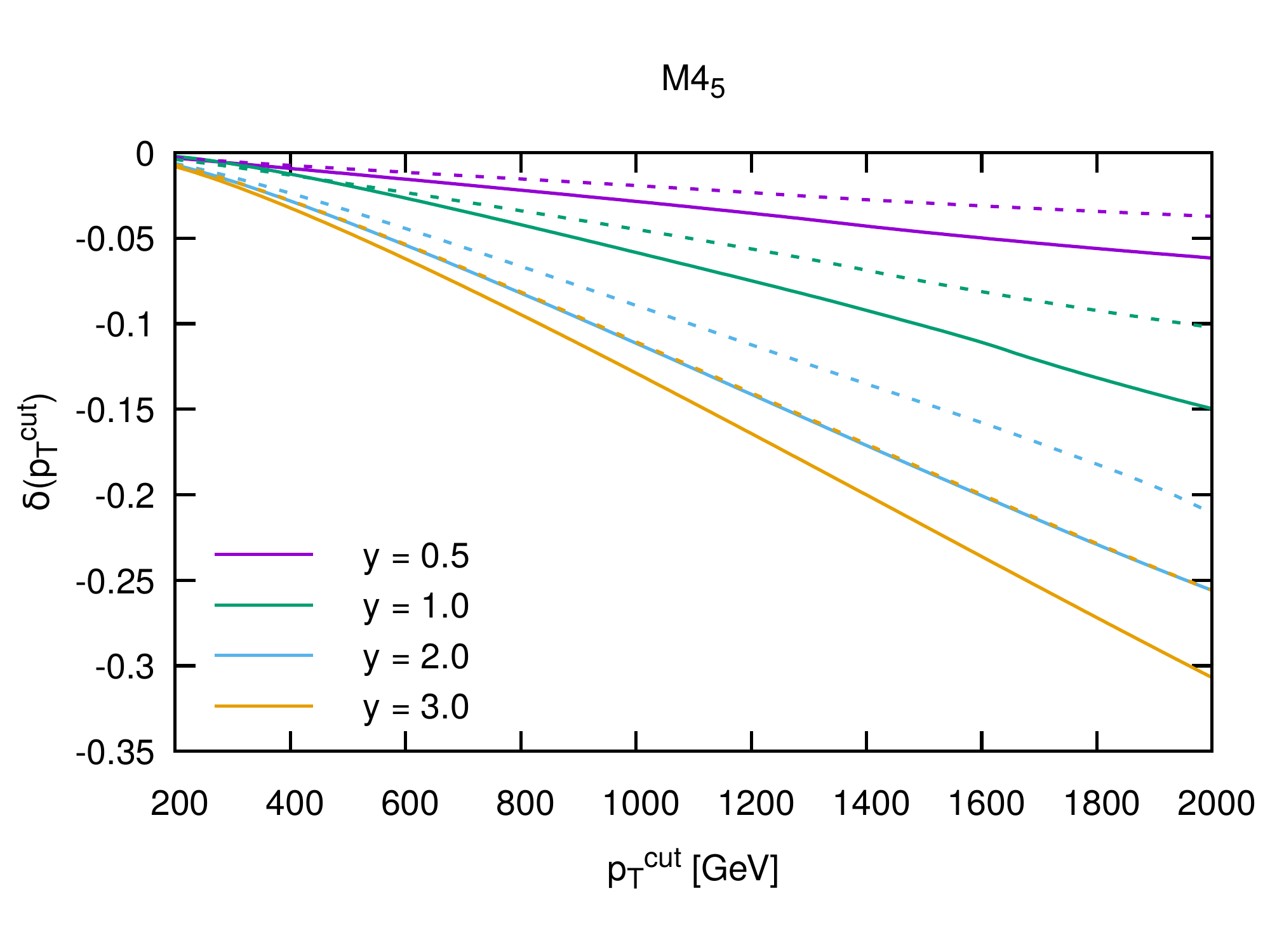}
\end{minipage}
  \begin{minipage}[r]{0.5\linewidth}
\includegraphics[width=.9\textwidth]{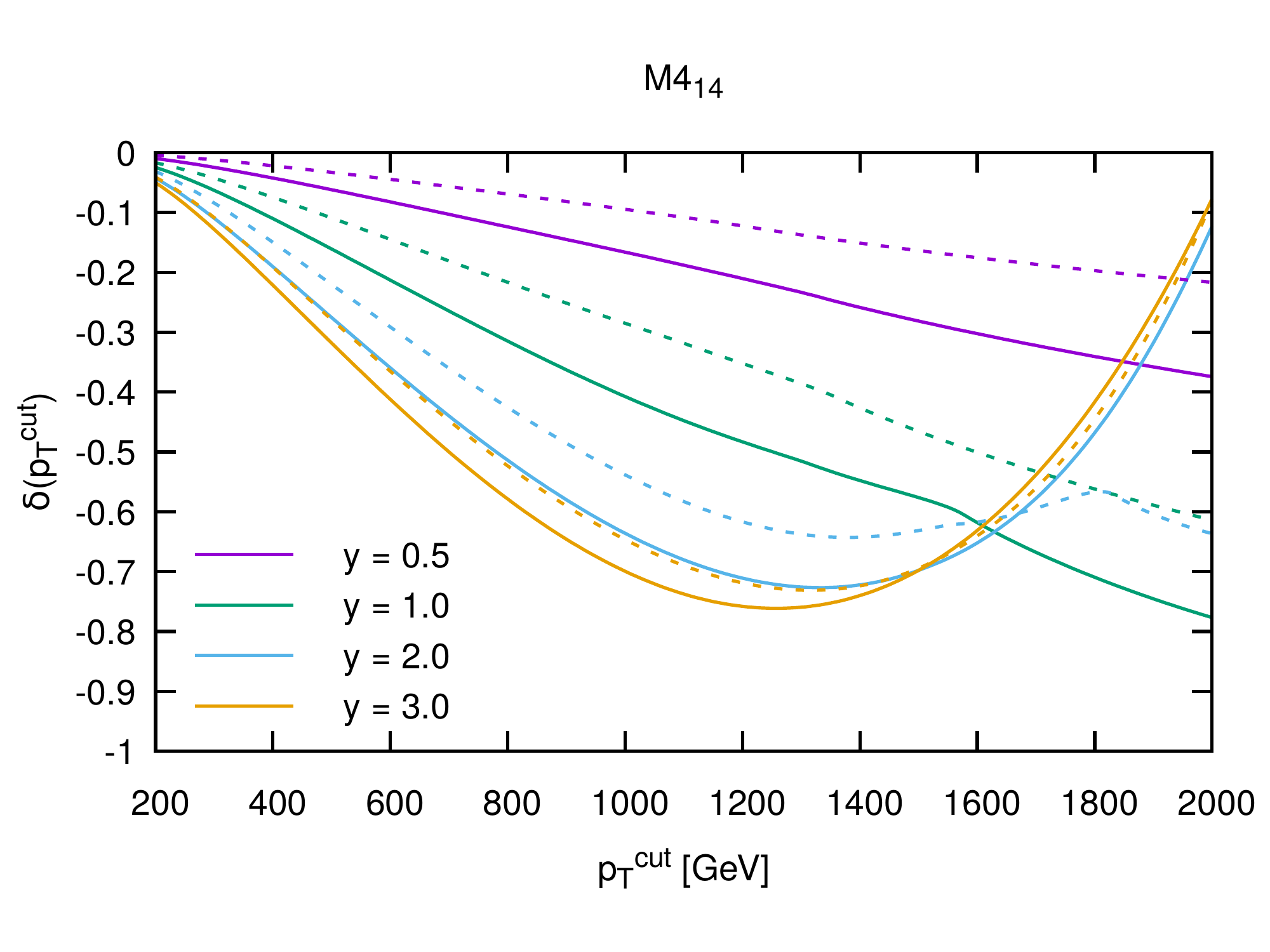}
\end{minipage}
\caption{The distribution $\delta(p_T^{\rm cut})$ for benchmark
  scenario~\ref{yvaries} and the four models considered in
  section~\ref{sec:2tps}.}
  \label{fig:bench-yvaries}
\end{figure}

Benchmark~\ref{yvaries} investigates the effects of varying $y$, the parameter
that determines the compositeness of the top quark, for two
quasi-degenerate vector-like quarks (the solid curves in
Fig.~\ref{fig:bench-yvaries}) and for a single top partner with
$M_{\Psi} =M_{\Psi{1}}$ (the dashed curves in
Fig.~\ref{fig:bench-yvaries}).  For singlet models the experimental constraints on $\kappa_t$ force $y$ to be less than one, so here we present results for fourplet models only, where a larger range of $y$ is allowed.
Fourplet models present a variety of features. In
$\mathbf{M4_5}$ the transverse momentum distribution is suppressed
compared to the SM due to a persistent negative interference between
the top and top-partner contributions, see Fig.~\ref{fig:yDep}. For
$\mathbf{M4_{14}}$ negative interference only dominates when
$y$ is not too big. With increasing $y$ and $\ptcut$, the interference
can become as big as the SM contribution, so the square of the
amplitude with the heavier top partner dominates, see
Eq.~\eqref{eq:M+++-lowpt}. As a consequence, $\delta(\ptcut)$ becomes
positive at large $\ptcut$.

\begin{figure}[htbp]
  \begin{minipage}[l]{0.5\textwidth}
  \includegraphics[width=.9\textwidth]{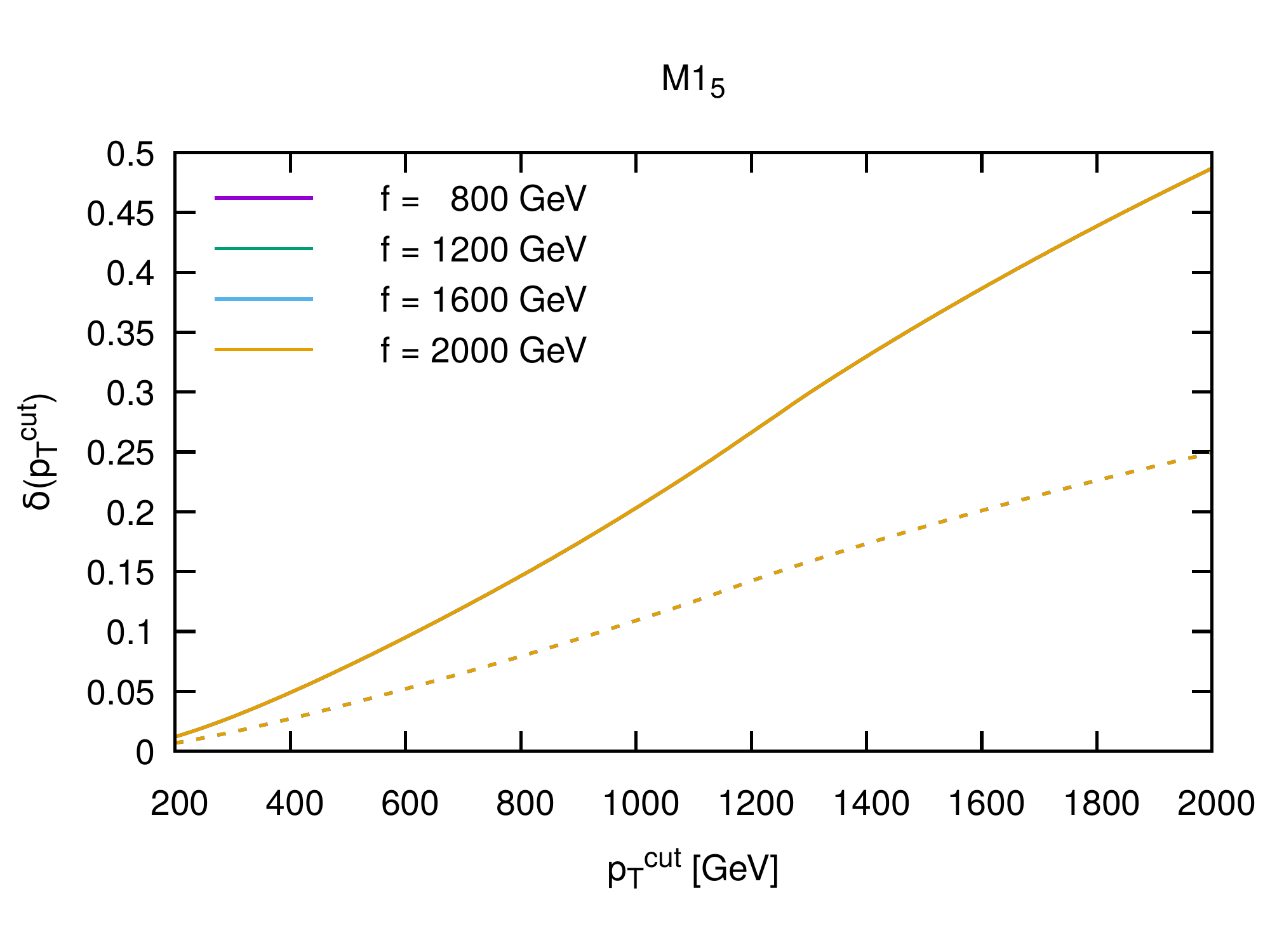}
  \end{minipage}
  \begin{minipage}[r]{0.5\textwidth}
  \includegraphics[width=.9\textwidth]{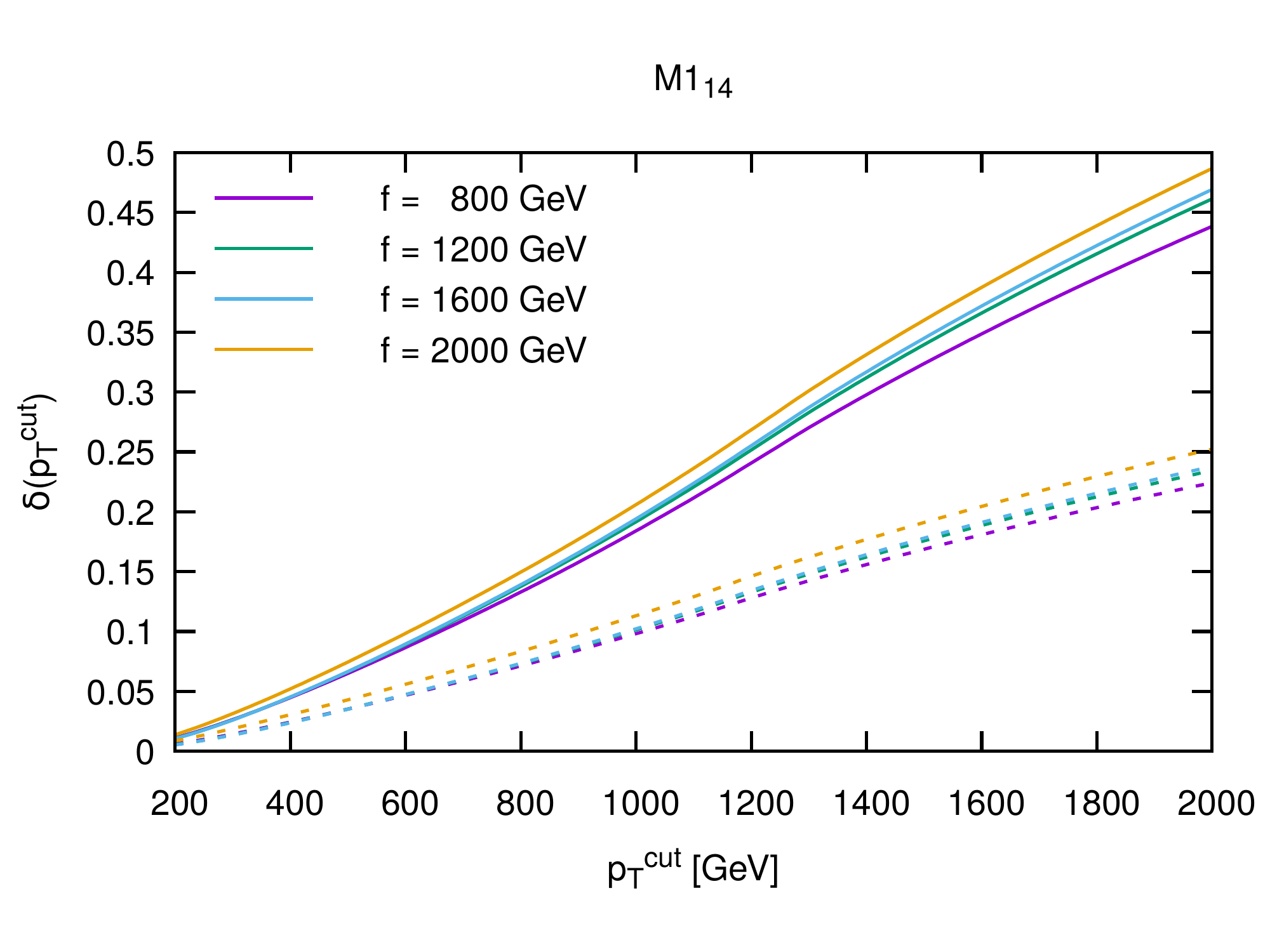}
  \end{minipage}
  \begin{minipage}[l]{0.5\linewidth}
  \includegraphics[width=.9\textwidth]{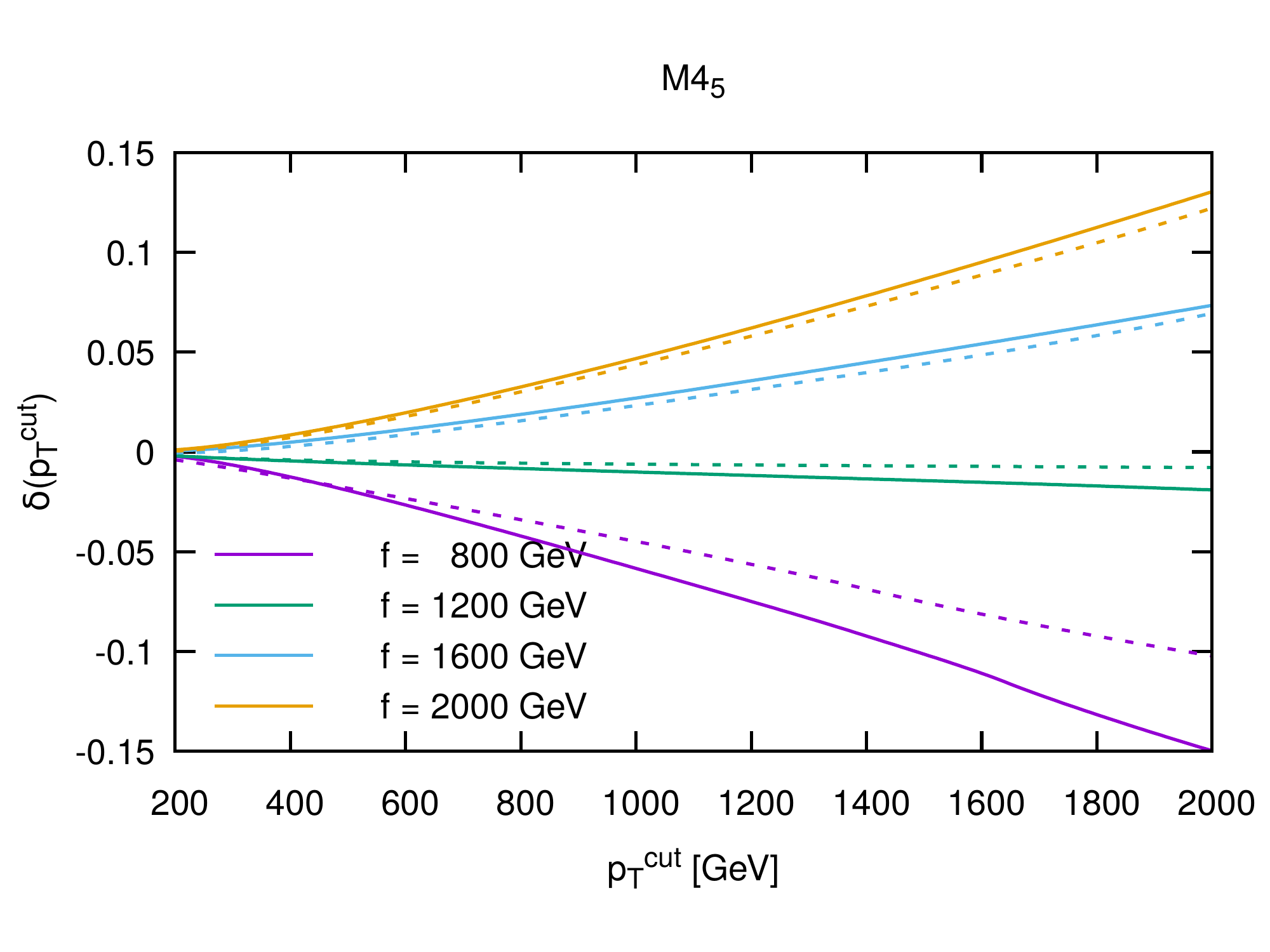}
\end{minipage}
  \begin{minipage}[r]{0.5\linewidth}
\includegraphics[width=.9\textwidth]{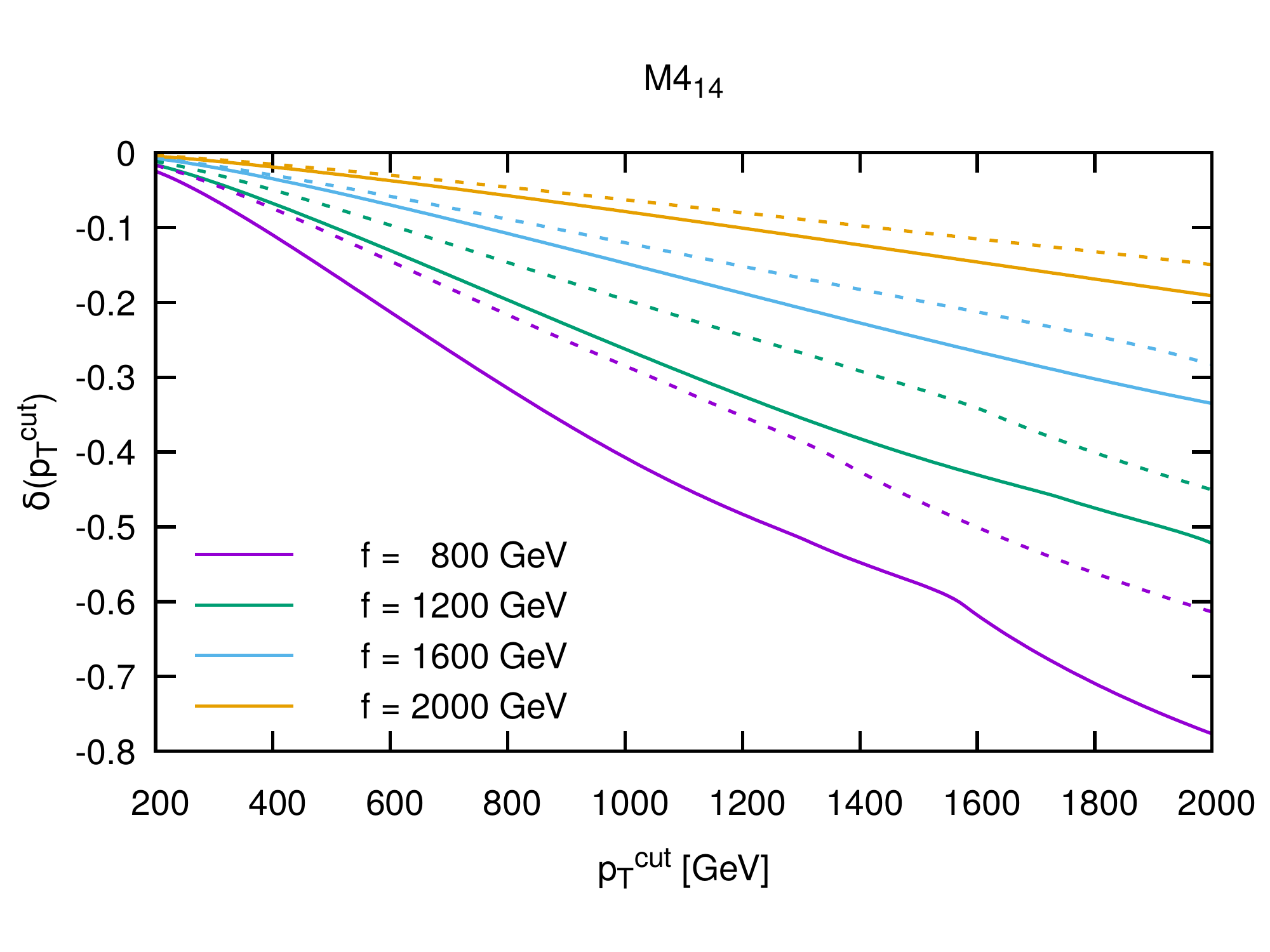}
\end{minipage}
\caption{The distribution $\delta(p_T^{\rm cut})$ for benchmark
  scenario~\ref{fvaries} and the four models considered in section~\ref{sec:2tps}.}
  \label{fig:bench-fvaries}
\end{figure}

Benchmark~\ref{fvaries} investigates the impact of varying the
compositeness scale $f$ in models with two top partners arising from
two vector-like multiplets of similar masses (the solid curves of
Fig.~\ref{fig:bench-fvaries}). For comparison we also consider the
corresponding curves with one top-partner only (the dashed curves of
Fig.~\ref{fig:bench-fvaries}), where $M_{\Psi}=\MPsi{1}$ and the same
values of $y$ and $f$. This is interesting because, as we have shown
in section~\ref{sec:2tps}, in this case the anomalous Yukawa coupling
of the heavier top partner can become larger than that of the lighter
top partner. We then plot $\delta(p_T^{\rm cut})$ as a function of
$p_T^{\rm cut}$ for each of the four models considered and for
selected values of $f$.
For singlet models, the considered values of the parameters lead to
deviations from the standard model that are not too large and are
largely independent of $f$. This can be appreciated for instance by
looking at the Yukawa couplings in the upper panel of
Fig.~\ref{fig:yukTops}, where for singlet models one sees a small
difference in Yukawa couplings when varying the compositeness scale.
It is also interesting to note that we observed the same behaviour in the studies of just one top-partner in Fig. \ref{fig:delta-sth2_0.1-pt_200-LHC14-singlet} and Fig. \ref{fig:delta-sth2_0.1-pt_600-LHC14-singlet} for the singlet top-partner models.
The deviations for the two top-partner case are approximately twice as
large as in the one top-partner case because, as can be seen from
Fig. \ref{fig:yukTops}, both top-partners have Yukawa couplings close
to the coupling in the one top-partner case for the parameters chosen
here.  The situation is more interesting with fourplet models where
for both $\mathbf{M4_5}$ and $\mathbf{M4_{14}}$ one observes negative
deviations from the SM result. This can be understood from the negative
  Yukawa couplings displayed for the fourplet models in
  Fig. \ref{fig:yukTops}. For $\mathbf{M4_5}$, the deviations from
the SM are moderate, but with a huge dependence on the compositeness
scale $f$, as can be inferred for instance by looking at the left
panel of Fig.~\ref{fig:fDep}. In particular, we see that when the
compositeness scales reaches the value of vector-like mass $\MPsi{1}$,
the deviation drops to zero, and becomes positive for larger values of
$\MPsi{1}$.  A large dependence on $f$ is also visible in
$\mathbf{M4_{14}}$ models due to the fact that negative anomalous
Yukawa couplings are larger for smaller values of $f$.

\begin{figure}[htbp]
  \begin{minipage}[l]{0.5\linewidth}
  \includegraphics[width=.9\textwidth]{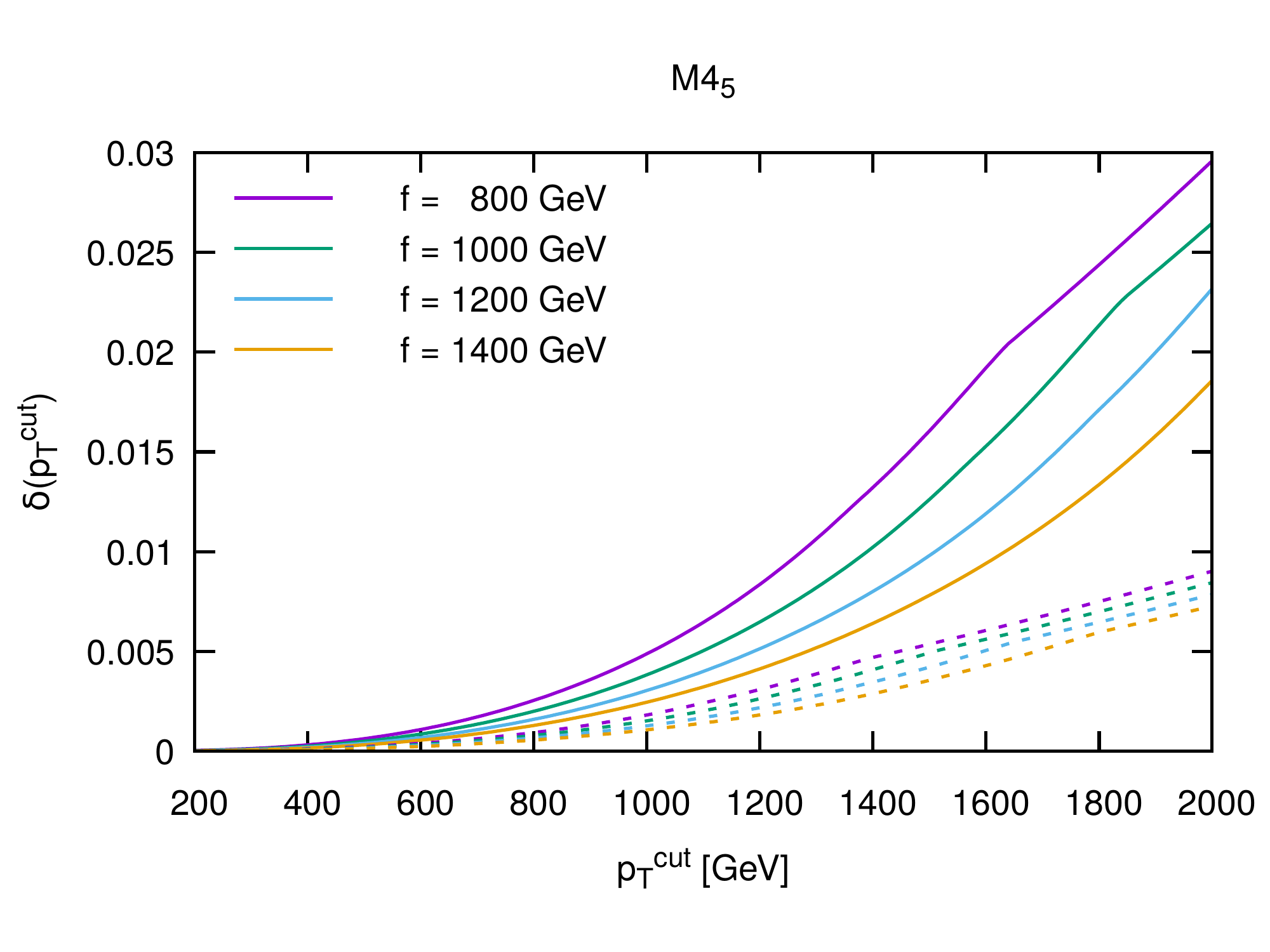}
\end{minipage}
  \begin{minipage}[r]{0.5\linewidth}
\includegraphics[width=.9\textwidth]{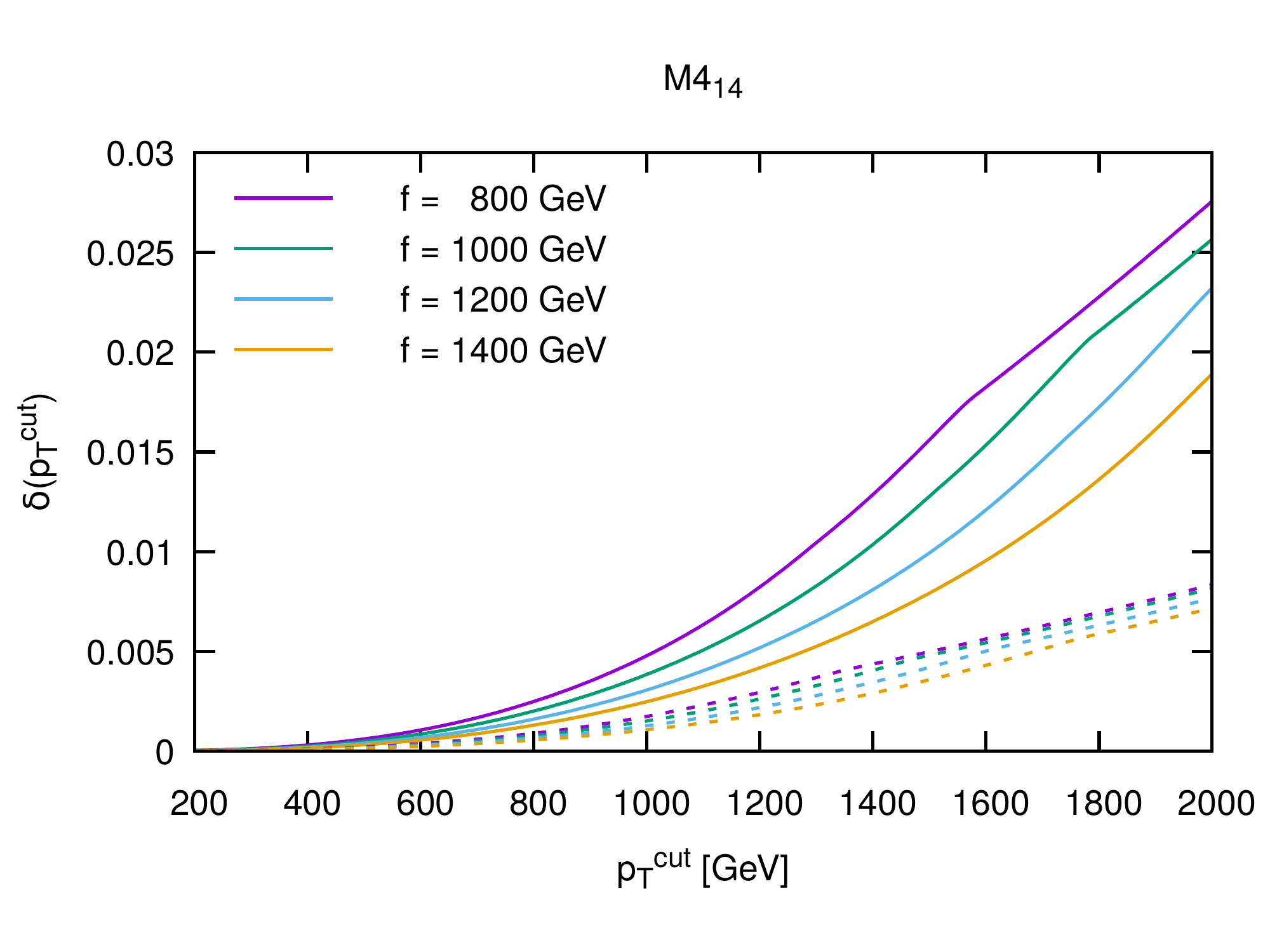}
\end{minipage}
\caption{The distribution $\delta(p_T^{\rm cut})$ for benchmark
  scenario~\ref{fvaries-cpodd} and the four models considered in section~\ref{sec:2tps}.\label{fig:bench-fvaries-cpodd}}
\end{figure}

Lastly,  in Fig.~\ref{fig:bench-fvaries-cpodd}, we investigate the impact of the CP-odd contributions arising in the
fourplet models. As for the one top-partner case we plot only the ratio between the
CP-odd contribution and the SM efficiency, because the CP-odd terms do not
interfere with the SM amplitude. The expected deviations from the SM
are quite small, less than 10\% for all considered values of
$\ptcut$ and slightly larger for $\mathbf{M4_{14}}$ than for
$\mathbf{M4_{5}}$.
Remarkably, as can be appreciated from Fig.~\ref{fig:CPo2}, deviations
are roughly twice as big for the two-top partner case than for the
case with one top-partner. This is a peculiarity of the fact that the
vector-like masses are very close to each other. If the larger
increases, then the CP-odd Yukawa coupling tends to that of a single
top partner. These effects can be better understood by observing the
dependence of the CP-odd couplings on the underlaying parameters in
Figs.~\ref{fig:CPo1} and~\ref{fig:CPo2}, where we see that the CP-odd
couplings in the two top-partner case are considerably larger than in
the one top-partner case for the parameter choices in
Fig. \ref{fig:bench-fvaries-cpodd}.

To summarise this section, models with two top partners exhibit a variety of
deviations from the SM in the $p_T$ spectrum of the Higgs boson or a
jet. The fact that the deviation depends strongly on $p_T^{\rm cut}$
suggests that the best way to exclude large fractions of parameter
space for composite Higgs models is a shape analysis of the $p_T$
distribution itself. Such an analysis would require appropriate modelling
of the irreducible SM background to Higgs production, including
detailed acceptance cuts and experimental systematic uncertainties for
the decay products of the Higgs. Though very interesting, this is
beyond the scope of the present work.
Another interesting study might be that of using the most recent
  LHC data on the Higgs $p_T$ cross
  section~\cite{Sirunyan:2018sgc,Aaboud:2018ezd,ATLAS:2018bsg,ATLAS:2018uso}
  to constrain the parameters of the various composite Higgs models studied here. However, to describe the
  Higgs cross section one would need to also include the modified coupling of the
  Higgs, namely to
  $\gamma \gamma$ and $ZZ$, relative to the decay channels considered there. This requires further studies for each
  composite Higgs model, and is again beyond the scope of what is presented
  here.

\section{Conclusions}

\noindent In this paper we have explored Higgs+Jet production in various composite Higgs scenarios at the LHC.
The presence of light top-partner states and deviations in the Higgs couplings due to its pseudo-Goldstone boson nature both make important contributions to this process.
It is already well known that in single-Higgs production the rate is insensitive to the mass spectrum of top-partner states, and thus studies of top-partner effects in Higgs+Jet production have been performed to highlight the sensitivity of this process to the top-partner mass spectrum.
In this paper we have extended the study of Higgs+Jet production in composite Higgs scenarios in several ways; we start by considering four different embeddings of the top and top-partner states in the global symmetry and study the corrections to the Yukawa couplings, we then add an additional light top-partner multiplet and study the Yukawa couplings and masses of the top quark and its partners, and lastly we study the effects of the different top-partner representations and the effects of an additional light top-partner multiplet on the Higgs+Jet rate.

In Section \ref{sec:background} we begin with an overview of the
$SO(5)/SO(4)$ composite Higgs model and the different top-partner
embeddings that we consider.  We also discuss the inclusion of
additional light top-partner states in the EFT.  In Section
\ref{sec:massyuk} we study the mass spectra and Yukawa couplings of
the top and top-partner states, and compare the results obtained with
different top-partner embeddings and with an additional light
top-partner.  An interesting result presented here is that in the
parameter space that we study the singlet top-partner models are more
constrained than the fourplet top-partner models due to how the
off-diagonal terms in the mass matrix scale with the decay constant
$f$.  The constraints are essentially on the allowed mixing angles,
and arise both from perturbativity of the coupling $y$ and from recent
bounds on the top Yukawa coupling from the measurement of $t\bar{t}h$.
We give an overview of the Higgs production process in Section
\ref{sec:HiggsProd}, both for single-Higgs and Higgs+Jet production.
Here we highlight the well-known and intriguing fact that in
single-Higgs production the rate calculated at leading order is
insensitive to the mass scale of the top-partners in the loop.  We
also discuss the $p_{T}$ regions in which the amplitude is sensitive
to the masses and couplings of the quarks in the loop.  In Sections
\ref{sec:onetpres} and \ref{sec:twotpres} we present our results for
scenarios with one and two light top-partners in the spectrum,
respectively, in terms of the net Higgs+Jet efficiency.  In the case
of just one light top-partner we clearly demonstrate the different
ways in which the efficiency scales with the top-partner mass, the
decay constant, and the mixing angles for the different top and
top-partner embeddings in $SO(5)$.  For some regions of parameter
space these deviation from the SM, as measured by the net Higgs+Jet
efficiency, are large and could present promising signals for future
collider studies.  With an additional light top-partner we show that
the deviations from the SM can be much larger than with just a single
top-partner, and that the best way to probe the parameter space of the
model using the Higgs+Jet signal would be through a shape analysis of
the $p_T$ distribution of the Higgs, or better the corresponding
efficiency.  We have also studied the contributions of the CP-odd
couplings on the Higgs+Jet rate, however while remaining within
constraints set by other experiments we find that these contributions
are typically small.
Through this work we have highlighted how at high transverse momentum the Higgs+Jet process could be used to study the top-partner spectrum in composite Higgs models, and how the results could provide insight as to the embedding of these states in the global symmetries of the strong sector.
Following on from this work a detailed study of how to search for these deviations at the LHC is required.

\appendix

\section{$\mathbf{CP}$-odd contribution to Higgs plus one jet in relevant limits }
\label{sec:CP-odd}

\noindent In this appendix we report the expression for the CP-odd
contribution to Higgs+Jet production and perform checks in three
relevant limits, following the strategy of
ref.~\cite{Banfi:2018pki}. In section~\ref{sec:decoupling} we discuss
the limit in which the mass of the fermion running in the loop is the
largest scale. In section~\ref{sec:soft} we consider the limit in
which the outgoing gluon is soft. And finally in
section~\ref{sec:collinear} we discuss the case in which the outgoing
parton is collinear to beam direction. This information constitutes an
important validation tool for our implementation of the calculation of
ref.~\cite{Grojean:2013nya}.

We first need the expression of the Born matrix element. Due to
conservation of angular momentum, the amplitude for the process
$gg\to h$ is non-zero only if the two gluons have opposite helicities.
The un-averaged matrix element squared for this process is
\begin{align}
| M_{gg\to h}|^2 &= \frac{(N_c^2 - 1) \alpha_s^2 \tilde{\kappa}^2 m_H^4}{4\pi^2 v^2} \ \left| \sum_{i=t,T^1,T^2} \mathcal M_{+-}^{i} \right|^2 \,.
\label{eq:squaredborn}
\end{align}
The index $i$ here refers to the particle running in the loop needed
to couple the gluons to the Higgs. The top quark contribution to the
above equation is
\begin{equation}
  \label{eq:Born-amplitude}
  \mathcal{M}^{i}_{+-} = m_i^2 C_0(m_H^2)\,.
\end{equation}
With this we can report the expression for the matrix element squared
for Higgs+Jet production in the various partonic channels contributing
to this process: $gg \to hg,~ q\bar q\to hg,~ qg\to hq, qh$.

The $gg \to hg$ amplitude can be expressed in terms of eight primitive
helicity amplitudes $\mathcal{M}_{h_1 h_2 h_3}$ corresponding to the
possible choices for each gluon helicity $h_i=\pm$. We use the
convention that the momenta of gluons $p_1$ and $p_2$ are incoming,
and that of gluon $p_3$ is outgoing, so that the Mandelstam variables,
in the convention of ref.~\cite{Grojean:2013nya}, are defined as
\begin{align}
s = (p_1 + p_2)^2\,, \quad
t =(p_1 - p_3)^2\,, \quad
 u = (p_1 - p_4)^2\,. 
\end{align}
The helicity amplitudes are then related to the full, un-averaged
amplitude squared via
\begin{align}
  | M_{gg\to Hg}|^2 &= \frac{N_c (N_c^2 - 1) \alpha_s^3 \tilde{\kappa}^2}{8\pi v^2}
  \sum_{h_1,h_2,h_3 = \pm} \left| \sum_{i=t,T^1,T^2}
    \mathcal M^i_{h_1 h_2 h_3} \right|^2 \,.
\label{eq:squaredamp}
\end{align}
After applying parity and crossing
symmetry, only four of the helicity amplitudes are independent, which
we take to be
$\mathcal M^{i}_{+++}, \mathcal M^{i}_{++-}, \mathcal
M^{i}_{-+-}, \mathcal M^{i}_{-++}$. 

The contributions to the helicity amplitudes due to loops containing a
fermion with mass $m$ and coupling to the Higgs
$\tilde\kappa$, are:
\begin{equation}
\label{eq:Mggg}
  \begin{split}
    \mathcal M^{i}_{+++} &= m_i^2  F_1(s,t,u)\,,    \\
    \mathcal M^{i}_{++-} &= m_i^2  F_1(s,u,t)\,,    \\
    \mathcal M^{i}_{-+-} &= m_i^2  F_2(s,t,u)\,,    \\
    \mathcal M^{i}_{-++} &= m_i^2  F_3(s,t,u)\,,    
  \end{split}
\end{equation}
where
\begin{align}
  \label{eq:Fi}
  F_1(s,t,u) & = \sqrt{\frac{t}{s u}}\left[G(s,t)-G(s,u)+G(t,u)\right]\,, \\
  F_2(s,t,u) & = - \frac{m_H^2}{ \sqrt{s t u}}\left[G(s,t)+G(s,u)+G(t,u)\right]\,, \\
  F_3(s,t,u) & = \sqrt{\frac{s}{t u}}\left[G(s,t)+G(s,u)-G(t,u)\right]\,,
\end{align}
and
\begin{equation}
  \label{eq:Gxy}
  G(x,y) = xy D_0(x,y)+ 2 x C_1(y) + 2 y C_1(x)\,.
\end{equation}

The functions $B_1, C_1, D_0$ are 1-loop basic scalar integrals. They
are functions of $(s,t,u)$, the mass of the particle in the loop, and
the Higgs mass; their definitions can be found in~\cite{Passarino:1978jh}. 

The other $pp \to hj$ subprocesses $(q\bar q \to h g,\, qg \to
hq,\, \bar q g \to h \bar q)$ are controlled by a third function,
the un-averaged amplitude squared
\begin{align}
  \sum | M_{q\bar{q}\to Hg}|^2(s,t,u)  &= \frac{2(N_c^2-1)\alpha_s^3\tilde{\kappa}^2}{\pi v^2 }\frac{t^2 + u^2}{s} \left| \sum_{i= t,T^{1},T^{2}} \mathcal M^i(q \bar{q} \to h g) \right|^2 \,.
\label{eq:squaredampquarks}
\end{align}
The amplitude for one fermion in the loop is given by
\begin{align}
  \mathcal M^{i}(q\bar q \to h g) &= m_i^2 C_1(s)\,.
\label{eq:qqbar}
\end{align}
We can get the amplitudes for the subprocesses $qg \to hq$ and
$ gq \to h q$ from the above expression by swapping the Mandelstam variable $s$
and $t$, and $s$ and $u$ respectively.

\subsection{Decoupling limit}
\label{sec:decoupling}

\noindent Here we give analytical expressions for the helicity amplitudes
introduced in section~\ref{sec:CP-odd} in the ``decoupling'' limit ($m^2\gg
m_H^2,s,|t|,|u|$) where $m$ is the mass of the fermion running in the loops. 
First, we give the expansion of the scalar integrals appearing in the
amplitudes:
\begin{equation}
  \label{eq:scalar-decoupling}
  \begin{split}
   B_1(q^2) \simeq \frac{q^2-m_H^2}{6 m^2}\,,\quad
  C_1(q^2) \simeq -\frac{1}{2 m^2}-\frac{q^2+m_H^2}{24 m^4}\,,\quad D_0(s,t)  \simeq \frac{1}{6m^4}\,.
  \end{split}
\end{equation}
This gives
\begin{equation}
\label{eq:M-decouple}
\begin{split}
  M^{i}_{+++} &\simeq-2 t \sqrt{\frac{t}{su}} \,,\qquad 
 M^{i}_{++-} \simeq -2u\sqrt{\frac{u}{st}}
  \,,\\
  M^{i}_{-+-} &\simeq 2 \frac{m_H^4 }{\sqrt{stu}} \,,\qquad 
  M^{i}_{-++} \simeq-2s\sqrt{\frac{s}{tu}} \,.  
\end{split}
\end{equation}
Similarly,
\begin{equation}
  \label{eq:Mq-decoupling}
  \mathcal{M}^{i}(q\bar q\to hg)\simeq\frac{(N_c^2-1)\alpha^3_s\tilde{\kappa}^2}{2\pi v^2}\frac{t^2+u^2}{s} \,,
\quad  \mathcal{M}^{i}(qg\to hq)\simeq-\frac{(N_c^2-1)\alpha^3_s\tilde{\kappa}^2}{2\pi v^2}\frac{s^2+u^2}{t}  \, ,
\end{equation}
\begin{equation}
 \mathcal{M}^{i}(gq\to hq)\simeq-\frac{(N_c^2-1)\alpha^3_s\tilde{\kappa}^2}{2\pi v^2}\frac{t^2+s^2}{u} \,.
\end{equation}

\subsection{Soft limit}
\label{sec:soft}

\noindent The soft limit $p_3 \to 0$ corresponds to 
\begin{equation}
  \label{eq:soft-limit}
  s \to m_H^2\,,\quad t,u\to 0\,. 
\end{equation}
In the soft limit amplitudes are proportional to the tree-level
amplitude $M_{-+}$, therefore we get a non-zero contribution only
from $\mathcal{M}_{-+-}$ and $\mathcal{M}_{-++}$.
Keeping the most relevant terms in this limit,~Eq.~\eqref{eq:Mggg} gives
\begin{equation}
  \label{eq:M-+-soft}
  \begin{split}
  \mathcal{M}^{i}_{-+-}& \simeq-\frac{ m_i^2 m_H^2}{\sqrt{s t u}}(st D_0(s,t) + s u D_0(s,u)+ t u D_0(t,u) +2 s C_1(t)+2 s C_1(u)) \,.
  \end{split}
\end{equation}
In the soft limit the relevant integral limits are
\begin{equation}
  \label{eq:soft-integrals}
  tC_0(t)\to 0\,,\quad uC_0(u)\to 0\,,\quad st D_0(s,t)\to 0\,,\quad us D_0(u,s)\to 0  \,,\quad ut D_0(u,t)\to 0\,,
\end{equation}
which gives
\begin{equation}
  \label{eq:M-+-soft-fine}
  \begin{split}
 \mathcal{M}^{i}_{-+-}  &\simeq-4m_i^2m_H^2\sqrt{\frac{s}{t u}} C_0(m_H^2)\simeq-(\sqrt{2})^3\sqrt{\frac{s}{tu}}\mathcal{M}^{i}_{+-} \,.
\end{split}
\end{equation}
Similarly, the other helicity amplitude~Eq.~\eqref{eq:Mggg} becomes
\begin{equation}
  \label{eq:M-++soft}
  \begin{split}
  \mathcal{M}^{i}_{-++}& \simeq m_i^2\sqrt{\frac{s}{t u}}(st D_0(s,t) + s u D_0(s,u)- t u D_0(t,u) +2 s C_1(t)+2 s C_1(u))
\,.
  \end{split}
\end{equation}
Evaluating again all scalar integrals in the soft limit we get
\begin{equation}
  \label{eq:M-++soft-fine}
  \begin{split}
 \mathcal{M}^{i}_{-++}  &\simeq 4 m_i^2m_H^2 \sqrt{\frac{s}{t u}} C_0(m_H^2)\simeq(\sqrt{2})^3\sqrt{\frac{s}{tu}}\mathcal{M}^{i}_{+-} \,.
\end{split}
\end{equation}
These expressions have to be compared with the universal behavior of
helicity amplitudes
\cite{Mangano:1990by,Dixon:1996wi}\footnote{The $\sqrt{2}$
  factors comes from the differing normalisation factors for gauge
  group generators $\operatorname{tr}[T^a T^b] = \delta^{ab}$ in the
  spinor helicity formalism, compared to the usual
  $\operatorname{tr}[T^a T^b] = \frac{1}{2}\delta^{ab}$. This is
  compensated by a relative $\sqrt{2}$ factor associated to the gauge
  coupling.},
\begin{equation}
  \label{eq:Msoft-universal}
  \begin{split}
     \mathcal{M}^{i}_{-+-} & = -(\sqrt{2})^3\frac{[ p_1 p_2]}{[ p_1 p_3] [ p_3 p_2]}  \mathcal{M}^{i}_{+-}\,,\\
    \mathcal{M}^{i}_{-++} & = (\sqrt{2})^3 \frac{\langle p_1 p_2\rangle}{\langle p_1 p_3
      \rangle \langle p_3 p_2 \rangle}  \mathcal{M}^{i}_{+-}\,. \\
  \end{split}
\end{equation}
Since we have not used the spinor-helicity formalism, it is not
immediate to rephrase our expressions in terms of helicity
products. However, for real momenta, spinor products are simply equal
to the square root of the relevant momentum invariant, up to a
phase. The universal soft factor has an implicit helicity set by the
helicity of the soft gluon, and so the choice of translating to angle
or square bracket spinor products is fixed by this. We then obtain
from Eq.~\eqref{eq:M-+-soft-fine} and Eq.~\eqref{eq:M-++soft-fine} that
$\mathcal{M}^{i}_{-+-}$ and
$\mathcal{M}^{i}_{-++}$ have the correct
behavior (i.e. Eq.~\eqref{eq:Msoft-universal}) in the soft limit, modulo an
overall phase that depends on the gluon helicity. This phase is the
same for all the particles running in the loop, and therefore can be factored out
of each helicity amplitude and will not contribute to the amplitude
squared.

\subsection{Collinear limits}
\label{sec:collinear}

\noindent We now consider the collinear limit $t\to 0$ where $p_1$ becomes collinear to $p_3$. Introducing the splitting fraction $z = \frac{m_H^2}{s}$, the invariants take the limiting values
\begin{equation}
  \label{eq:u->0}
  t\to 0\,,\quad s =\frac{m_H^2}{z}\,,\quad u\to -\frac{1-z}{z}m_H^2\,.
\end{equation}
In this limit $tC_0(t) \to 0$, whereas $sC_0(s)$ and $uC_0(u)$ stay finite. For the box integrals, we have
\begin{equation}
  \label{eq:Box-collinear}
  su D_0(s,u)\to 2\left[sC_0(s)+uC_0(u)-m_H^2 C_0(m_H^2)\right]\,,\quad stD_0(s,t)\to 0\,,\quad tu D_0(t,u)\to 0\,. 
\end{equation}
In this limit we get
\begin{equation}
  \label{eq:M-+-collinear}
  \begin{split}
    \mathcal{M}^{i}_{-+-}&\simeq-\frac{2m_i^2 m_H^2(m_H^2+s+u)}{\sqrt{stu}}C_0(m_H^2) \simeq \frac{4 m_i^2 m_H^2 z}{\sqrt{(1-z)}\sqrt{-t}}C_0(m_H^2)  \,.
  \end{split}
\end{equation}
Similarly, for the other helicity configuration we obtain
\begin{equation}
  \label{eq:M-++collinear}
  \begin{split}
    \mathcal{M}^{i}_{-++}&\simeq\frac{2m_i^2 \sqrt{s} (m_H^2+s-u)}{\sqrt{tu}}C_0(m_H^2) \simeq -\frac{4 m_i^2 m_H^2 }{z\sqrt{(1-z)}\sqrt{-t} }C_0(m_H^2) \,.
  \end{split}
\end{equation}
Now in the collinear case the limit depends on the helicity of each collinear leg. This means that there are two more possibilities to consider, and therefore we should also look at the limit of the two helicity
amplitudes $ \mathcal{M}^{i}_{++-}$ and $ \mathcal{M}^{i}_{+++}$. For the first we have 
\begin{equation}
  \label{eq:M++-collinear}
  \begin{split}
     \mathcal{M}^{i}_{++-} &\simeq \frac{2m_i^2\sqrt{u} (m_H^2-s+u)}{\sqrt{st}}C_0(m_H^2) \simeq -\frac{4m_i^2 m_H^2 (1-z)^{3/2}}{ z\sqrt{-t}}C_0(m_H^2)\,
  \end{split}
\end{equation}
and for the second we have
\begin{equation}
  \label{eq:M+++collinear}
  \begin{split}
   \mathcal{M}^{i}_{+++} &\simeq\frac{2m_i^2\sqrt{t} (-m_H^2+s+u)}{\sqrt{su}}C_0(m_H^2) \simeq 0\,.
  \end{split}
\end{equation}
Collecting all results we have
\begin{equation}
  \label{eq:M-coll-all}
  \begin{split}
 \mathcal{M}^{i}_{-++} & \simeq  \frac{-(\sqrt{2})^3}{z\sqrt{(1-z)}\sqrt{-t}}\mathcal{M}^{i}_{+-}\,,\\
 \mathcal{M}^{i}_{-+-}&\simeq  \frac{z(\sqrt{2})^3}{\sqrt{(1-z)}\sqrt{-t}}\mathcal{M}^{i}_{+-}\,, \\
 \mathcal{M}^{i}_{++-} & \simeq
\frac{-(1-z)^2(\sqrt{2})^3}{z\sqrt{(1-z)}\sqrt{-t}}\mathcal{M}^{i}_{+-}\,, \\
 \mathcal{M}^{i}_{+++}& \simeq 0\,.
\end{split}
\end{equation}
To check the correctness of the above limits, we have to translate our
conventions for the helicity and the splitting fraction into those
available in the literature, in which all momenta are considered to be
outgoing. First, we need to flip the helicity of each incoming
particle. Additionally, the relation of $z$ to the momenta is
different when the collinear gluons are outgoing. One can switch
between the two cases by making the replacement $z \to \frac{1}{z}$.
Adopting the usual convention of associating negative momentum signs
to angle spinors we expect the behaviour
\cite{Mangano:1990by,Dixon:1996wi}
\begin{equation}
 \label{eq:M-coll-universal}
 \begin{split}
\frac{\mathcal{M}^{i}_{-++}}{\mathcal{M}^{i}_{+-}}& \simeq \operatorname{Split}_{+}\left(-1^-,3^+;\frac{1}{z}\right)= \frac{-(\sqrt{2})^3}{z\sqrt{1-z}\langle p_1
     p_3\rangle}\,, \\
\frac{\mathcal{M}^{i}_{-+-}}{\mathcal{M}^{i}_{+-}}&\simeq \operatorname{Split}_{+}\left(-1^-,3^-;\frac1z\right)= \frac{z(\sqrt{2})^3}{\sqrt{1-z}[p_1 p_3]} \,, \\
\frac{\mathcal{M}^{i}_{++-}}{\mathcal{M}^{i}_{+-}}& \simeq \operatorname{Split}_{+}\left(-1^+,3^-;\frac 1z\right)= \frac{-(1-z)^2(\sqrt{2})^3}{z\sqrt{1-z}\langle p_1
     p_3\rangle}\,,\\
\frac{\mathcal{M}^{i}_{+++}}{\mathcal{M}^{i}_{+-}}& \simeq \operatorname{Split}_{+}\left(-1^+,3^+;\frac 1z\right) = 0\,.
\end{split}
\end{equation}
We must now translate Eq.~\eqref{eq:M-coll-all} to helicity language. The
translation from Mandelstam variables to spinor invariants is similar
to the soft case, although the helicity consideration is slightly
subtler. As the three legs of the splitting amplitude are collinear,
we no longer have information about the contribution from each
individual leg, as the helicity spinors become proportional. Instead
what matters is the overall (outgoing) helicity of the three, which
governs whether it is appropriate to translate to angle or square
brackets, and with this consideration we indeed find the correct
momentum dependence. However, this is not relevant in the end because,
up to an overall phase
$[p_1 p_3]\sim \langle p_1 p_3\rangle\sim \sqrt{-t} $.

\section*{Acknowledgements}

\noindent
AB is grateful to Giuliano Panico and Andrea Wulzer for useful
discussions on composite Higgs models. We also thank Alexander Belyaev
and Sebastian J\"ager for pointing out relevant issues on
perturbativity and CP-odd contributions.  BMD acknowledges the
financial support from the Slovenian Research Agency (research core
funding No. P1-0035 and J1-8137). SK acknowledges the studentship
jointly funded by a Weizmann-UK Making Connections grant and the
School of Mathematical and Physical Sciences at the University of
Sussex. The work of AB is partly supported by Science and Technology
Facility Council under the grant ST/P000738/1.  WK acknowledges the
financial support form the Royal Thai Government.

\bibliographystyle{JHEP}
\bibliography{current}

\end{document}

%% file: FeynTikZ.tex
\usepackage{tikz}
\usetikzlibrary{arrows,positioning,shapes,mindmap}
\usetikzlibrary{decorations.pathmorphing}
\usetikzlibrary{decorations.markings}

\makeatletter
\renewcommand{\boxed}[1]{\textcolor{orange}{%
\tikz[baseline={([yshift=-0ex]current bounding box.center)}] \node [rectangle, minimum width=0ex,draw] {\normalcolor\m@th$\displaystyle#1$};}}
 \makeatother

\makeatletter
\def\CenterMiddlePoint#1#2{%
  \expandafter\ifx\csname tikz@special@arrow@end#2\endcsname\relax
    \pgfsetarrowsend{#2}
  \else%
    \pgfsetarrowsend{\csname tikz@special@arrow@end#2\endcsname}%
  \fi%
  \pgf@x=0pt\pgf@y=0pt%
  \expandafter\csname pgf@arrow@left@#2\endcsname%
  \pgf@y=\pgf@x\pgf@x=0pt%
  \expandafter\csname pgf@arrow@right@#2\endcsname%
  \pgftransformxshift{0.5\pgf@y-0.5\pgf@x}%
  \pgftransformxscale{#1}
  \pgflowlevelsynccm%
  \pgflowlevelobj{}{\pgf@endarrow}%
}
\makeatother

\tikzstyle{every picture}+=[remember picture]
\pgfdeclaredecoration{asines}{initial}
{
    \state{initial}[
        width=+0pt,
        next state=sine,
        persistent precomputation={\pgfmathsetmacro\matchinglength{
            \pgfdecoratedinputsegmentlength / int(\pgfdecoratedinputsegmentlength/\pgfdecorationsegmentlength)}
            \setlength{\pgfdecorationsegmentlength}{\matchinglength pt}
        }] {}
    \state{sine}[width=\pgfdecorationsegmentlength]{
        \pgfpathsine{\pgfpoint{0.25\pgfdecorationsegmentlength}{0.5\pgfdecorationsegmentamplitude}}
        \pgfpathcosine{\pgfpoint{0.25\pgfdecorationsegmentlength}{-0.5\pgfdecorationsegmentamplitude}}
        \pgfpathsine{\pgfpoint{0.25\pgfdecorationsegmentlength}{-0.5\pgfdecorationsegmentamplitude}}
        \pgfpathcosine{\pgfpoint{0.25\pgfdecorationsegmentlength}{0.5\pgfdecorationsegmentamplitude}}
}
    \state{final}{}
}
\tikzset{
fermion/.style={very thick,draw=black, line cap=round, postaction={decorate},
    decoration={markings,mark=at position 0.65 with {\arrow[black]{triangle 45}}}},
photon/.style={very thick, line cap=round,decorate, draw=black,
    decoration={asines,amplitude=4pt, segment length=6pt}},
boson/.style={very thick, line cap=round,decorate, draw=black,
    decoration={asines,amplitude=4pt,segment length=8pt}},
gluon/.style={very thick,line cap=round, decorate, draw=black,
    decoration={coil,aspect=1,amplitude=3pt, segment length=8pt}},
scalar/.style={dashed,very thick,line cap=round, decorate, draw=black},
ghost/.style={dotted,very thick,line cap=round, decorate, draw=black},
->-/.style={decoration={
  markings,
  mark=at position 0.6 with {\arrow{>}}},postaction={decorate}}
 }

\makeatletter
\tikzset{
    position/.style args={#1 degrees from #2}{
        at=(#2.#1), anchor=#1+180, shift=(#1:\tikz@node@distance)
    }
}
\makeatother